\documentclass[12pt]{article}
\usepackage{cite,epsfig,amssymb,amsmath,epsfig,chngcntr}

\counterwithin{equation}{section}

\newcommand{\mycomm}[1]{\hfill\break $\phantom{a}$\kern-3.5em{\tt===$>$ \bf #1}\hfill\break}
\newcommand{\mycommA}[1]{\hfill\break $\phantom{a}$\kern-3.5em{\tt   $>$ \bf #1}\hfill\break}

\newcommand{\be}{\begin{equation}}
\newcommand{\ee}{\end{equation}}
\newcommand{\ba}{\begin{eqnarray}}
\newcommand{\ea}{\end{eqnarray}}

\def\MSbar{\hbox{\tiny ${\overline{\rm MS}}$}}

\def\IR{\hbox{\tiny IR}}
\def\DY{\hbox{\tiny DY}}
\def\GB{\hbox{\tiny GB}}
\def\BZ{\hbox{\tiny BZ}}
\def\QD{\hbox{\tiny QD}}

\def\Mink{\hbox{\tiny Mink}}
\def\Eucl{\hbox{\tiny Eucl}}

\def\DIS{\hbox{\tiny DIS}}
\def\APT{\hbox{\tiny APT}}

\def\sing{\hbox{\tiny sing.}}

\def\PV{\hbox{\tiny PV}}

\def\UV{\hbox{\tiny UV}}

\catcode`\@=11 
\def\lsim{\mathrel{\mathpalette\@versim<}}
\def\gsim{\mathrel{\mathpalette\@versim>}}
\def\@versim#1#2{\vcenter{\offinterlineskip
        \ialign{$\m@th#1\hfil##\hfil$\crcr#2\crcr\sim\crcr } }}
\catcode`\@=12 

\begin{document}

\begin{titlepage}

\begin{flushright}
\begin{tabular}{l}
Cavendish-HEP-07/06\\
CPHT-RR 144.0907\\
Edinburgh 2007/24\\
\end{tabular}
\end{flushright}
\vspace{1.5cm}

\begin{center}
{\Large \bf
A dispersive approach to Sudakov resummation}

\vspace{.6cm}

{\sc Einan Gardi}${}^{1,2,3}$
and
{\sc Georges Grunberg}${}^{4}$
\\[0.5cm]
\vspace*{0.1cm} ${}^1${\it
Cavendish Laboratory, University of Cambridge, \\J J Thomson
Avenue,
Cambridge, CB3 0HE, UK} \\[0.2cm] 
\vspace*{0.1cm} ${}^2${\it
Department of Applied Mathematics \& Theoretical Physics,\\
Wilberforce Road, Cambridge CB3 0WA,~UK}\\[0.2cm]
\vspace*{0.1cm} ${}^3${\it
School of Physics, The University of Edinburgh,\\
Edinburgh EH9 3JZ, Scotland, UK \footnote{Address after October 1st 2007.}}\\[0.2cm]
\vspace*{0.1cm} ${}^4${\it
 Centre de Physique Th\'eorique, \'Ecole
Polytechnique, CNRS ,\\
        91128 Palaiseau Cedex, France} \\[.6cm]

\vskip0.7cm
{\bf Abstract:\\[10pt]} \parbox[t]{\textwidth}{
We present a general all--order formulation of Sudakov resummation in QCD in terms of dispersion integrals. We show that the Sudakov exponent can be written as a dispersion integral over spectral density functions, weighted by characteristic functions that encode information on power corrections. The characteristic functions are defined and computed analytically in the large--$\beta_0$ limit. The spectral density functions encapsulate the non-Abelian nature of the interaction. They are defined by the time--like discontinuity of specific effective charges (couplings) that are directly related to the familiar Sudakov anomalous dimensions and can be computed order--by--order in perturbation theory. The dispersive approach provides a realization of Dressed Gluon Exponentiation, where Sudakov resummation is enhanced by an internal resummation of running--coupling corrections. We establish all--order relations between the scheme--invariant Borel formulation and the dispersive one, and address the difference in the treatment of power corrections. We find that in the context of Sudakov resummation the infrared--finite--coupling hypothesis is of special interest because the relevant coupling can be uniquely identified to any order, and may have an infrared fixed point already at the perturbative level. We prove that this infrared limit is universal: it is determined by the cusp anomalous dimension. To illustrate the formalism we discuss a few examples including B-meson decay spectra, deep inelastic structure functions and Drell--Yan or Higgs production.
\vskip.2cm
}
  \vskip1cm
\end{center}


\end{titlepage}

{\tableofcontents}

\newpage

\section{Introduction}
\setcounter{equation}{0}

An accurate theoretical description of inclusive differential cross sections and decay spectra is essential for many aspects of collider and flavor physics; it is therefore one of the primary goals of QCD perturbation theory.
The main challenge in describing differential distributions is associated with kinematic regions where there is a large hierarchy of scales, as occurs for example in the production of a jet with a large energy and a small mass.
Near the exclusive limit of phase space, the so called
Sudakov region, only soft and collinear radiation is kinematically allowed, and because of the dynamical enhancement of
such radiation, multiple gluon emission becomes important.
In this way the hierarchy of scales translates into large logarithms that appear with increasingly high powers in perturbation theory and spoil the convergence of the expansion.
Even a qualitative description of the distribution in this region require all--order resummation.
Classical examples where the Sudakov region has been thoroughly studied include deep inelastic structure functions for
$x\to 1$~\cite{Sterman:1986aj,Catani:1989ne,Vogt:2000ci,Forte:2002ni,Gardi:2002xm}, Drell--Yan and Higgs production near the partonic threshold
\cite{Sterman:1986aj,Catani:1989ne,Catani:1990rr,Catani:2003zt,Sterman:2006hu,Eynck:2003fn,Moch:2005ba,Moch:2005ky,Laenen:2005uz,Ravindran:2005vv}
or at small transverse momentum (see e.g.~\cite{Collins:1984kg,Laenen:2000ij,Bozzi:2007pn,Ravindran:2006bu,Bolzoni:2006ky}), event--shape distributions~\cite{Catani:1992ua,Korchemsky:1999kt,Korchemsky:2000kp,Gardi:2001ny,Gardi:2002bg,GM,Berger:2004xf,Dokshitzer:1997iz,Dokshitzer:1998kz,Dokshitzer:1998pt,Dasgupta:2003iq},
heavy--quark  fragmentation~\cite{Cacciari:2002xb,Cacciari:2001cw,Gardi:2005yi,Aglietti:2006yf,Neubert:2007je},
and inclusive B decays into a light quarks~\cite{Korchemsky:1994jb,Bauer:2000yr,Bosch:2004th,Gardi:2004ia,Gardi:2005yi,Andersen:2005bj,Andersen:2005mj,Andersen:2006hr,Aglietti_et_al,Aglietti:2006yb,Gardi:2007jx}, e.g. $\bar{B}\to X_s \gamma$ or ${\bar B}\to X_u l\bar{\nu}$.

From a theoretical perspective the Sudakov limit is very interesting. Despite the fact that one is dealing with a complex multi--scale problem in a non-Abelian gauge theory, one can in fact control the dominant perturbative corrections to all orders and resum the series. The simplification of the expansion in the Sudakov limit is a direct reflection of the factorization property of infrared (soft and collinear) singularities.
In infrared safe quantities the singularity associated with soft and collinear radiation (integrated over phase space) cancel exactly, order by order in $\alpha_s$, with infrared singularities in virtual corrections.
The structure of these singularities and their cancellation is encoded in the resummation formalism: the large logarithms are nothing but the finite remainder in the sum of the real and virtual contributions, which are separately divergent.

Sudakov resummation is best formulated in moment (Mellin) space where the multi--gluon phase space factorizes.  In moment space the combined effect of soft and collinear radiation and the corresponding virtual corrections appears an exponential Sudakov factor. The Sudakov factor sums up \emph{to all orders}
the dominant radiative corrections that are enhanced by powers of $\ln N$ (where $N$ is the moment index),
while neglecting corrections that are suppressed by powers of $N$
for $N\to \infty$.

While Sudakov resummation is designed to deal with logarithmic singularities alone, infrared sensitivity appears also through power--suppressed effects, ${\cal O}\left((\Lambda/m)^p\right)$~\cite{Korchemsky:1994is,Webber:1994cp,DMW,Akhoury:1995sp,Manohar:1994kq,Beneke:1995pq,Dokshitzer:1997ew,Gardi:1999dq,Gardi:2000yh,Gardi:2001ny}, where $m$ is the hard scale and where $p$ is an integer.
In the Sudakov region, power corrections are particularly important: whereas perturbative corrections are enhanced at $N\to \infty$ by powers of the logarithm $\ln N$, power corrections are enhanced by \emph{powers of $N$}, taking the form ${\cal O}\left((N\Lambda/m)^p\right)$. This parametric enhancement of non-perturbative corrections implies that for any fixed coupling (or a fixed hard scale) the $N\to \infty$ limit itself is strictly beyond the reach of resummed perturbation theory.
Put differently, upon ignoring these power corrections one can only expect the large--$N$ resummation to be a valid approximation in a restricted range: $N\ll m/\Lambda$.

Quite remarkably, perturbation theory itself is sensitive~\cite{Korchemsky:1994is} to the presence of these power--suppressed corrections~\cite{Gardi:2006jc}. This sensitivity appears through infrared renormalons \cite{Beneke:1998ui,Beneke:2000kc}, factorially increasing coefficients that emerge out of the integral over the running coupling. Despite the fact that the Sudakov exponent can be uniquely computed to any logarithmic accuracy, the series as a whole diverges. The understanding that renormalons are an inherent part of the Sudakov
exponent~\cite{Korchemsky:1994is,Beneke:1995pq,Dokshitzer:1997ew} has prompted the development of a new resummation formalism, Dressed Gluon Exponentiation  \cite{Gardi:2001ny,DGE,Gardi:2002bg};
for a recent review, see \cite{Gardi:2006jc}. DGE combines Sudakov resummation with an additional ``internal'' resummation of running--coupling corrections\footnote{``Internal'' refers here to the fact that the argument of the relevant logarithms involves loop--momenta that are integrated over, rather than external scales. Resummation of running--coupling corrections has been extensively studied in the past, primarily in the context of single--scale quantities~\cite{Beneke:1998ui,Beneke:2000kc,Beneke:1994qe,Ball:1995ni,DMW,Brodsky:1997vq,Grunberg:1998ix,Neubert:1994vb,Gardi:1999dq}.
See in particular the discussion in section 7 of  \cite{Grunberg:1998ix} and in \cite{Gardi:1999dq}. For a related early discussion of Sudakov resummation for the electron form factor in QED see also  \cite{Grunberg:1982fw}.}.
The internal resummation is based on an all--order calculation of the exponentiation kernel in the large--$\beta_0$ limit, which exposes the renormalons and thus opens the way for including non-perturbative power corrections.

DGE has been applied and successfully compared with data in a variety of inclusive distributions: event--shape distributions~\cite{Gardi:2001ny,Gardi:2002bg}, deep inelastic structure functions~\cite{Gardi:2002xm}, heavy quark fragmentation~\cite{Cacciari:2002xb} and inclusive decay spectra~\cite{Gardi:2004ia,Andersen:2005bj,Andersen:2005mj,Andersen:2006hr,Gardi:2007jx}.
It has proven to provide a good description of these distributions over the entire Sudakov region, significantly extending the range of applicability of resummed perturbation theory. In its first application, to event--shape distribution, DGE has opened the way for a quantitative description of the two--jet region, which had been inaccessible to analytic methods before.
In heavy--quark fragmentation and inclusive decay spectra DGE provides a viable alternative to the conventional approach where the corresponding distributions (``heavy--quark fragmentation function'' and ``shape function'', respectively) have been parametrized by
some ad hoc functional forms. Further details can be found in recent review talks~\cite{Gardi:2006jc,Gardi:2007jx} and in the original publications.

Technically these results have been obtained by trading~\cite{Gardi:2001ny}
the integration over the running coupling in the Sudakov exponent for a scheme--invariant~\cite{Grunberg:1992hf} Borel integration. The Borel transform has been computed analytically in the large--$\beta_0$ limit, allowing to identify the renormalon singularities.
The perturbative sum is then defined by the Principal Value prescription and  power corrections are parametrized based on the renormalon ambiguities. While the original calculations of event--shape distributions in Refs.~\cite{Gardi:2001ny,Gardi:2002bg} have been performed at next--to--leading logarithmic (NLL) accuracy\footnote{A proper definition of the QCD scale $\Lambda$ corresponding to the ``gluon bremsstrahlung'' coupling~\cite{Catani:1990rr} (see Eq.~(104) in~\cite{Gardi:2001ny}) is sufficient to account for the
entire non--Abelian  correction, ${\cal O}(C_A/\beta_0)$, at this order; $\Lambda$ is fixed by the NLO correction to the cusp anomalous dimension~\cite{Korchemsky:1987wg,Korchemsky:1988si}.}  it was observed that the Borel formulation is in fact completely general, and can accommodate subleading corrections with any logarithmic accuracy, provided that the relevant Sudakov anomalous dimensions
are known. This has been consequently used in other applications~\cite{Gardi:2002xm,Gardi:2005yi,Andersen:2005bj,Gardi:2006jc} to perform calculations with formal NNLL accuracy while preserving the pattern of renormalon singularities found in the large--$\beta_0$ limit.

In this paper we establish an alternative formulation of Sudakov resummation
that captures the same class of radiative corrections as well as the renormalon structure,
using the dispersive approach\footnote{A dispersive representation of the Sudakov exponent was first considered in the original DGE paper~\cite{Gardi:2001ny}, see Eq. (43) there. In that paper, however, the discussion was limited to NLL accuracy.  }.
Following~\cite{Grunberg:2006gd,Grunberg:2006ky} we show that the Sudakov exponent can be written, \emph{to any logarithmic accuracy}, as a dispersive integral of the time--like discontinuity of specific non-Abelian effective charges, integrated with respect to characteristic functions that are defined (and computed analytically) in the large--$\beta_0$ limit.
This way the exponent in the \emph{non-Abelian theory} is written in full analogy with the way renormalon resummation has been formulated in Refs.~\cite{DMW,Beneke:1994qe,Ball:1995ni} at the level of a single dressed gluon (the large--$\beta_0$ limit) with one important difference: the  ``effective'' coupling is \emph{computed} order--by--order in the non-Abelian theory.

Refs.~\cite{Grunberg:2006gd,Grunberg:2006ky} emphasized the fact that Sudakov resummation leaves some freedom (see also \cite{Friot:2007fd}) that does not allow to make conclusive statements about power corrections. Indeed, in general there is a conflict between the heuristic argument for power corrections in expressions involving integration over the running coupling (e.g. \cite{Korchemsky:1994is}) and renormalon--based arguments~\cite{Beneke:1995pq}. 
It was observed that, at the \emph{perturbative} level, the resummation of Sudakov logarithms can be implemented in infinitely many ways, each of which consisting of an integral mapping of a perturbative object (the `running Sudakov coupling') onto a fixed target: the `Sudakov exponent'. To each mapping corresponds a different kernel. In Ref.~\cite{Grunberg:2006gd} it was proposed to fix this ambiguity and select the unique resummation scheme by requiring that in the large--$\beta_0$ limit, the  `running Sudakov coupling' coincides with a dressed gluon propagator.
The specified scheme so determined is the dispersive approach, where the above conflict does not arise. In the present paper we further develop this approach, establishing the connection with the physical Sudakov anomalous dimensions.

We show that this dispersive formulation is in close correspondence with the scheme--invariant Borel formulation and establish the relations between the two.
We further show that the effective charges appearing in the dispersive formulation are uniquely defined to any order in perturbation theory and
stand in one-to-one correspondence with the relevant Sudakov anomalous dimensions.
These effective charges provide a (process--dependent) generalization of the concept of the ``gluon bremsstrahlung'' coupling, which was defined in~\cite{Catani:1990rr} at the next--to--leading order.
Moreover, we find that the dispersive formulation is closely related to the joint resummation formalism of Ref.~\cite{Laenen:2000ij}, where the effective charges are given a direct diagrammatic interpretation in terms of ``webs''~\cite{Gatheral:1983cz}.

The dispersive approach proposed here is particularly suited to the implementation of the infrared finite coupling hypothesis of Ref.~\cite{DMW}. This possibility is of special interest in the context of Sudakov resummation, since the relevant non-Abelian coupling can be identified to any order, and furthermore, may have an infrared fixed--point already at the perturbative level. In this case the coupling may have a causal analyticity structure~\cite{Gardi:1998rf,Gardi:1998qr,Gardi:1998ch}, free of any Landau singularities.

We proceed as follows: in Sec.~\ref{sec:setup} we
recall some general features of the Sudakov limit and define the basic elements used in Sudakov resummation. Then, in Sec.~\ref{sec:kernel} we analyze the structure of the exponentiation kernel in momentum space. We first use general considerations of infrared factorization and renormalization--group invariance to identify the physical Sudakov anomalous dimensions.  We then employ all--order resummation in the large--$\beta_0$ limit arriving at a dispersive representation of the kernel.
Next, in Sec.~\ref{sec:general_dispersive} we generalize the dispersive
representation to the full non-Abelian theory. We show that this leads to a unique  (yet process--dependent) generalization of the concept of a ``gluon bremsstrahlung'' coupling, the ``Sudakov effective charges'', which are computed order--by--order in perturbation theory.
We then analyze the evolution of these effective charges, finding that they may have a perturbative infrared fixed point. We further prove that the corresponding fixed--point value is process--independent, and compute its Banks--Zaks expansion.
Then in Sec.~\ref{sec:dispersive_exponent} we use the tools of the previous sections to derive a general dispersive representation of the Sudakov exponent.
We also present explicit results for the characteristic functions in several different processes and analyze their properties. In Sec.~\ref{sec:Borel} we explain the relation between the dispersive formulation and the scheme invariant Borel formulation. Next, in Sec.~\ref{sec:PC} we address power corrections. Finally, in Sec.~\ref{sec:conc} we summarize our conclusions.

\section{General set-up: Sudakov resummation for inclusive distributions~\label{sec:setup}}

Consider an inclusive infrared--safe distribution, $d\Gamma(m^2,r)/dr$ where the $r\to 0$ limit is characterized by Sudakov logarithms,
\begin{equation}
\label{d_Gamma_expansion}
\frac{d\Gamma(m^2,r)}{dr}=\delta (r) \Big(1+{\cal O}(\alpha_s)\Big)\,+\,C_R\,\frac{\alpha_s}{\pi}\left\{
\left[-\frac{\ln(r)}{r}+\frac{b_1-d_1}{r}
\right]_{+}+\text{regular\,terms}\right\}+\cdots
\end{equation}
whose maximal power grows with order as ${\alpha_s}^n \ln^{2n-1}(r)/r$ due to multiple soft and collinear radiation. The notation $b_i$ and $d_i$ used for the coefficients is associated with their phase--space origin; this will be elucidated in the next section. The integration prescription $[\,]_{+}$ is defined by\footnote{Differentiation with respect to $u$ can be used to generate powers of $\ln(r)$. See Eq. (6.4) in \cite{Gambino:2006wk} for a more general definition.}
\begin{equation}
\label{plus}
\int_0^1 dr F(r) \left[\frac{1}{r^{1+u}}\right]_{+}=\int_0^1dr\,
\Big(F(r)-F(0)\Big)\,\frac{1}{r^{1+u}}\,,
\end{equation}
where $F(r)$ is a smooth test function.
This prescription accounts for the divergent virtual corrections, which cancel against the singularity generated when integrating the real--emission contributions near $r=0$.

In order to describe the distribution at small $r$ we need to resum all the singular terms in (\ref{d_Gamma_expansion}) to any order. To this end it is convenient to work in moment space, where the multi--gluon phase space factorizes. The moments are given by:
\begin{equation}
\label{moments}
 \Gamma_N(m^2)=\int_0^1dr\,(1-r)^{N-1}\,\,\frac{d\Gamma (m^2,r)}{dr}\,,
\end{equation}
where the plus prescription in (\ref{d_Gamma_expansion}) guarantees that the moments are finite, {\it cf.} (\ref{Exponent}) below.
Once resummation has been performed analytically in moment space one recovers the distribution in momentum space by an inverse Mellin transformation:
\begin{equation}
\label{inv_Mellin}
 \left.\frac{d\Gamma (m^2,r)}{dr}\right\vert_{\rm resummed}= \int_{-i\infty}^{i\infty}\frac{dN}{2\pi i}\,(1-r)^{-N} \left.\Gamma_N(m^2)\right\vert_{\rm resummed},
\end{equation}
where the integration contour runs parallel to the imaginary axis in the complex $N$ plane, to the right of the singularities of the integrand.

We assume that the large--$N$ limit singles out infrared singularities associated with two physical scales, which we call the jet scale --- momenta of order ${\cal O}(m^2r)$ --- and the soft scale --- momenta of order ${\cal O}(m^2r^2)$.  We therefore have a double hierarchy of scales:
\begin{equation}
\label{hierarchy}
m^2\gg m^2r\gg m^2r^2 \qquad  \longleftrightarrow \qquad m^2\gg m^2/N\gg m^2/N^2.
\end{equation}
While we are interested in large $N$, in order to apply perturbation theory we assume that all three scales are within the perturbative regime, namely $m^2/N^2\gg \Lambda^2$.

This general scenario, with minor variations, is encountered in many different applications, for example:
\begin{itemize}
\item{} Inclusive B decays into light quarks~\cite{Korchemsky:1994jb,Bauer:2000yr,Bosch:2004th,Gardi:2004ia,Gardi:2005yi,Andersen:2005bj,Andersen:2005mj,Andersen:2006hr,Aglietti_et_al,Aglietti:2006yb,Gardi:2007jx}, e.g. $\bar{B}\to X_s \gamma$ or\footnote{Note that in case of the semileptonic decay $b\to ul \bar{\nu}$, resummation is applied to the \emph{triple differential width}~\cite{Andersen:2005mj}. For simplicity of the notation we do not write here explicitly derivatives with respect to other kinematic variables.} ${\bar B}\to X_u l\bar{\nu}$.
These decays are characterized by jet--like kinematics having
a large hierarchy between the energy of the hadronic system $X$
and its mass, or equivalently, between the two lightcone components
of the jet, $p^{\pm}\equiv E\mp|\vec{p}|$, namely $p^-\gg p^+$.  Thus, in (\ref{hierarchy}) the hard scale is the b--quark mass
or the large lightcone component of the jet ($p^-$), while $r$ is the ratio between the small and large components $r=p^+/p^-$.
The intermediate scale is the jet mass $p^+p^-\sim m^2 r$.
The `soft' scale $m^2r^2$ is associated with soft gluons that
couple of the b quark, putting
it slightly off its mass shell prior to the decay.
In perturbation theory the soft subprocess
is therefore associated with the momentum distribution
of the b quark inside an initial on-shell b quark, see \cite{Gardi:2005yi} and Refs. therein.
Beyond perturbation theory this function is replaced by the momentum
distribution of the b quark in a \emph{meson}, which differs from its
perturbative counterpart by power corrections.
\item{} Inclusive B production in $e^+e^-$ annihilation, see e.g.~\cite{Cacciari:2002xb,Cacciari:2001cw,Gardi:2005yi,Aglietti:2006yf,Neubert:2007je}. Here $r=1-x$ where $x=2E_B/Q$, the energy fraction of the detected B meson and $Q$ is the center--of--mass energy. At a difference with (\ref{hierarchy}) the three scales are $Q^2\gg Q^2(1-x)\gg m_b^2(1-x)^2$, where $m_b$ is the b--quark mass. The soft scale $m_b^2r^2$ is associated with the heavy quark fragmentation process (which, at the perturbative level, is
 similar to b decay~\cite{Gardi:2005yi}).
\item{} Drell--Yan or Higgs production in hadronic collisions in the DIS  scheme~\cite{Sterman:1986aj,Catani:1989ne,Catani:1990rr,Catani:2003zt,Sterman:2006hu,Eynck:2003fn,Moch:2005ba,Moch:2005ky,Laenen:2005uz}, where $m=Q$ is the mass of the produced pair (or $m_H$, in case of Higgs production),
    $r=1-Q^2/\hat{s}$ is the fraction of energy carried by soft gluon radiation into the final state, and $\hat{s}$ is the partonic center--of--mass energy.
    Here the soft scale is genuinely associated with the Drell--Yan process while the jet mass scale enters via the quark distribution function in the DIS factorization scheme (the $x\to 1$ limit of deep inelastic structure functions) and it can be traded for $\mu_F^2$ upon using the $\overline{\rm MS}$ factorization scheme.
\item{} Event--shape distributions
in $e^+e^-$ annihilation, near the two--jet limit~\cite{Catani:1992ua,Korchemsky:1999kt,Korchemsky:2000kp,Gardi:2001ny,Gardi:2002bg,GM,Berger:2004xf,Dasgupta:2003iq}. Here $r$ is identified with a shape variable that vanishes in the two--jet limit,
for example $1-T$ where $T$ is the thrust variable or $C/6$ where $C$ is the ``$C$ parameter''; the hard scale $m$ is the
center--of--mass energy in the collision.
Because jet observables are less inclusive~\cite{Nason:1995hd,Gardi:2001ny,Gardi:2002bg,Dasgupta:2003iq}, the formalism we develop applies only to a limited logarithmic accuracy. Non-inclusive effects can be accommodated in a class of event--shape variables (where Eqs.~(\ref{Exponent}) and (\ref{R1}) below are valid~\cite{Catani:1992ua}) but this goes beyond the scope of the present work.
\end{itemize}
As we shall see the same resummation formalism applies in all these
examples. Moreover, the coefficients of the
logarithms have a certain degree of universality: first, the
leading logarithms are always related to the cusp anomalous
dimension~\cite{Korchemsky:1987wg,Korchemsky:1988si},
${\cal A}(\alpha_s)$ in (\ref{R1}) below; second, the physics on the
jet--mass scale $m^2r$ is the same in all these examples: it is associated with the production of a quark jet with a
constrained invariant mass of ${\cal O}(m^2r)$.
The corresponding Sudakov logarithms are the same as in deep inelastic structure functions in the $x\to 1$ limit; they
are generated, in all these examples,
by the anomalous dimension function ${\cal B}(\alpha_s)$ in (\ref{R1}).
In contrast, the physics on the soft scale $m^2r^2$ is rather different; this will be reflected in the resummation formalism through a
\emph{process--dependent} Sudakov anomalous dimension
${\cal D}(\alpha_s)$.

Owing to the factorization property of soft and collinear radiation the moments (\ref{moments}) can be written as follows:
\begin{equation}
\label{fact_Mellin}
\Gamma_N(m^2)=H(m^2)\,\times\,{\rm Sud}(m^2,N)+\Delta \Gamma_N(m^2)\,,
\end{equation}
where the Sudakov factor ${\rm Sud}(m^2,N)$ sums up to all orders in perturbation theory \emph{all the terms that diverge for $N\to \infty$}, $H(m^2)$ is a hard coefficient function that includes constant ($N$--independent) terms at each order in $\alpha_s$; such corrections arise from virtual
diagrams, proportional to $\delta(r)$. Finally, $\Delta \Gamma_N(m^2)={\cal O}(1/N)$ includes residual real--emission corrections that fall at large $N$ as $1/N$ (up to logarithms).
In this paper we are interested in the Sudakov factor ${\rm Sud}(m^2,N)$.
Both $H(m^2)$ and $\Delta \Gamma_N(m^2)$ can be computed order by order in $\alpha_s$ by ``matching'' the resummation formula with the fixed--order expansion. We shall not discuss them further here.

The general formula~\cite{Sterman:1986aj,Catani:1989ne}~(see also \cite{Vogt:2000ci,Catani:1992ua,Cacciari:2001cw,Gardi:2005yi,Andersen:2005bj,Andersen:2005mj,Andersen:2006hr,Aglietti_et_al})
for the Sudakov factor is:
\begin{align}
\label{Exponent}
{\rm Sud}(m^2,N)=\exp\left\{C_R\int_0^1\frac{dr}{r} \Big[(1-r)^{N-1}-1\Big] R(m^2,r)\right\},
\end{align}
where the subtraction of 1 accounts for the virtual corrections encapsulated in the plus prescription in (\ref{d_Gamma_expansion}), and the momentum--space kernel, which is fully defined by the real--emission contributions that are singular for $r\to 0$, takes the general form:
\begin{align}
\label{R1}
C_R\frac{R(m^2,r)}{r}=\frac{1}{r}\,\left[\int_{r^2m^2}^{r m^2}\frac{dk^2}{k^2}{\cal A}\left(\alpha_s(k^2)\right)+{\cal B}\left(\alpha_s(r m^2)\right)-{\cal D}\left(\alpha_s(r^2 m^2)\right)\right],
\end{align}
where ${\cal A}$, ${\cal B}$ and ${\cal D}$ are Sudakov anomalous dimensions that can be computed order--by--order in $\alpha_s$.
${\cal A}(\alpha_s)$ is the universal cusp anomalous dimension~\cite{Korchemsky:1987wg,Korchemsky:1988si}, whose expansion, in the $\overline{\rm MS}$ scheme is known to three-loop order~\cite{Moch:2004pa}:
\begin{eqnarray}
\label{A_cusp} {\cal A}\big(\alpha_s\big)&=&
\frac{C_R}{\beta_0}\bigg[\bar{a}
+a_2
\bar{a}^2
+a_3
\bar{a}^3
+\cdots\bigg],
\end{eqnarray}
where $\bar{a}\equiv \beta_0
\alpha_s^{\MSbar}/\pi$.
The other Sudakov anomalous dimensions appearing in Eq.~(\ref{R1})
are
\begin{align}
\label{calB}
\begin{split}
&{\cal B}(\alpha_s)=\frac{C_R}{\beta_0}\bigg[b_1
\bar{a}+b_2\bar{a}^2+b_3\bar{a}^3+\cdots\bigg],
\end{split}
\end{align}
which is associated the the jet mass scale $m^2r$, and
\begin{align}
\label{calD}
\begin{split}
{\cal D}(\alpha_s)=&\frac{C_R}{\beta_0}\bigg[d_1
\bar{a}+d_2\bar{a}^2+d_3\bar{a}^3+\cdots\bigg],
\end{split}
\end{align}
which is associated with the soft scale $m^2r^2$. The known expansion coefficients of these anomalous dimensions are summarized in Appendix \ref{sec:coef}. As already mentioned, ${\cal D}$ is process dependent. In Appendix \ref{sec:coef} we quote the coefficients in two examples, one of Drell--Yan (or Higgs) production and the second of inclusive B decay (or heavy--quark fragmentation --- the anomalous dimension is the same~\cite{Gardi:2005yi}). These two examples will be used throughout this paper.

Next, note that $C_R$, the overall coefficients of all the anomalous dimensions entering (\ref{R1}) in a given process, depends on the color charge of the hard parton(s) that radiates. In inclusive B decay or heavy--quark fragmentation $C_R=C_F$, while for Drell--Yan production $C_R=-2C_F$ and for Higgs production by gluon--gluon fusion $C_R=-2C_A$.

Finally, it is useful to introduce some alternative definitions
for the Sudakov factor using the Laplace weight instead of the Mellin one. The first observation is that at large $N$ one can make the following approximation of the real--emission weight factor in (\ref{Exponent})
\[
(1-r)^{N-1}\simeq {\rm e}^{-Nr}\,,
\]
which only modifies the exponent by ${\cal O}(1/N)$ terms.
Next one notes that as far as the real--emission part is concerned the integral over $r$ can be extended to $r=\infty$, well beyond the physical phase space $r=1$: the resulting modification of the  Sudakov exponent is exponentially small, ${\cal O}({\rm e}^{-N})$. This, however, does not apply to the virtual terms, where any change of the upper limit of integration is reflected in a modification of the constant ${\cal O}(N^0)$ term for $N\to \infty$. Let us therefore define:
\begin{align}
\label{Exponent_Laplace_with_Mellin_const}
\widetilde{\rm Sud}(m^2,N)  =\exp\left\{C_R\int_0^{\infty}\,\frac{dr}{r} \Big[{\rm e}^{-Nr}-{\theta(r<1)}\Big] R(m^2,r)\right\}\,,
\end{align}
which satisfies
\begin{equation}
\ln\left(\widetilde{\rm Sud}(m^2,N)\right)  =\ln\left( {\rm Sud}(m^2,N)\right)  +{\cal O}(1/N).
\end{equation}
The ${\cal O}(1/N)$ contributions by which Eq.~(\ref{Exponent_Laplace_with_Mellin_const}) differs from Eq.~(\ref{Exponent}), can be compensated in the expression for the moments by a different remainder function. The physical moments $\Gamma_N(m^2)$ can be obtained by ``matching'' with the fixed--order result:
\begin{equation}
\label{fact_Laplace_with_Mellin_const}
\Gamma_N(m^2)=H(m^2)\,\times\,\widetilde{\rm Sud}(m^2,N)+\widetilde{\Delta \Gamma}_N(m^2);
\end{equation}
where $H(m^2)$ is the same hard function as in (\ref{fact_Mellin}), while $\widetilde{\Delta \Gamma}_N(m^2)$ can be determined to fixed order from the perturbative expansion of $\Gamma_N (m^2)$.

It is useful to further define a Sudakov factor where the subtraction term too is integrated to $r=\infty$:
\begin{align}
\label{Exponent_Laplace}
\overline{\rm Sud}(m^2,N)  =\exp\left\{C_R\int_0^{\infty}\,\frac{dr}{r} \Big[{\rm e}^{-Nr}-{\rm e}^{-r}\Big] R(m^2,r)\right\}\,.
\end{align}
The suppression of the weight factor ${\rm e}^{-Nr}-{\rm e}^{-r}$ guarantees convergence for $r\to \infty$.
Again, the manipulations used in obtaining (\ref{Exponent_Laplace}) from (\ref{Exponent})
do not change the $N\to \infty$ divergent terms, which all emerge from the $r\to 0$ limit. Here however, in contrast with (\ref{Exponent_Laplace_with_Mellin_const}), the constant terms ${\cal O}(N^0)$ for $N\to \infty$ \emph{are} modified.

Note that the moment--space exponent of Eq.~(\ref{Exponent}) is composed of harmonic sums
(powers of the $\Psi(N)$ function, which differ from $\ln(N)$ by constants and inverse powers of $N$) whereas the Laplace version, Eq.~(\ref{Exponent_Laplace}), involves strictly powers of $\ln(N)$, with no additional constants nor ${\cal O}(1/N)$ terms. Eq.~(\ref{Exponent_Laplace_with_Mellin_const}) contains powers of $\ln(N)$ and constant terms, with no additional ${\cal O}(1/N)$ terms~\cite{Grunberg:2006hg,Grunberg:2006gd}. Both (\ref{Exponent_Laplace}) and (\ref{Exponent}), but not (\ref{Exponent_Laplace_with_Mellin_const}), are defined such that for $N=1$ the exponent exactly vanishes so the Sudakov factor becomes~$1$.

Both the constant term, ${\cal O}(N^0)$, and the ${\cal O}(1/N)$ contributions by which Eq.~(\ref{Exponent_Laplace}) differs from Eq.~(\ref{Exponent}), can be compensated in the expression for the moments by a different multiplicative hard function and a different remainder function:
\begin{equation}
\label{fact_Laplace}
\Gamma_N(m^2)=\overline{H}(m^2)\,\times\,\overline{\rm Sud}(m^2,N)+\overline{\Delta \Gamma}_N(m^2);
\end{equation}
similarly to ${H}(m^2)$ and ${\Delta \Gamma}_N(m^2)$ in (\ref{fact_Mellin}), $\overline{H}(m^2)$ and $\overline{\Delta \Gamma}_N(m^2)$ in (\ref{fact_Laplace}) can be determined to fixed order from the perturbative expansion of $\Gamma_N (m^2)$. Finally, Eq.~(\ref{fact_Laplace}), similarly to (\ref{fact_Laplace_with_Mellin_const}) and (\ref{fact_Mellin}),
  can be inserted into the inverse Mellin transformation~(\ref{inv_Mellin}) to generate the resummed distribution in momentum space.

\section{The exponentiation kernel~\label{sec:kernel}}
\setcounter{footnote}{1}

The key to constructing an effective resummation technique is the
understanding of the all--order structure of the kernel.
The purpose of this section is to analyze the momentum--space kernel (\ref{R1}),
first using the general considerations of infrared factorization and renormalization--group invariance, and then by considering the all--order calculation of the kernel in the large--$\beta_0$ limit.

\subsection{Infrared factorization and renormalization--group invariance}

Let us begin by noting that Sudakov logarithms in infrared and collinear safe distributions arise from different regions of phase space, involving different physical scales.
As follows for (\ref{d_Gamma_expansion}), in the perturbative expansion of the kernel,
\begin{align}
\label{LO}
C_R\frac{R(m^2,r)}{r} =C_R\left[-\frac{\ln(r)}{r}+\frac{b_1-d_1}{r}\right]\frac{\alpha_s(m)}{\pi}+\cdots
\end{align}
the two are mixed together. In contrast, upon taking one derivative ({\it cf.} Eq.~(\ref{R1}))
\begin{align}
\label{dR}
\frac{dR(m^2,r)}{d\ln m^2}\,=\,{\cal J}(r m^2)-{\cal S}(r^2 m^2),
\end{align}
the dependence upon the `jet' ($r m^2$) and `soft' ($r^2 m^2$) scales is nicely disentangled into two different functions corresponding to the two subprocess, ${\cal J}$ and ${\cal S}$ respectively. These are the physical `jet' and `soft' Sudakov anomalous dimensions. They are related to the ones appearing in (\ref{R1}) by the following relations:
\begin{align}
\begin{split}
\label{Sud_anom_dim}
C_R\ {\cal J}(\mu^2)={\cal A}(\alpha_s(\mu^2))+\frac{d{\cal B}(\alpha_s(\mu^2))}{d\ln \mu^2}
\\
C_R\ {\cal S}(\mu^2)={\cal A}(\alpha_s(\mu^2))+\frac{d{\cal D}(\alpha_s(\mu^2))}{d\ln \mu^2}.
\end{split}
\end{align}
We emphasize that ${\cal J}(\mu^2)$ and ${\cal S}(\mu^2)$ are renormalization--group invariant functions, while their conventional decomposition into the cusp ${\cal A}$, ${\cal B}$ and ${\cal D}$ (which are renormalization--scheme--dependent quantities) is specific to the $\overline{\rm MS}$ scheme.

\subsection{The physical interpretation of the Sudakov anomalous dimensions }

The functions ${\cal J}(\mu^2)$ and ${\cal S}(\mu^2)$ have a clear physical
meaning: they govern the scale--dependence of certain physical quantities
(the `physical evolution kernels' or `physical anomalous dimensions') in the
Sudakov limit. Let us illustrate this statement using a couple of examples.

\vspace*{10pt}
\noindent
\underline{Deep inelastic structure functions}\\
The scale--dependence of the (flavour non-singlet) deep inelastic structure function $F_2$ can be expressed in terms of $F_2$ itself, yielding the following evolution equation (see e.g.~Refs.~\cite{Grunberg:1982fw,Catani:1996sc,van_Neerven:1999ca}):
\begin{equation}
\label{F_2_evolution}
\frac{dF_2(x,Q^2)}{d\ln Q^2}\,=\,\int_x^1 \frac{dz}{z}\,K_{\DIS (F_2)}(x/z,Q^2)\,F_2(z,Q^2).
\end{equation}
$K_{\DIS (F_2)}(x,Q^2)$ is the \emph{physical evolution kernel}; it is renormalization--group invariant. Defining moments by
\begin{equation}
\widetilde{F}_2(N,Q^2)=\int_{0}^{1} dx \,\,x^{N-1} \,F_2(x,Q^2)/x\,,
\end{equation}
Eq.~(\ref{F_2_evolution}) implies that the moment--space physical evolution kernel,
the `physical anomalous dimension', is:
\begin{equation}
\label{DIS_kernel_def}
\widetilde{K}_{\DIS (F_2)}(N,Q^2) \equiv \int_{0}^{1} \,dx\,  x^{N-1} \,K_{\DIS (F_2)}(x,Q^2)/x= \frac{d\ln \widetilde{F}_2(N,Q^2)}{d\ln Q^2}.
\end{equation}

Let us now consider the $x\to 1$ limit. In this limit the evolution of the structure function takes a simple from~\cite{Forte:2002ni,Gardi:2002xm} (see Eq. (13) in \cite{Gardi:2006jc}):
\begin{eqnarray}
\label{DIS_physical}
 \!\!\!\!\!\!\frac{d\ln \widetilde{F}_2(N,Q^2)}{d\ln Q^2}
&=& C_F \int_0^1\! dx\frac{ x^{N-1}-1}{1-x} {\cal
J}\left((1-x)Q^2\right)+\frac{d\ln H(Q;\mu_F)}{d\ln Q^2}+{\cal O}(1/N)\,
\end{eqnarray}
where the first term, which includes the $N\to \infty$ divergent corrections to all orders, is controlled by the Sudakov anomalous dimension ${\cal J}(\mu^2)$ of Eq.~(\ref{Sud_anom_dim}). The constant term can be written~\cite{Friot:2007fd} in terms of the quark electromagnetic form factor ${\cal F}(Q^2)$~\cite{Sen:1981sd,Mueller:1979ih,Collins:1980ih,Magnea:1990zb}:
\begin{align}
\label{DIS_constant}
\begin{split}
\frac{d\ln H(Q;\mu_F)}{d\ln Q^2}&=\frac{d\ln \left({\cal F}(Q^2)\right)^2}{d\ln Q^2}\,+\,C_F\int_0^{Q^2}\frac{d\mu^2}{\mu^2}{\cal J}(\mu^2)\\
&={\mathbb G}\left(1,\alpha_s(Q^2),\varepsilon=0\right)\,+\,{\cal B}\left(\alpha_s(Q^2)\right)\,,
\end{split}
\end{align}
where each of the two terms in the first line is separately infrared divergent, but the divergence cancels in the sum~\cite{Friot:2007fd}; in the second line the result is expressed in terms of ${\cal B}$ of Eq.~(\ref{calB}) above and  ${\mathbb G}\left(Q^2/\mu^2,\alpha_s(\mu^2),\varepsilon\right)$, which is the finite part of ${d\ln \left({\cal F}(Q^2)\right)^2}/{d\ln Q^2}$ as defined in Ref.~\cite{Magnea:1990zb} using dimensional regularization.

Comparing (\ref{DIS_kernel_def}) and (\ref{DIS_physical}) we therefore find the following relation in momentum space:
\begin{align}
\label{K_x}
\begin{split}
K_{\DIS (F_2)}(x,Q^2)&=C_F \frac{{\cal J}\left((1-x)Q^2\right)}{1-x}\,+\,\frac{d\ln \left({\cal F}(Q^2)\right)^2}{d\ln Q^2}\,\delta(1-x)+{\cal O}\left((1-x)^{0}\right)\\
&=C_F \left[\frac{{\cal J}
\left((1-x)Q^2\right)}{1-x}\right]_{+} + \left(\underbrace{
\frac{d\ln \left({\cal F}(Q^2)\right)^2}{d\ln Q^2}\,+\,C_F\int_0^{Q^2}\frac{d\mu^2}{\mu^2}{\cal J}(\mu^2)
}_{\text{infrared \,\,finite}}\right)\,\delta(1-x)\\&\hspace*{50pt} +\,{\cal O}\left((1-x)^{0}\right)\,,
\end{split}
\end{align}
where the plus prescription is defined in (\ref{plus}) and the ${\cal O}\left((1-x)^{0}\right)$
term neglected here is integrable for $x\to 1$ (although divergent: it contains powers of $\ln(1-x)$).
We thus find that $C_F\,{{\cal J}\left((1-x)Q^2\right)}/({1-x})$ is the leading term in the expansion of the physical evolution kernel $K_{\DIS (F_2)}(x,Q^2)$ in the \hbox{$x\to 1$} limit with fixed jet mass $(1-x)Q^2$. The next term in this expansion, proportional to $\delta(1-x)$, is comprised of purely virtual corrections associated with the quark form factor. This term is infrared divergent, but as indicated in the second line in (\ref{K_x}), the singularity cancels exactly upon integrating over~$x$ with the divergence of the integral of $C_F {{\cal J} \left((1-x)Q^2\right)}/({1-x})$ near $x\to 1$.

\vspace*{10pt}
\noindent
\underline{Drell--Yan cross section}\\
The cross section of the Drell--Yan process, $h_a+h_b\to e^+e^- +X$, is:
\begin{equation}
\label{DY}
\frac{d\sigma}{dQ^2}= \frac{4\pi\alpha_{\rm em}^2}{9Q^2s}\,
\sum_{i,j}\int_0^1 \frac{dx_i}{x_i} \frac{dx_j}{x_j}
\,f_{i/h_a}(x_i,\mu_F)\, f_{j/h_b}(x_j,\mu_F) \,g_{ij}(\tau,Q^2,\mu_F^2)
\end{equation}
where $s$ is the hadronic center--of--mass energy,
$\hat{s}=x_ix_js$ is the partonic one,
$Q^2$ is the squared mass of the lepton pair and $\tau=Q^2/\hat{s}$. The partonic threshold $\tau\to 1$ is characterized by Sudakov logarithms.

Let us define the Mellin transform of the quark--antiquark partonic cross section in (\ref{DY}),
e.g. in electromagnetic annihilation
$g_{q\bar{q}}(\tau,Q^2,\mu_F^2)=e_q^2  \left[\delta(1-\tau)+{\cal O}(\alpha_s)\right]$, by
\begin{align}
\label{G_N_def}
\begin{split}
&G_{q\bar{q}}(N,Q^2,\mu_F^2)\equiv \int_0^1d\tau\, \tau^{N-1}\,g_{q\bar{q}}(\tau,Q^2,\mu_F^2)\,.
\end{split}
\end{align}
Considering the Sudakov limit one has
\begin{align}
\label{G_N_Sud}
\begin{split}
&G_{q\bar{q}}(N,Q^2,\mu_F^2)\simeq H_{\DY}(Q^2,\mu_F^2)\,{\rm Sud}_{\DY}(N,Q^2,\mu_F^2) +{\cal O}(1/N)
\end{split}
\end{align}
with
\begin{align}
\label{Sud_DY}
\begin{split}
&{\rm Sud}_{\DY}
=\exp\left\{2\int_0^1 d\tau \frac{\tau^{N-1}-1}{1-\tau}
\,\left[\int_{\mu_F^2}^{(1-\tau)^2Q^2}\!\!\!\!\!\!\!\!\!\!{\cal A}\left(\alpha_s(k^2)\right)\frac{dk^2}{k^2}+{\cal D}_{\DY}\left(\alpha_s((1-\tau)^2 Q^2)\right)\right]
\right\}\,.
\end{split}
\end{align}
Upon taking the logarithmic derivative we identify the \emph{physical anomalous dimension} ${\cal S}_{\DY}$ of Eq.~(\ref{Sud_anom_dim}):
\begin{align}
\begin{split}
\label{dG_N}
\frac{d\ln G_{q\bar{q}}(N,Q^2,\mu_F^2)}{d\ln Q^2}&\equiv \widetilde{K}_{\DY}(N,Q^2)\equiv
\int_{0}^{1} {d\tau} \tau^{N-1} K_{\DY}(\tau,Q^2)\\&
\simeq
2C_F\int_0^1d\tau \frac{\tau^{N-1}-1}{1-\tau} \, {\cal S}_{\DY}\left(Q^2(1-\tau)^2\right)+\frac{d\ln H_{\DY}(Q^2,\mu_F^2)}{d\ln Q^2}+{\cal O}(1/N)\,,
\end{split}
\end{align}
where $K_{\DY}(\tau, Q^2)$ is the physical Drell--Yan evolution kernel defined for arbitrary $\tau$;
in the second line we considered the large--$N$ limit where Eqs.~(\ref{G_N_Sud}) and
(\ref{Sud_DY}) apply and then used Eq.~(\ref{Sud_anom_dim}).

The constant term can be expressed in terms of the analytically--continued electromagnetic quark form factor~\cite{Friot:2007fd} (see also \cite{Eynck:2003fn,Moch:2005ky,Laenen:2005uz}):
\begin{equation}
\label{DY_constant}
 \frac{d\ln H_{\DY}(Q^2,\mu_F^2)}{d\ln Q^2}=
\frac{d\ln \left|\left({\cal F}(-Q^2)\right)\right|^2}{d\ln Q^2}\,+\,C_F\int_0^{Q^2}\frac{d\mu^2}{\mu^2}{\cal S}_{\DY}(\mu^2)\,,
\end{equation}
where the infrared singularities cancel in the sum, as in (\ref{DIS_constant}).
Thus, in momentum space we have:
\begin{align}
\label{K_x_DY}
\begin{split}
 K_{\DY}(\tau,Q^2)&=
2C_F\frac{
{\cal S}_{\DY}\left(Q^2(1-\tau)^2\right)}{1-\tau}
+
\frac{d\ln \left|\left({\cal F}(-Q^2)\right)\right|^2}{d\ln Q^2}\,\delta(1-\tau)+{\cal O}\left((1-\tau)^0\right)
\\
&=
2C_F\left[\frac{
{\cal S}_{\DY}\left(Q^2(1-\tau)^2\right)}{1-\tau}
\right]_{+} \\&
+
\left(
\underbrace{
\frac{d\ln \left|\left({\cal F}(-Q^2)\right)\right|^2}{d\ln Q^2}\!+\!
C_F\int_0^{Q^2}\frac{d\mu^2}{\mu^2}{\cal S}_{\DY}(\mu^2)}_{\text{infrared\,\,finite}}
\right)\,\delta(1-\tau)+{\cal O}\left((1-\tau)^0\right)
\end{split}
\end{align}
We see that the Sudakov anomalous dimension ${\cal S}_{\DY}$ controls the leading term in the expansion of the physical Drell--Yan kernel (\ref{dG_N}) near threshold. The  $\tau\to 1$
limit is taken such that $Q(1-\tau)$, corresponding to the total energy carried by soft gluons to the final state, is kept fixed.
The next term in this expansion, proportional to $\delta(1-\tau)$, is determined by the quark form factor, analytically--continued to the time--like axis. This purely virtual term is infrared singular, but upon performing an integral over $\tau$ this singularity cancels with the one generated by integrating the real-emission term $2C_F {\cal S}_{\DY}\left(Q^2(1-\tau)^2\right)/({1-\tau})$ near $\tau\to 1$.

\subsection{The large--$\beta_0$ limit: a dispersive representation of the kernel}

Upon disentangling the `hard', `jet' and `soft' scales in (\ref{dR}) we have significantly reduced the complexity of the problem: we are now dealing with two independent single--scale quantities, ${\cal J}(r m^2)$ and ${\cal S}(r^2 m^2)$. At this point we need additional tools to examine the all--order structure of these two anomalous dimensions. As these objects define the exponentiation \emph{kernel}, an approximation based on a \emph{single} dressed gluon appears most natural~\cite{Gardi:2001ny}; multiple emission is taken into account by exponentiation.
In this section we shall therefore consider the single--dressed--gluon approximation, the large--$\beta_0$ limit, postponing the discussion of multiple emission
to Sec.~\ref{sec:dispersive_exponent} where we shall perform exponentiation in moment space.

A convenient way to compute the Sudakov anomalous dimensions in the single--dressed--gluon approximation is the dispersive technique\footnote{The dispersive technique is a general method to perform resummation of running--coupling corrections.
This technique was developed in Refs.~\cite{DMW,Beneke:1994qe,Ball:1995ni}, with no specific consideration of the Sudakov limit; it was used to study power corrections in a variety of applications. A well--known example is the average thrust~\cite{Webber:1994cp,DMW,Gardi:1999dq}.
In Ref.~\cite{Gardi:2001ny}  the dispersive technique was used to compute the Sudakov exponentiation kernel for the thrust distribution based on the results of Ref.~\cite{Gardi:2000yh} for the characteristic function. Similar calculations using the dispersive technique were later done for other event--shape variables including the heavy--jet mass~\cite{Gardi:2002bg}, the $C$ parameter~\cite{GM} and angularities~\cite{Berger:2004xf}, as well as for heavy quark fragmentation~\cite{Cacciari:2002xb}.}. A brief description of the method is given in Appendix~\ref{sec:Renormalon_sum_techniques}.
In order to compute the real--emission contribution to some physical quantity to all orders within this approximation one first evaluates the single gluon emission diagrams with an \emph{off-shell gluon}, $k^2=\mu^2\neq 0$. The result of this ${\cal O}(\alpha_s)$ calculation defines the ``characteristic function'' ${\cal F}(\mu^2/m^2,r)$. This function is then integrated with
the discontinuity of the coupling, generating the all--order sum.
Here we are interested specifically in the singular contributions for $r\to 0$. In taking this limit we expect to identify the dependence on the two scale $rm^2$ and $r^2m^2$ corresponding to the anomalous dimensions ${\cal J}$ and ${\cal S}$, respectively.

\subsubsection*{Dispersive representation of the kernel}

The result for the characteristic function at small $r$ takes the form:
\begin{align}
\label{calF_general}
C_R \,\frac{\alpha_s}{\pi} \,{\cal F} (\epsilon,r) = C_R \,\frac{\alpha_s}{\pi} \,\underbrace{\frac{1}{r}\bigg[{\cal F}_{\cal J}\left(\epsilon/r\right)\,-{\cal F}_{\cal S}\left(\epsilon/r^2\right)\bigg]}_{{\cal F}_{\sing}(\epsilon,\,r)} \,\,+\,\, {\text{regular\, terms}\, }\,,
\end{align}
where $\epsilon\equiv \mu^2/m^2$, the ratio between the gluon virtuality $\mu^2$ and the hard scale in the process, $m^2$.
The singular terms are distinguished by considering the limits
$\lim_{r\to 0} \left(r {\cal F} (\epsilon,r)\right)$ with either fixed $\epsilon/r$ or fixed $\epsilon/r^2$.
The terms that are regular in these two limits are irrelevant for Sudakov resummation. A detailed example of how ${\cal F}_{\sing}(\epsilon,\,r)$ is constructed by considering these two limits can be found in Sec.~3 of~Ref.~\cite{GM}, where the kernel of the $C$ parameter was computed (note the difference in complexity(!) between the full ${\cal F}(\epsilon,\,r)$ and ${\cal F}_{\sing}(\epsilon,\,r)$). Below we shall summarize the results for ${\cal F}_{\sing}(\epsilon,\,r)$ in some examples.

The central point is that Eq.~(\ref{calF_general}) exhibits the general property of infrared factorization summarized by Eq. (\ref{dR}): ${\cal F} (\epsilon,r)$, which is a function of two variables can be decomposed in this limit into a sum of two functions of a single variable: one depends on the jet mass scale and the other on the soft scale.
Such a separation of scales cannot be done in an ordinary fixed--order calculation: it requires keeping an infrared regulator\footnote{As alternatives to the gluon mass regulator one can work in D-dimensions or use the Borel technique (see Appendix~\ref{sec:Renormalon_sum_techniques}).} in place.

With a single gluon emission the contribution to the physical quantity itself and to the exponentiation kernel are the same (see the arguments leading to Eq.~(\ref{F_DIS}) below).
We can therefore use (\ref{calF_general}) to construct the \emph{resummed} kernel in the large--$\beta_0$ limit as a dispersive integral according to Eq.~(\ref{RQ}):
\begin{align}
\label{Kernel_dispersive}
\begin{split}
&\left.C_R\frac{R(m^2,r)}{r} \right\vert_{\rm large \,\,\beta_0}\,
=\,\frac{C_R}{\beta_0}\,\int_0^{\infty}\frac{d\mu^2}{\mu^2} \rho_V(\mu^2)  \bigg[{\cal F}_{\sing} (\mu^2/m^2,r)-{\cal F}_{\sing} (0,r)\bigg]\,,
\end{split}
\end{align}
where we neglected the power corrections associated with the ``analytization'' of the Landau singularity; we shall revisit this issue in Sec.~\ref{sec:PC} and Appendix \ref{sec:Mink_integral}.
Here the spectral density function $\rho_V(\mu^2)$, as defined by
Eq.~(\ref{eq:discontinuity}), stands for the timelike discontinuity
of the coupling in the V-scheme, where
\begin{equation}
\label{Lambda_V}
\Lambda_V^2= \Lambda^2
{\rm e}^{{5}/{3}}\,,
\end{equation}
where $\Lambda^2$ is defined in the $\overline{\rm MS}$ scheme.
As we shall see below (Eq.~(\ref{Lambda_GB}))
this is also the large--$\beta_0$ limit
of the ``gluon bremsstrahlung'' coupling.

\subsubsection*{Dispersive representations of the Sudakov anomalous dimensions}

Eq.~(\ref{Kernel_dispersive}) with ${\cal F}_{\sing} (\mu^2/m^2,r)$ of (\ref{calF_general}) constitute
the dispersive representation of the Sudakov evolution kernel
in the large--$\beta_0$ limit; it involve two physical scales.
We now wish to construct similar representations for the `jet' and `soft'
Sudakov anomalous dimensions (\ref{Sud_anom_dim}) each involving just one scale.

According to the leading--order result, Eq.~(\ref{LO}),
${\cal F}_{\sing}(\mu^2/m^2, r)$ must reach a finite limit for $\mu^2\to 0$, namely:
\begin{align}
\label{calF_LO}
{\cal F}_{\sing}(0, r) =-\frac{\ln(r)}{r}+\frac{b_1-d_1}{r}\,.
\end{align}
This, together with the functional form
of ${\cal F}_{\sing}(\mu^2/m^2, r)$ summarized by (\ref{calF_general}), implies that the two functions ${\cal F}_{\cal J}$ and ${\cal F}_{\cal S}$ are separately \emph{logarithmically divergent} for $\mu^2\to 0$,
\begin{align}
\label{calF_small_epsilon}
{\cal F}_{\cal J}(\epsilon)=-\ln(\epsilon) +b_1+{\cal O}(\epsilon)\,;\qquad \quad
{\cal F}_{\cal S}(\epsilon)&=-\ln(\epsilon) +d_1+{\cal O}(\epsilon)\,.
\end{align}
Therefore, Eq.~(\ref{Kernel_dispersive}), as written, cannot be split into the `jet' and `soft' contributions.

An alternative representation of the kernel can be obtained using integration-by-parts; following~\cite{DMW} we define the timelike coupling $a_V^{\Mink}(\mu^2)$ through
\begin{equation}
\label{eq:int_discontinuity}
\rho_V(\mu^2)=\frac{d a_V^{\Mink}(\mu^2)}{d\ln\mu^2}\,;\qquad\quad
 a_V^{\Mink}(\mu^2)\equiv -\int_{\mu^2}^{\infty} \frac{dm^2}{m^2}\rho_V(m^2),
\end{equation}
obtaining
\begin{align}
\label{R_int_by_parts}
\begin{split}
\left.C_R\frac{R(m^2,r)}{r}\right\vert_{\rm large \,\,\beta_0}\,
& = \frac{C_R}{\beta_0}\, \int_0^{\infty}\frac{d\mu^2}{\mu^2} a_V^{\Mink}(\mu^2)  \,{\cal \dot{F}_{\sing}} (\mu^2/m^2,r) \\
& =\,\frac{C_R}{\beta_0}\,\frac{1}{r}\,\int_0^{\infty}\frac{d\mu^2}{\mu^2} a_V^{\Mink}(\mu^2)  \,\,\bigg[\dot{\cal F}_{\cal J}\left(\frac{\mu^2}{r m^2}\right)
-\dot{\cal F}_{\cal S}\left(\frac{\mu^2}{r^2 m^2}\right)\bigg]\,,
\end{split}
\end{align}
where we use the usual convention,
\begin{align}
\label{calF_dot}
\dot{\cal F} (\epsilon,r) \equiv \frac{d{\cal F}(\epsilon,r)}{d\ln 1/\epsilon}
\end{align}
and similarly
\begin{equation}
\label{d_cal_F}
\dot{\cal F}_{\cal J}(\epsilon)\equiv -\epsilon\,\frac{d}{d\epsilon}\,{\cal F}_{\cal J}(\epsilon);\qquad
\dot{\cal F}_{\cal S}(\epsilon)\equiv-\epsilon\,\frac{d}{d\epsilon}\,{\cal F}_{\cal S}(\epsilon)\,,
\end{equation}
so
\begin{equation}
\label{cal_F_from_d_cal_F}
{\cal F}_{\cal J}(\epsilon)=\int_{\epsilon}^{\infty}
\frac{dy}{y}\dot{\cal F}_{\cal J}(y);\qquad
{\cal F}_{\cal S}(\epsilon)=\int_{\epsilon}^{\infty}
\frac{dz}{z}\dot{\cal F}_{\cal S}(z)\,.
\end{equation}
In the second line of Eq.~(\ref{R_int_by_parts}) we inserted ${\cal \dot{F}_{\sing}} (\mu^2/m^2,r)$ according to~(\ref{calF_general}).

Infrared safety of (\ref{R_int_by_parts}) implies that ${\cal \dot{F}_{\sing}} (\mu^2/m^2,r)$ should vanish for $\mu^2\to 0$.
Obviously this is not true for the separate
terms $\dot{\cal F}_{\cal J}$ and $\dot{\cal F}_{\cal S}$ in the square brackets, which do not vanish for $\mu^2\to 0$ but rather tend to a constant; according to (\ref{calF_small_epsilon}) they obey
\begin{equation}
\label{dotF_equality}
\dot{\cal F}_{\cal J}(0)=\dot{\cal F}_{\cal S}(0)=1\,.
\end{equation}
Thus, the integral (\ref{R_int_by_parts}) cannot be simply split into two finite integrals one corresponding to the `jet' contribution and one to the `soft'. According to (\ref{dR}), however, such a split must be possible upon taking one logarithmic derivative. Indeed using  (\ref{R_int_by_parts}) one gets
\begin{align}
\label{dR1}
  \left.\frac{1}{r}\,\frac{dR(m^2,r)}{d\ln m^2}\right\vert_{\rm large \,\,\beta_0}\,
  = &\,\frac{1}{\beta_0}\,\int_0^{\infty}\frac{d\mu^2}{\mu^2} \rho_V(\mu^2)  \dot{\cal F}_{\sing} (\mu^2/m^2,r)\,,
\end{align}
and therefore
\begin{align}
\label{dR2}
\left.\frac{dR(m,r)}{d\ln m^2}\right\vert_{\rm large \,\,\beta_0}\,&=&\,\frac{1}{\beta_0}\,\int_0^{\infty}\frac{d\mu^2}{\mu^2} \rho_V(\mu^2)  \bigg[\dot{\cal F}_{\cal J}\Big(\frac{\mu^2}{r m^2}\Big)-\dot{\cal F}_{\cal S}\Big(\frac{\mu^2}{r^2 m^2}\Big)\bigg]\,,
\end{align}
which, in virtue of (\ref{dotF_equality}), can be split into two finite integrals by subtracting $\dot{\cal F}_{\cal J}(0)=1$ and adding $\dot{\cal F}_{\cal S}(0)=1$ inside the square brackets.
By comparing with (\ref{dR}), Eq.~(\ref{dR2}) implies that the `jet' and `soft' anomalous dimensions have the following dispersive representation in the large--$\beta_0$ limit:
\begin{align}
  \label{Jet}
  \left.{\cal J}(k^2)\right\vert_{\rm large \,\,\beta_0}\,
  = &\,\frac{1}{\beta_0}\,\int_0^{\infty}\frac{d\mu^2}{\mu^2} \rho_V(\mu^2)  \left[\dot{\cal F}_{\cal J}(\mu^2/k^2)-\dot{\cal F}_{\cal J}(0)\right],
\end{align}
and
\begin{align}
  \label{Soft}
  \left.{\cal S}(k^2)\right\vert_{\rm large \,\,\beta_0}\,
  = &\,\frac{1}{\beta_0}\,\int_0^{\infty}\frac{d\mu^2}{\mu^2} \rho_V(\mu^2)  \left[\dot{\cal F}_{\cal S}(\mu^2/k^2)-\dot{\cal F}_{\cal S}(0)\right],
\end{align}
where we used (\ref{dotF_equality}). Thus $\dot{\cal F}_{\cal J}$ and $\dot{\cal F}_{\cal S}$  can be identified as the characteristic functions associated to the `jet'  and `soft' Sudakov anomalous dimensions ${\cal J}$ and  ${\cal S}$, respectively. The corresponding representations in term of $a_V^{\Mink}(\mu^2)$ are:
\begin{align}
  \label{Jet-1}
  \left.{\cal J}(k^2)\right\vert_{\rm large \,\,\beta_0}\,
  = &\,\frac{1}{\beta_0}\,\int_0^{\infty}\frac{d\mu^2}{\mu^2} a_V^{\Mink}(\mu^2) \ddot{\cal F}_{\cal J}(\mu^2/k^2),
\end{align}
and
\begin{align}
  \label{Soft-1}
  \left.{\cal S}(k^2)\right\vert_{\rm large \,\,\beta_0}\,
  = &\,\frac{1}{\beta_0}\,\int_0^{\infty}\frac{d\mu^2}{\mu^2} a_V^{\Mink}(\mu^2)  \ddot{\cal F}_{\cal S}(\mu^2/k^2),
\end{align}
where
\begin{equation}
\ddot{\cal F}_{\cal J}(\epsilon)=-\epsilon\,\frac{d}{d\epsilon}\,\dot{\cal F}_{\cal J}(\epsilon)\,;\qquad\quad
\ddot{\cal F}_{\cal S}(\epsilon)=-\epsilon\,\frac{d}{d\epsilon}\,\dot{\cal F}_{\cal S}(\epsilon).
\end{equation}
These representations of the Sudakov anomalous dimensions will be generalized in Sec.~\ref{sec:general_dispersive} to full QCD.

\subsubsection*{Dispersive representation of Sudakov anomalous dimensions as limits of
the full physical evolution kernels}

Eq.~(\ref{Jet}) and (\ref{Soft}) can be obtained as $r\rightarrow 0$ limits of the corresponding finite--$r$ dispersive representations of the physical evolution kernels in momentum space. We illustrate this statement (and identify the precise limit to be taken) with the examples of deep inelastic structure functions and the Drell--Yan cross section. This will also provide an illustration of the statement preceding  Eq.~(\ref{Kernel_dispersive}) above.\\

\noindent
\underline{Deep inelastic structure functions}\\
Let us consider the large--$\beta_0$ limit but \emph{finite $x$} dispersive representation \cite{DMW} of the physical evolution kernel. We note that at the level of the partonic calculation,
the convolution on the r.h.s of (\ref{F_2_evolution}) becomes trivial in the large--$\beta_0$ limit since the ${\cal O}(\alpha_s)$ corrections to $F_2(z,Q^2)$ generate terms that are subleading by
powers of $1/\beta_0$. Therefore, in this limit $K_{\DIS (F_2)}(x,Q^2)$ is directly proportional to the partonic ${dF_2(x,Q^2)}/{d\ln Q^2}$, so
\begin{align}
  \label{finite-r-DIS}
  \left.(1-x)K_{\DIS (F_2)}(x,Q^2)\right\vert_{\rm large \,\,\beta_0}\,
  = &\,\frac{C_F}{\beta_0}\,\int_0^{\infty}\frac{d\mu^2}{\mu^2} \rho_V(\mu^2)  \left[(1-x)\left(\dot{\cal F}(\mu^2/Q^2,x)-\dot{\cal F}(0,x)\right)\right],
\end{align}
where ${\cal F}(\mu^2/Q^2,x)$ is the standard notation for the characteristic function corresponding to $F_2(x,Q^2)$ (see Eq. (4.27) in~\cite{DMW}). In Eq.~(\ref{finite-r-DIS}) we multiplied both sides by a factor of $r=1-x$ in order to get a \emph{finite limit}
for the integrand for $x\to 1$ (see Eq.~(\ref{DIS_jet}) below), in agreement with the general result Eq.~(\ref{calF_general}).
Upon taking this limit under the integral with a fixed invariant mass $W^2=Q^2(1-x)/x$
one obtains
\begin{align}
  \label{small-r-DIS}
\lim_{\begin{array}{l}\vspace*{-7pt}\\
{x\to 1}\hspace*{-30pt}\\
\text{fixed}\,W^2\hspace*{-80pt}
\end{array}}
 \left\{\left. (1-x)K_{\DIS (F_2)}(x,Q^2)\right\vert_{\rm large \,\,\beta_0}\right\}\,
  = &\,\frac{C_F}{\beta_0}\,\int_0^{\infty}\frac{d\mu^2}{\mu^2} \rho_V(\mu^2)  \left[\dot{\cal F}_{\DIS}(\mu^2/W^2)-\dot{\cal F}_{\DIS}(0)\right],
\end{align}
where we substituted
\begin{align}
\label{DIS_F}
\dot{\cal F}_{\DIS}(\mu^2/W^2)=\lim_{\begin{array}{l}\vspace*{-7pt}\\
{x\to 1}\hspace*{-30pt}\\
\text{fixed}\,W^2\hspace*{-80pt}
\end{array}}
\left\{(1-x)\dot{\cal F}(\mu^2/Q^2,x)\right\}\,.
\end{align}
Eq.~(\ref{small-r-DIS}) shows in particular that the specified limit of $\left. (1-x)K_{\DIS (F_2)}(x,Q^2)\right\vert_{\rm large \,\,\beta_0}$ does exist. This (large--$\beta_0$) result is consistent with the more general result  Eq.~(\ref{K_x}) (valid also at {\em finite} $\beta_0$). Moreover,   Eq.~(\ref{K_x}) allows to connect directly the momentum-space physical quantity $K_{\DIS (F_2)}(x,Q^2)$ in the Sudakov limit with the physical Sudakov anomalous dimension, i.e. the exponentiation kernel.
Comparing with Eq.~(\ref{K_x}),  we thus deduce  Eq.~(\ref{Jet}), with the identification
\begin{align}
\label{F_DIS}
\dot{\cal F}_{\cal J}(\mu^2/k^2)=\dot{\cal F}_{\DIS}(\mu^2/k^2).
\end{align}
The explicit result~\cite{DGE,Friot:2007fd} will be given in Eq.~(\ref{DIS_jet}) below.
\\

\noindent
\underline{Drell--Yan cross section}\\
In quite a similar way, we start from the large--$\beta_0$ limit but \emph{finite $\tau$} dispersive representation of the quark--antiquark partonic cross section
$g_{q\bar{q}}(\tau,Q^2,\mu_F^2)$ in (\ref{DY}), which was calculated
in~Ref.~\cite{DMW}. By taking a logarithmic derivative with respect to $Q^2$ we obtain the
large--$\beta_0$ result for the physical Drell--Yan evolution kernel (\ref{dG_N}),
\begin{align}
  \label{finite-r-DY}
  \left.(1-\tau)K_{\DY}(\tau,Q^2)\right\vert_{\rm large \,\,\beta_0}\,
  = &\,\frac{C_F}{\beta_0}\,\int_0^{\infty}\frac{d\mu^2}{\mu^2} \rho_V(\mu^2)  \left[(1-\tau)\left(\dot{\cal F}_{\DY}(\mu^2/Q^2,\tau)-\dot{\cal F}_{\DY}(0,\tau)\right)\right],
\end{align}
where ${\cal F}_{\DY}(\mu^2/Q^2,\tau)$ is the characteristic function corresponding to
$g_{q\bar{q}}(\tau,Q^2,\mu_F^2)$ and where we multiplied by a factor of $r=1-\tau$.
Next, by taking the $\tau\to 1$ limit under the integral while fixing the total energy radiated into the final state, $E_{\DY}=Q(1-\tau)$, we obtain:
\begin{align}
  \label{small-r-DY}
\lim_{\begin{array}{l}\vspace*{-7pt}\\
{\tau\to 1}\hspace*{-30pt}\\
\text{fixed}\,E_{\DY}\hspace*{-80pt}
\end{array}}
 \left\{\left. (1-\tau)K_{\DY}(\tau,Q^2)\right\vert_{\rm large \,\,\beta_0}\right\}\,
  = &\,\frac{C_F}{\beta_0}\,\int_0^{\infty}\frac{d\mu^2}{\mu^2} \rho_V(\mu^2)  \left[2\dot{\cal F}_{\DY}(\mu^2/E_{\DY}^2)-2\dot{\cal F}_{\DY}(0)\right],
\end{align}
where we substituted
\begin{align}
\label{DY_F}
2\dot{\cal F}_{\DY}(\mu^2/E_{\DY}^2)=\lim_{\begin{array}{l}\vspace*{-7pt}\\
{\tau\to 1}\hspace*{-30pt}\\
\text{fixed}\,E_{\DY}\hspace*{-80pt}
\end{array}}
\left\{(1-\tau)\dot{\cal F}_{\DY}(\mu^2/Q^2,\tau)\right\},
\end{align}
and the limit can be shown~\cite{Friot:2007fd} to exist, see Eq.~(\ref{Fr-x-scaling-DY}) below.
Comparing with Eq.~(\ref{K_x_DY}),  we deduce  Eq.~(\ref{Soft}) for $\cal S=\cal S_{\DY}$, with the identification:
\begin{align}
\label{F_DY}
\dot{\cal F}_{\cal S}(\mu^2/k^2)=\dot{\cal F}_{\DY}(\mu^2/k^2).
\end{align}

\subsection{Characteristic functions of Sudakov anomalous dimensions: results\label{sec:calF_results}}

So far we have analyzed the properties of the physical Sudakov anomalous dimensions and their dispersive formulation in the large--$\beta_0$ limit based on general considerations. Let us now summarize the results for the characteristic functions based on explicit calculations.
These calculations have been usually done considering a specific distribution at finite $r$,
computing the full characteristic function ${\cal F}(\epsilon,r)$ by evaluating the squared matrix
element with an off-shell gluon, and only at the end identifying the terms that are singular
at $r\to 0$ as demonstrated
in Eqs. (\ref{DIS_F}) and (\ref{DY_F}) above. There is however an alternative:
it was shown in \cite{DGE} that there exists a systematic approximation ---
where one lightcone component
of the gluon is taken small together with its transverse momentum squared and its virtuality ---
that allows a direct calculation of the $r\to 0$ singular terms, namely ${\cal F}_{\cal J,\, S}$.
The calculation of the soft characteristic function ${\cal F}_{\cal S}$ can be
simplified further, taking all the gluon momentum components to be small.
This is the standard soft approximation albeit with an off--shell gluon. Here
the hard partons that emit the radiation can be replaced by Wilson lines.
The soft function can then be obtained through the renormalization of a corresponding
Wilson--loop operator defined by the trajectories of the colored hard
partons~\cite{Korchemskaya:1992je,Korchemsky:1992xv,Gardi:2005yi,Gardi:2006jc}.

A compilation of the expressions for the characteristic functions in a few examples is given
in Table~\ref{table:calF}. The results for the derivative
$\dot{\cal F}_{\cal J,\, S}$, defined by Eq.~(\ref{d_cal_F}) or (\ref{cal_F_from_d_cal_F}), and the corresponding
Borel function, defined by Eq.~(\ref{B_relation_dotSJ}) below,
are summarized in Table~\ref{table:calF_B}.

\begin{table}[htb]
  \centering
  \begin{tabular}{|l|l|c|}
    \hline
    &&\vspace*{-13pt}\\
     process        & & ${\cal F}_{\cal J}(y=\epsilon/r)^{\qquad}$
    \\ \hline\hline
    &&\vspace*{-6pt}\\
    \begin{tabular}{l}
    jet function\\
    (e.g. DIS)
    \end{tabular}
    & ${\cal J}$            & ${\displaystyle{\theta(y<1)\left[
    -\ln(y)-\frac{3}{4}+\frac{1}{2}\,y
+\frac{1}{4}\,y^2\right]}}$ \\
    &&\vspace*{-6pt}\\
    \hline\hline
    &&\vspace*{-13pt}\\
     process        & & ${\cal F}_{\cal S}(z=\epsilon/r^2)^{\qquad}$
    \\ \hline
    &&\vspace*{-6pt}\\
    \begin{tabular}{l}
    B decay; HQ  \\
    Fragmentation
    \end{tabular}
    & ${\cal S}_{\QD}$     &
    ${\displaystyle
    \ln\left(1+1/z\right) +\frac{1}{1+z}
    }$     \\
    &&\vspace*{-6pt}\\
    \hline
    &&\vspace*{-6pt}\\
    \begin{tabular}{l}
    Drell--Yan\,; \\
    $gg \to$ Higgs
    \end{tabular}
    & ${\cal S}_{\DY}$ & ${\displaystyle \theta\left({z}<1/4\right)
    \left[2\ln\frac{1+\sqrt{1-4z}}{2}-\ln(z)\right]
    }$  \\
    &&\vspace*{-6pt}\\
    \hline
    &&\vspace*{-12pt}\\
    \begin{tabular}{l}
    $e^+e^-\to \text{jets}$\\
    C parameter\\ ($r=c=C/6$)
    \end{tabular}
    & ${\cal S}_{c}$       & ${\displaystyle -{\theta (z<4)}\ln(z)+{\theta(z>4)}\left[2\, {\tanh}^{-1}\left(\sqrt{1-4/z}\right)-\ln(z)\right]
      }$   \\
    &&\vspace*{-12pt}\\\hline
    &&\vspace*{-12pt}\\
    \begin{tabular}{l}
    $e^+e^-\to \text{jets}$\\
    Thrust\\ ($r=t=1-T$)
    \end{tabular}
     & ${\cal S}_{t}$       & ${\displaystyle -{\theta (z<1)}\ln(z)}$ \vspace*{-12pt}\\
    &&\\\hline
  \end{tabular}
  \caption{Results for the momentum--space characteristic
  functions  of some inclusive distributions.}
  \label{table:calF}
  \end{table}

\begin{table}[htb]
  \centering
  \begin{tabular}{|l|l|c|c|}
    \hline
    &&&\vspace*{-13pt}\\
     process        &  & $\dot{\cal F}_{\cal J}(y=\epsilon/r)^{\qquad}$ & ${\mathbb B}_{\cal J}(u)$
    \\ \hline\hline
    &&&\vspace*{-6pt}\\
    \begin{tabular}{l}
    jet function\\
    (e.g. DIS)
    \end{tabular}
    & ${\cal J}$            & ${\displaystyle{\theta(y<1)\left[1-\frac{y}{2}-\frac{y^2}{2}\right]}}$ & ${\displaystyle \frac{1}{2}\left(\frac{1}{1-u}+\frac{1}{1-u/2}\right)}$\\
    &&&\vspace*{-6pt}\\
    \hline\hline
    &&&\vspace*{-13pt}\\
     process        &  & $\dot{\cal F}_{\cal S}(z=\epsilon/r^2)^{\qquad}$ & ${\mathbb B}_{\cal S}(u)$
    \\ \hline
    &&&\vspace*{-6pt}\\
    \begin{tabular}{l}
    B decay; HQ  \\
    Fragmentation
    \end{tabular}
    & ${\cal S}_{\QD}$     & ${\displaystyle 1-\frac{1}{\left(1+1/z\right)^2}}$   & ${\displaystyle
    (1-u)\frac{\pi u}{\sin \pi u}}$  \\
    &&&\vspace*{-6pt}\\
    \hline
    &&&\vspace*{-6pt}\\
    \begin{tabular}{l}
    Drell--Yan\,; \\
    $gg \to$ Higgs
    \end{tabular}
    & ${\cal S}_{\DY}$ & ${\displaystyle \frac{\theta\left({4z}<1\right)}{\sqrt{1-4z}}}$ & ${\displaystyle \frac{\Gamma^2(1-u)}{\Gamma(1-2u)}}$ \\
    &&&\vspace*{-6pt}\\
    \hline
    &&&\vspace*{-12pt}\\
    \begin{tabular}{l}
    $e^+e^-\to \text{jets}$\\
    C parameter\\ ($r=c=C/6$)
    \end{tabular}
    & ${\cal S}_{c}$       & ${\displaystyle 1-\frac{\theta(z>4)}{\sqrt{1-4/z}}}$  & ${\displaystyle \frac{\Gamma^2(1+u)}{\Gamma(1+2u)}}$ \\
    &&&\vspace*{-12pt}\\\hline
    &&&\vspace*{-12pt}\\
    \begin{tabular}{l}
    $e^+e^-\to \text{jets}$\\
    Thrust\\ ($r=t=1-T$)
    \end{tabular}
     & ${\cal S}_{t}$       & ${\displaystyle 1-\theta(z>1)}$ & 1\vspace*{-12pt}\\
    &&&\\\hline
  \end{tabular}
  \caption{Results for the logarithmic derivatives of momentum--space characteristic
  functions of some inclusive distributions and the corresponding Borel functions (\ref{B_relation_dotSJ}) .}
  \label{table:calF_B}
\end{table}

\vspace*{10pt}
\noindent
\underline{Inclusive B decay and heavy--quark fragmentation}\\
Let us consider the characteristic function corresponding to the exponentiation kernel in inclusive B decays. The simplest calculation is of the photon--energy spectrum $d\Gamma(m_b^2,x)/dx$ in
radiative B decays ${\bar B}\to X_s \gamma$,
where $x=2E_\gamma/m_b$. This spectrum was computed in the large--$\beta_0$ limit in \cite{Gardi:2004ia}.
The original calculation was done using the Borel representation of the dressed gluon propagator. In the dispersive formulation we obtain the following singular terms for $r=1-x\to 0$:
\begin{align}
\label{calF}
{\cal F}_{\sing} (\epsilon,r) = \,\frac{1}{r}\,\Bigg\{\underbrace{\theta (\epsilon<r)\left[\,
-\ln(\epsilon/r)-\frac{3}{4}+\frac{1}{2}\frac{\epsilon}{r}
+\frac{1}{4}\frac{\epsilon^2}{r^2}\right]}_{{\cal F}_{\cal J}(\epsilon/r)}\,-\underbrace{\left[\ln\left(1+r^2/\epsilon\right)\,+\,
\frac{1}{1+\epsilon/r^2}\right]}_{{\cal F}_{\cal S_{\QD}}(\epsilon/r^2)}\Bigg\}\,
,
\end{align}
with $\epsilon=\mu^2/m_b^2$, where we have already split the result according to (\ref{calF_general}) and identified the two functions ${\cal F}_{\cal J}(\epsilon/r)$ and ${\cal F}_{\cal S_{\QD}}(\epsilon/r^2)$. Equivalently, for the derivative we have:
\begin{align}
\label{calF_dot_QD}
\dot{\cal F}_{\sing} (\epsilon,r) \equiv \frac{d{\cal F}_{\sing}(\epsilon,r)}{d\ln 1/\epsilon} =\frac{1}{r}\Bigg\{
\underbrace{\,\theta (\epsilon<r)\left[\,
1-\frac{1}{2}\frac{\epsilon}{r}
-\frac{1}{2}\frac{\epsilon^2}{r^2}\right]}_{\dot{\cal F}_{\cal J}(\epsilon/r)}\,-\underbrace{\left[\frac{1}{1+\epsilon/r^2}
+\frac{\epsilon/r^2}
{(1+\epsilon/r^2)^2}\right] }_{\dot{\cal F}_{\cal S_{\QD}}(\epsilon/r^2)}\Bigg\}
\,.
\end{align}
Here the superscript `QD' stands for Quark Distribution, indicating that this particular soft function is related to the longitudinal momentum distribution of an off--shell b quark inside an on--shell b-quark state. The same result for the singular terms (\ref{calF}) was obtained for the triple differential rate in $b\to X_u l \bar{\nu}$; here $r$ is defined as the ratio between the two lightcone components $p^+/p^-$ of the $X_u$ jet, $p^{\pm}=E \mp |\vec{p}|$, and the hard scale is $p^-$.

Next we comment that both ${\cal F}_{\cal J}(\epsilon/r)$ and ${\cal F}_{\cal S}(\epsilon/r^2)$ can be extracted from the calculation of heavy--quark fragmentation in $e^+e^-$ annihilation, Eq.~(54) in \cite{Cacciari:2002xb}. The fact that ${\cal F}_{\cal S}(\epsilon/r^2)$ is the same for heavy quark decay and fragmentation follows from the general equality of the two Sudakov anomalous dimensions, proven in~\cite{Gardi:2005yi}.

Finally, the quark distribution function at finite mass and arbitrary momentum fraction has been
computed in the large--$\beta_0$ limit in Ref.~\cite{Gardi:2005yi} using the Borel regularization.
Repeating the calculation with a massive gluon and identifying the terms that are singular
for $r\to 0$ we recover ${\cal F}_{{\cal S}_{\QD}}(\epsilon/r^2)$ of (\ref{calF}).

\vspace*{10pt}
\noindent
\underline{The jet function}\\
As already mentioned in the introduction, the jet function appears in many different inclusive distributions where Sudakov logs emerge from a constraint on the invariant mass of a jet in the final state~\cite{DGE}. The simplest example is deep--inelastic structure functions, where ${\cal J}$ is the only source of Sudakov logarithms~\cite{Gardi:2002xm}. According to Eq.~(\ref{DIS_F}),
the `jet' characteristic function is obtained upon considering the $x\to 1$ limit of the $F_2(x,Q^2)$ characteristic function (Eq.~(4.27) in~\cite{DMW}) with a fixed invariant mass $Q^2(1-x)/x$. The result is~\cite{DGE,Friot:2007fd}:
\begin{align}
\label{DIS_jet}
\begin{split}
{\cal F}_{\cal J}(\epsilon/(1-x))&=\lim_{\begin{array}{l}
{x\to 1}\hspace*{-30pt}\\
\text{fixed}\,\epsilon\,/(1-x)\hspace*{-80pt}
\end{array}}
\left\{(1-x){\cal F}(\epsilon,x)\right\}\\
&=\theta (\epsilon<1-x)\left[-\ln\left(\frac{\epsilon}{1-x}\right)-\frac{3}{4}+
\frac{1}{2}
\frac{\epsilon}{1-x}
+\frac{1}{4}\frac{\epsilon^2}{(1-x)^2}\right]\,,
\end{split}
\end{align}
where $\epsilon=\mu^2/Q^2$. As anticipated, this result is \emph{identical}
to the jet--function contribution to various infrared safe observables including
 inclusive B decay (see~\cite{Gardi:2004ia}), Eq.~(\ref{calF}) above with $r=1-x$, and a range of distributions in
$e^+e^-$ annihilation into hadrons~\cite{DGE}: event--shape variables ---
e.g. Eq.~(16) in \cite{Gardi:2002bg} and Sec.~3 in \cite{GM},
single particle inclusive cross section --- Eq.~(28) in \cite{DGE},
and heavy--quark production cross section.

\vspace*{10pt}
\noindent
\underline{The Drell--Yan soft function}\\
The Drell--Yan (or Higgs) production case offers the basic example of a `soft' function.
Using in Eq.~(\ref{DY_F}) above the explicit expression from Ref.~\cite{DMW} for the Drell--Yan characteristic function
${\cal F}_{\DY}(\epsilon,\tau)$ with $\epsilon=\mu^2/Q^2$, one
obtains~\cite{Friot:2007fd}:
\begin{align}
\label{Fr-x-scaling-DY}
{\cal F}_{{\cal S}_{\DY}}(\epsilon/(1-\tau)^2)&=\frac{1}{2}\lim_{\begin{array}{l}
\tau\rightarrow 1 \hspace*{-35pt}\\
\epsilon/(1-\tau)^2 \,\,\text{fixed}\hspace*{-80pt}
\end{array}
}
\left\{(1-\tau){\cal F}_{\DY}(\epsilon,\tau)\right\}\\
&=
2\,\textrm{tanh}^{-1}\left(\sqrt{1-\frac{4\epsilon}{(1-\tau)^2}}\right)
\,\theta\left({4\epsilon}/{(1-\tau)^2}<1\right)
\,.
\end{align}
Upon defining $z\equiv \epsilon/(1-\tau)^2$,
\begin{align}
\label{calF_DY_z}
{\cal F}_{{\cal S}_{\DY}}(z)= \,2\,\textrm{tanh}^{-1} \left(\sqrt{1-4z}\right)\,\theta(z<1/4)=
\,\left[2\ln\frac{1+\sqrt{1-4z}}{2}-\ln(z)\right]\,\theta(z<1/4)
\end{align}
and therefore
\begin{align}
\label{Fr-x-scaling-DY1}
\dot{\cal F}_{{\cal S}_{\DY}}(\epsilon/(1-\tau)^2)=
\frac{\theta\left({4\epsilon}/{(1-\tau)^2}<1\right)}
{\sqrt{1-{4\epsilon}/{(1-\tau)^2}}}
\,.
\end{align}
The same result has been obtained in Ref.~\cite{Beneke:1995pq} (see Eq. (4.6) there) and
Ref.~\cite{DGE} (see Eq. (69) and the discussion following it) by considering directly the
renormalon calculation in the soft approximation.
Finally, note that in the Drell--Yan case the second derivative
 $\ddot{\cal F}_{{\cal S}_{\DY}}$ is singular
and therefore one must use (\ref{Soft}) rather than (\ref{Soft-1}).

\vspace*{10pt}
\noindent
\underline{Event--shape distributions}\\
Beyond the universal jet function discussed above, event--shape distributions in
$e^+e^-$ annihilation are sensitive to large--angle soft gluon emission from the
back--to--back recoiling quarks. The soft function strongly
depends on the way the shape variable weighs the parton momenta.
This is reflected in shape--variable dependent subleading
Sudakov logarithms as well as power corrections.

Before considering specific examples, a general comment is due concerning the not--completely--inclusive nature of event--shape variables: in contrast with the
inclusive cross sections and decay spectra discussed above, event--shape variables do distinguish at some level between an off-shell gluon and the final--state particles to which it decays.
Differences arise already in the large--$\beta_0$ limit.
The characteristic functions for event--shape variables are defined in the inclusive approximation, neglecting any such differences.

Let us first recall the result in the case of the thrust
(or the heavy jet mass) which are particularly simple. Starting from the exponentiation
kernel in the large--$\beta_0$ limit
(e.g. Eq. (20) in \cite{Gardi:2002bg}) we have, in full analogy with
Eq.~(\ref{R_int_by_parts}),
\begin{align}
\label{R_int_by_parts_Thrust}
\begin{split}
\left.C_R\frac{R(m^2,t)}{t}\right\vert_{\rm large \,\,\beta_0}\,
& = \frac{C_R}{\beta_0}\, \int_0^{\infty}
\frac{d\epsilon}{\epsilon} a_V^{\Mink}(\epsilon Q^2)
\,
\underbrace{\frac{1}{t}\,\left[1-\frac{1}{2}\frac{\epsilon}{t}-
\frac{1}{2}
\frac{\epsilon^2}{t^2}\right]\,\theta(\epsilon<t)\,
\theta(\epsilon>t^2)}_{{\cal \dot{F}_{\sing}} (\epsilon,t)} \\
\end{split}
\end{align}
where $\epsilon=\mu^2/Q^2$, $Q^2$ is the center--of--mass energy
and $t=1-T$, where $T$ is the thrust variable. For thrust $C_R=2C_F$.
In order to extract the separate `jet' and `soft' contributions to
the characteristic function we use the identity:
\[
\,\theta(\epsilon<t)\,
\theta(\epsilon>t^2)=
\theta(\epsilon<t)\,-\,\Big(1\,-\,
\theta(\epsilon>t^2)\Big)
\]
and note that the terms proportional to $\epsilon/t$ or
$\epsilon^2/t^2$ in (\ref{R_int_by_parts_Thrust}) contribute only at the $\epsilon=t$ phase--space
limit (they are power suppressed at the $\epsilon=t^2$ limit).
This leads to the following decomposition of
${\cal \dot{F}_{\sing}} (\epsilon,t)$ in
(\ref{R_int_by_parts_Thrust})
\begin{align}
{\cal \dot{F}_{\sing}} (\epsilon,t)=
\,\frac{1}{t}\,\,\Bigg\{\underbrace{
\theta(\epsilon<t)\,\left[1-\frac{1}{2}\frac{\epsilon}{t}-
\frac{1}{2}
\frac{\epsilon^2}{t^2}\right]}_{\dot{\cal F}_{{\cal J}}(\epsilon/t)}-
\,\underbrace{\Big(1\,-\,
\theta(\epsilon>t^2)\Big)}_{\dot{\cal F}_{{\cal S}_{t}}(\epsilon/t^2)}
\Bigg\}\,,
\end{align}
where we neglected regular contributions that are irrelevant for Sudakov resummation.
Here we recognize the characteristic function of the
jet from (\ref{calF_dot_QD}) and (\ref{DIS_jet}) while for the soft
contribution we have:
\begin{align}
\label{dF_thrust}
&\dot{\cal F}_{{\cal S}_{t}}(\epsilon/t^2)= 1-\theta(\epsilon/t^2>1)
\,=\,\theta(\epsilon/t^2<1)\,.
\end{align}

In a similar way we can write down the characteristic function of the $C$ parameter
using the results of~Ref.~\cite{GM}.
Upon neglecting regular contributions, we obtain (see Eq. (3.5) in~\cite{GM}):
\begin{align}
\begin{split}
{\cal \dot{F}_{\sing}} (\epsilon,c)&=
\,\theta(\epsilon<c)\,
\theta(\epsilon>4c^2)
\,\frac{1}{c}\,\,\Bigg\{1-\frac{1}{2}\frac{\epsilon}{c}-
\frac{1}{2}\frac{\epsilon^2}{c^2}+\frac{4c^2/\epsilon}{
\sqrt{1-4c^2/\epsilon}\,
\left(1+\sqrt{1-4c^2/\epsilon}\right)}
\Bigg\}\\
&\simeq\,\frac{1}{c}\,\,
\Bigg\{\underbrace{\theta(\epsilon<c)\left[1-\frac{1}{2}\frac{\epsilon}{c}-
\frac{1}{2}\frac{\epsilon^2}{c^2}\right]}_{\dot{\cal F}_{{\cal J}}(\epsilon/c)}-
\underbrace{\left[1-
\frac{\theta(\epsilon>4c^2)}{\sqrt{1-4c^2/\epsilon}}\right]}_{\dot{\cal F}_{{\cal S}_{c}}(\epsilon/c^2)}
\Bigg\}
\end{split}
\end{align}
with $c=C/6$, where $C$ is the conventional normalization of the $C$ parameter
(see \cite{GM}).
Again we recognize the familiar jet function and find to be:
\begin{align}
\label{dF_c}
&\dot{\cal F}_{{\cal S}_{c}}(\epsilon/c^2)=
1-\frac{\theta(\epsilon/c^2>4)}{\sqrt{1-4c^2/\epsilon}}\,.
\end{align}
Note the similarity~\cite{Gardi:2006jc} with the case of the Drell--Yan soft function, Eq.~(\ref{Fr-x-scaling-DY1}).
Also here the second derivative $\ddot{\cal F}_{{\cal S}_{c}}$ is singular,
so one can only use (\ref{Soft}).

\section{From the large--$\beta_0$ limit to the non-Abelian theory~\label{sec:general_dispersive}}

The dispersive technique is a convenient way to derive the all--order result in the large--$\beta_0$ limit. But, in fact, it is much more than that: a dispersive representation, when it exists, summarizes the analytic structure of the observable in the complex momentum plane as well as its infrared sensitivity.  Moreover, it is particularly well--suited to parametrizing power corrections by means of infrared--finite coupling.

The purpose of the present section is to show that the dispersive representation of the Sudakov anomalous dimensions,  Eqs.~(\ref{Jet}) and (\ref{Soft}), which emerges out of the calculation in the large--$\beta_0$ limit, can be readily generalized to the full theory. In this generalization the characteristic functions, which have been computed analytically  in the large--$\beta_0$ limit, are kept fixed, while the V--scheme coupling $a_V^{\Mink}$, which was so far identified only to NLO in the large--$\beta_0$ limit, is replaced by specific effective charges that are defined order-by-order in QCD and capture the non-Abelian nature of the interaction. After presenting the general formalism we will examine a few examples where corrections are known to the NNLO and beyond. The investigation of the evolution of these effective charges leads to interesting observations.

\subsection{A general dispersive representation of Sudakov anomalous dimensions~\label{sec:general_disparsive_formulae}}

The generalization of Eqs.~(\ref{Jet}) and (\ref{Soft})
(as well as~(\ref{Jet-1}) and (\ref{Soft-1}))
beyond the large--$\beta_0$ limit is straightforward:
\begin{align}
  \label{J_Sud_anom_dim_conv}
  \begin{split}
  {\cal J}(k^2)\,
  = &\,\frac{1}{\beta_0}\,\int_0^{\infty}\frac{d\mu^2}{\mu^2} \rho_{\cal J}(\mu^2)  \left[\dot{\cal F}_{\cal J}(\mu^2/k^2)-\dot{\cal F}_{\cal J}(0)\right]\\
  = &\,\frac{1}{\beta_0}\,\int_0^{\infty}\frac{d\mu^2}{\mu^2} a_{\cal J}^{\Mink}(\mu^2) \ddot{\cal F}_{\cal J}(\mu^2/k^2),
  \end{split}
 \end{align}
and
\begin{align}
  \label{S_Sud_anom_dim_conv}
  \begin{split}
 {\cal S}(k^2)\,
 = &\,\frac{1}{\beta_0}\,\int_0^{\infty}\frac{d\mu^2}{\mu^2} \rho_{\cal S}(\mu^2)  \left[\dot{\cal F}_{\cal S}(\mu^2/k^2)-\dot{\cal F}_{\cal S}(0)\right]\\
 = &\,\frac{1}{\beta_0}\,\int_0^{\infty}\frac{d\mu^2}{\mu^2} a_{\cal S}^{\Mink}(\mu^2)  \ddot{\cal F}_{\cal S}(\mu^2/k^2)\,,
 \end{split}
\end{align}
where, in full analogy with (\ref{eq:int_discontinuity}),
\begin{equation}
\label{eq:int_discontinuity_JS}
\rho_{\cal J,\, S}(\mu^2)=\frac{d a_{\cal J,\, S}^{\Mink}(\mu^2)}{d\ln\mu^2}\,;\qquad\quad
 a_{\cal J,\, S}^{\Mink}(\mu^2)\equiv -\int_{\mu^2}^{\infty} \frac{dm^2}{m^2}\rho_{\cal J,\, S}(m^2)\,,
\end{equation}
and, similarly to (\ref{eq:discontinuity}),  $\rho_{\cal J,\, S}(\mu^2)$ correspond to the timelike discontinuities of some ``Euclidean'' effective charges, originally defined for spacelike momenta:
\begin{align}
  \label{Eucl-Mink}
  a_{\cal J,\, S}^{\Eucl}( k^2)\,
  &= \int_0^{\infty}\frac{d\mu^2}{\mu^2} \,a_{\cal J,\, S}^{\Mink}(\mu^2) \, \frac{\mu^2/k^2}{(1+\mu^2/k^2)^2}
  = -\int_0^{\infty}\frac{d\mu^2}{\mu^2+k^2} \,\rho_{\cal J,\, S}(\mu^2)\,,
\end{align}
where we ignored potential Landau singularities in $a_{\cal J,\,S}^{\Eucl}( k^2)$ that would violate the physical analytic properties of these effective charges ({\it cf.} Eq.~(\ref{eq:simple-dispersive})). Of course, such singularities may appear at any given order in perturbation theory.
We shall return to this issue in Secs.~\ref{sec:ECH_evolution} and~\ref{sec:PC}.

Let us note in passing that while the Minkowskian representation
of the Sudakov anomalous dimensions, Eqs. (\ref{J_Sud_anom_dim_conv}) and (\ref{S_Sud_anom_dim_conv}), can always be written, in general \emph{there are no equivalent Euclidean representations} consisting of integrals over the effective charges $a_{\cal J,\, S}^{\Eucl}(k^2)$.
This fact reflects the timelike nature of these quantities.

Obviously, we require that in the large--$\beta_0$ limit $\rho_{\cal S}(\mu^2)$ and $\rho_{\cal J}(\mu^2)$ would both coincide with the discontinuity of the V--scheme coupling, $\rho_V(\mu^2)$ of Eq.~(\ref{eq:discontinuity}). In Sec.~\ref{ECH:order_by_order} we will
show, order-by-order in perturbation theory, that a generalization according to (\ref{J_Sud_anom_dim_conv}) and (\ref{S_Sud_anom_dim_conv}) is indeed possible, and then further characterize these effective charges. We stress that the order-by-order analysis is entirely independent of the assumed analytic structure, as Landau singularities can only modify the dispersion relations (\ref{Eucl-Mink}) by power--suppressed terms.

\subsection{From anomalous dimensions to effective charges:
order--by--order relations~\label{ECH:order_by_order}}

We wish to show that the perturbative expansion of the anomalous dimensions, ${\cal S}$ and~${\cal J}$, and consequently their conventional decomposition in ${\overline{\rm MS}}$ into the cusp ${\cal A}$, ${\cal B}$ and ${\cal D}$, Eq.~(\ref{Sud_anom_dim}), are effectively translated by (\ref{J_Sud_anom_dim_conv}) and (\ref{S_Sud_anom_dim_conv}), order-by-order in perturbation theory, into an expansion of the corresponding effective charges, $a_{\cal J,\, S}^{\Mink}(\mu^2)$. Thus, by virtue of the dispersion relations (\ref{Eucl-Mink}), they uniquely translate into the expansion of the Euclidean effective charges, $a_{\cal J,\, S}^{\Eucl}(\mu^2)$.

Defining  the expansion coefficients of $a_{\cal J,\, S}^{\Eucl}(\mu^2)$ in the $\overline{\rm MS}$ scheme
\begin{align}
\begin{split}
\label{Eucl-expansions}
  a_{\cal J}^{\Eucl}(\mu^2)&=\bar{a}
+ j_2^E \bar{a}^2
+j_3^E
\bar{a}^3
+j_4^E
\bar{a}^4
+\cdots\\
  a_{\cal S}^{\Eucl}(\mu^2)&=\bar{a}
+s_2^E
\bar{a}^2
+ s_3^E
\bar{a}^3
+ s_4^E
\bar{a}^4
+\cdots,
\end{split}
\end{align}
where $\bar{a}\equiv \beta_0 \alpha_s^{\MSbar}(\mu^2)/\pi$, we have
\begin{align}
\begin{split}
\label{Mink-expansions}
 a_{\cal J}^{\Mink}(\mu^2)&=\bar{a}
+ j_2^E
\bar{a}^2
+\left[ j_3^E-\frac{\pi^2}{3}\right]
\bar{a}^3
+\left[ j_4^E-\pi^2\left(j_2^E+\frac{5}{6}\delta\right)\right]
\bar{a}^4
+\cdots\,,\\
 a_{\cal S}^{\Mink}(\mu^2)&=
\bar{a}
+ s_2^E
\bar{a}^2
+\left[ s_3^E-\frac{\pi^2}{3}\right]
\bar{a}^3
+\left[ s_4^E-\pi^2\left(s_2^E+\frac{5}{6}\delta\right)\right]
\bar{a}^4
+\cdots\,,
\end{split}
\end{align}
where we used (\ref{Eucl-Mink}) and the perturbative expansion of the running coupling in the $\overline{\rm MS}$ scheme, where:
\begin{align}
\label{a_zmu}
\begin{split}
\bar{a}(z\mu^2)\,=\,\bar{a}(\mu^2)-\ln(z)\bar{a}^2(\mu^2) \,&+\left(\ln^2(z)-\delta\ln(z)\right)\bar{a}^3(\mu^2)\\
+&\left(-\ln^3(z)+\frac{5}{2}\delta\ln^2(z)-\delta_2\ln(z)\right)
\bar{a}^4(\mu^2)
+\cdots
\end{split}
\end{align}
where the $\beta$ function is given by
\begin{equation}
\label{beta_MSbar}
\frac{d\bar{a}}{d\ln \mu^2}=\,-\bar{a}^2\Big[1+\delta\,\bar{a}+\delta_2\,
\bar{a}^2+\cdots\Big],
\end{equation}
with $\delta=\beta_1/{\beta_0}^2$ and $\delta_n=\beta_n/{\beta_0}^{n+1}$.

Next we note that Eqs. (\ref{J_Sud_anom_dim_conv}) and (\ref{S_Sud_anom_dim_conv}) can be written as
\begin{align}
\label{J_Sud_anom_dim_conv1}
{\cal J}(\mu^2)&=\,\frac{1}{\beta_0}\,\int_0^{\infty}\frac{dy}{y}\,
a_{\cal J}^{\Mink}(y\mu^2)\,\ddot{\cal F}_{\cal J}(y)\,
\end{align}
and
\begin{align}
\label{S_Sud_anom_dim_conv1}
{\cal S}(\mu^2)&=\,\frac{1}{\beta_0}\,
\int_0^{\infty}\frac{dz}{z}
a_{\cal S}^{\Mink}(z\mu^2)\,\ddot{\cal F}_{\cal S}(z)\,.
\end{align}
Substituting (\ref{a_zmu}) into (\ref{Mink-expansions}) we get:
\begin{align}
\label{a_zmu2_expansion1}
\begin{split}
&a_{\cal J}^{\Mink}(y\mu^2)\,=\bar{a}(\mu^2)+\left(j_2^E-\ln(y)\right)\bar{a}^2(\mu^2)
\\&
+\left[j_3^E-\frac{\pi^2}{3}+\ln^2(y)
-2\ln(y)j_2^E-\delta\ln(y)\right]\bar{a}^3(\mu^2)
+
\bigg[-\ln^3(y)+\left(3 j_2^E+\frac{5}{2}\delta\right)\ln^2(y)
\\&+\Big(-2 j_2^E \delta-3j_3^E +\pi^2-\delta_2\Big)
\ln(y)
+j_4^E-\pi^2\left(j_2^E+\frac{5}{6}\delta\right)\bigg]
\bar{a}^4(\mu^2)
+\cdots,
\end{split}
\end{align}
and a similar expression for $a_{\cal S}^{\Mink}(z\mu^2)$ where $j_n^E$ are replaced by $s_n^E$.
We observe that upon inserting these expansions into (\ref{J_Sud_anom_dim_conv1}) and (\ref{S_Sud_anom_dim_conv1}) we would need to evaluate the log-moments of   $\ddot{\cal F}_{\cal J}(y)$ and  $\ddot{\cal F}_{\cal S}(z)$, respectively,
\begin{align}
\label{JkSk_def}
\begin{split}
J_k&\equiv \int_0^{\infty}\frac{dy}{y}
\,\ln^k(y)\,\ddot{\cal F}_{\cal J}(y)\,,\\
S_k&\equiv \int_0^{\infty}\frac{dz}{z}
\,\ln^k(z)\,\ddot{\cal F}_{\cal S}(z)\,.
\end{split}
\end{align}
The first few log-moments in some examples are given in
Table~\ref{table:Jk_Sk} (see discussion below).
\begin{table}[h]
  \centering
  \begin{tabular}{|l|l|c|c|c|c|c|c|}
    \hline
     process        &  & $J_0$  & $J_1=b_1$& $J_2$ & $J_3$ &
     $J_4$& $J_5$
    \\ \hline\hline
    \begin{tabular}{l}
    jet function\\
    (e.g. DIS)
    \end{tabular}
    & ${\cal J}$           & 1& $-3/4$& 5/4& $-27/8$& 51/4& $-495/8$
    \\\hline\hline
     process        &  & $S_0$  & $S_1=d_1$ & $S_2$ & $S_3$
     & $S_4$ & $S_5$
    \\ \hline\hline
    \begin{tabular}{l}
    B decay; HQ  \\
    Fragmentation
    \end{tabular}
    & ${\cal S}_{\QD}$     &1& 1& $\pi^2/3$& $\pi^2$& $7\pi^4/15$ &
    $7\pi^4/3$ 
      \\
    \hline
    \begin{tabular}{l}
    Drell--Yan\,; \\
    $gg \to$ Higgs
    \end{tabular}
    & ${\cal S}_{\DY}$ &1& 0&
    $-\pi^2/3 $& $12 \zeta_3$& $-3\pi^4/5$ &
    $-40 \pi^2 \zeta_3+720 \zeta_5$
    \\
    \hline
    \begin{tabular}{l}
    $e^+e^-\to \text{jets}$\\
    C parameter
    \end{tabular}
    & ${\cal S}_{c}$      & 1& 0& $-\pi^2/3 $& $-12 \zeta_3$& $-3\pi^4/5$
    & $40 \pi^2 \zeta_3-720 \zeta_5$
    \\\hline
    \begin{tabular}{l}
    $e^+e^-\to \text{jets}$\\
    Thrust
    \end{tabular}
     & ${\cal S}_{t}$       &1 & 0& 0& 0& 0& 0
     \vspace*{-12pt}\\
    &&&&&&&\\\hline
  \end{tabular}
  \caption{Summary of results for the first few log-moments
  of the characteristic functions, $J_k$ and $S_k$ with $k=0$ to 5,
  of some inclusive distributions ({\it cf.} Table \ref{table:calF_B}).
  The log-moments are defined in Eq.~(\ref{JkSk_def}).
They can be most
  easily extracted using (\ref{B_relation_dotSJ})
  by expanding the Borel functions in Table \ref{table:calF_B}.
  }\label{table:Jk_Sk}
\end{table}

Using Eq. (\ref{a_zmu2_expansion1}), Eq. (\ref{J_Sud_anom_dim_conv1})  yields the following expansions:
\begin{align}
\label{expansions1}
\begin{split}
C_R\ {\cal J}(\mu^2)&=
\frac{C_F}{\beta_0}\bigg[\bar{a}
+\left(j_2^E-J_1\right)
\bar{a}^2
+\left(j_3^E-\frac{\pi^2}{3}+J_2-2J_1 j_2^E-J_1\delta\right)
\bar{a}^3
+\cdots\bigg]\\
&=\frac{C_R}{\beta_0}\left[\left(\bar{a}
+a_2
\bar{a}^2
+a_3
\bar{a}^3
+\cdots\right)-
\left(\bar{a}^2+\delta\bar{a}^3+\cdots\right)\left(b_1+2b_2\bar{a}+\cdots
\right)\right]\\
&={\cal A}(\alpha_s(\mu^2)+\frac{d\bar{a}(\mu^2)}{d\ln\mu^2}
\frac{d{\cal B}(\alpha_s(\mu^2))}{d\bar{a}}\,,
\end{split}
\end{align}
where we have identified the expansions of the anomalous dimensions ${\cal A}(\alpha_s(\mu^2))$ and ${\cal B}(\alpha_s(\mu^2))$
of Eqs. (\ref{A_cusp}) and (\ref{calB}), respectively. At order ${\cal O}(\bar{a}^2)$ we find $b_1=J_1$.
Eq.~(\ref{expansions1}) yields the following expressions for
the coefficients for the Euclidean effective charge
$a_{\cal J}^{\Eucl}$ in (\ref{Eucl-expansions}):
\begin{align}
\label{j2_to_4_general}
\begin{split}
j_2^E &= a_2\\
j_3^E &= a_3-J_2+2 b_1 a_2+\frac{1}{3}\pi^2-2b_2\\
j_4^E &= a_4+6b_1^2 a_2+\left(2 a_2\delta-3 J_2-6 b_2+3 a_3\right)b_1
+\left(-3 a_2-\frac{5}{2}\delta\right)J_2+J_3
\\&\hspace*{20pt}+\frac{5}{6}\pi^2\delta-3b_3+\pi^2a_2-2\delta b_2\,,
\end{split}
\end{align}
where the log-moments $J_k$ are defined in (\ref{JkSk_def})
and $a_n$ and $b_n$ are the $\overline{\rm MS}$
coefficients of (\ref{A_cusp}) and (\ref{calB}), respectively.

Similarly (\ref{S_Sud_anom_dim_conv1}) yields:
\begin{align}
\label{expansions2}
\begin{split}
C_R\ {\cal S}(\mu^2)&=
\frac{C_F}{\beta_0}\bigg[\bar{a}
+\left(s_2^E-S_1\right)
\bar{a}^2
+\left(s_3^E-\frac{\pi^2}{3}+S_2-2S_1s_2^E-S_1\delta\right)
\bar{a}^3
+\cdots\bigg]\\
&=\frac{C_R}{\beta_0}\left[\left(\bar{a}
+a_2
\bar{a}^2
+a_3
\bar{a}^3
+\cdots\right)-
\left(\bar{a}^2+\delta\bar{a}^3+\cdots\right)\left(d_1+2d_2\bar{a}+\cdots
\right)\right]\\
&={\cal A}(\alpha_s(\mu^2)+\frac{d\bar{a}(\mu^2)}{d\ln\mu^2}
\frac{d{\cal D}(\alpha_s(\mu^2))}{d\bar{a}}\,,
\end{split}
\end{align}
where the expansions of the anomalous dimensions ${\cal A}(\alpha_s(\mu^2))$ and ${\cal D}(\alpha_s(\mu^2))$ of Eqs. (\ref{A_cusp}) and (\ref{calD}), respectively, have been identified.  At order ${\cal O}(\bar{a}^2)$
 we find $d_1=S_1$.
Eq.~(\ref{expansions2}) yields the following coefficients for
$a_{\cal S}^{\Eucl}$ in (\ref{Eucl-expansions})
\begin{align}
\label{s2_to_4_general}
\begin{split}
s_2^E &= a_2\\
s_3^E &= a_3-S_2+2 d_1 a_2+\frac{1}{3}\pi^2-2d_2\\
s_4^E &= a_4+6d_1^2 a_2+\left(2 a_2\delta-3 S_2-6 d_2+3 a_3\right)d_1
+\left(-3 a_2-\frac{5}{2}\delta\right)S_2+S_3
\\&\hspace*{20pt}+\frac{5}{6}\pi^2\delta-3d_3+\pi^2a_2-2\delta d_2\,,
\end{split}
\end{align}
where the process--dependent log-moments $S_k$ are defined in (\ref{JkSk_def}) and $a_n$ and the process--dependent $d_n$ are the $\overline{\rm MS}$ coefficients of (\ref{A_cusp}) and (\ref{calD}), respectively.

Obviously, such an identification can
be done to arbitrarily high order: according to (\ref{J_Sud_anom_dim_conv}) and (\ref{S_Sud_anom_dim_conv}), the information contained in ${\cal J}$ and ${\cal S}$ at a given order in the perturbative expansion is sufficient to determine $a_{\cal J}^{\Eucl}$ and $a_{\cal S}^{\Eucl}$, as well as $a_{\cal J}^{\Mink}$ and $a_{\cal S}^{\Mink}$, to that order.

Finally, note that following the identification we made above
we can write the small--$\epsilon$ expansion of the characteristic functions (\ref{calF_small_epsilon}) as
\begin{align}
\label{calF_small_epsilon1}
{\cal F}_{\cal J}(\epsilon)&=-\ln(\epsilon) +J_1+{\cal O}(\epsilon)=-\ln(\epsilon)+\int_0^{\infty}\frac{dy}{y}\,\ln(y)\, \ddot{\cal F}_{\cal J}(y) \,+\, {\cal O}(\epsilon)
\\
{\cal F}_{\cal S}(\epsilon)&=-\ln(\epsilon) +S_1+{\cal O}(\epsilon)
=-\ln(\epsilon) +\int_0^{\infty}\frac{dz}{z}\,\ln(z)\, \ddot{\cal F}_{\cal S}(z) \,+\, {\cal O}(\epsilon)
\end{align}
where the constant terms $b_1=J_1$ and $d_1=S_1$ were fixed by the first log--moment of the characteristic function itself.

\subsubsection*{Cusps and NLO universality of the
``gluon bremsstrahlung'' coupling}

We observe that the effective charges $a_{\cal J,\,S}^{\Eucl}(\mu^2)$ are closely related to the cusp anomalous dimension~\cite{Korchemsky:1987wg,Korchemsky:1988si}:
\begin{equation}
\label{NLO_universality}
{\cal A} \simeq \frac{C_R}{\beta_0} a_{\cal S}^{\Eucl}\simeq
\frac{C_R}{\beta_0} a_{\cal J}^{\Eucl}\simeq \frac{C_R}{\beta_0}
\left[\bar{a}+  a_2 \bar{a}^2+\cdots\right]\,;
\end{equation}
their expansions start deviating~\cite{Grunberg:2006gd,Grunberg:2006ky} only at the NNLO.
We emphasize that the NLO universality of the effective charges
holds despite the fact that the source of the soft radiation
can be quite different:
it applies to the entire class of inclusive distributions
 discussed in this paper, including B decay spectra,
event--shape distributions, deep inelastic structure functions, single--particle inclusive cross sections, Drell--Yan production and heavy quark fragmentation.
To NLL accuracy, the only way by which the non-Abelian nature
of the interaction affects the exponent is through the cusp
anomalous
dimension (see e.g.~\cite{Korchemskaya:1992je,Korchemsky:1992xv,Gardi:2005yi}),
and therefore this effect is independent of the details of
the configuration of color sources emitting the radiation.

This NLO universality explains and justifies a posteriori some results obtained in the past in several occasions~\cite{Catani:1990rr,DMW,Gardi:2001ny}. In particular, in~\cite{Catani:1990rr} it was proposed to absorb the NLO term into the definition of the coupling, to define the ``gluon bremsstrahlung'' coupling. This can be formulated as a rescaling of the dimensional transmutation scale $\Lambda$ as follows (see also Eq. (60) in~\cite{Catani:1990rr}):
\begin{equation}
\label{Lambda_GB}
\Lambda_{\GB}^2= \Lambda^2 \,\exp{\left\{a_2^{\MSbar}\right\}}=\Lambda^2
 \,\exp{\left\{{\displaystyle \frac {5}{3}}  +
{\displaystyle {\frac{\mathit{C_A}}{\beta_{0}} \left({\displaystyle
\frac {1}{3}}  - {\displaystyle \frac {\pi ^{2}}{12}} \right)\,}}\right\}}\,,
\end{equation}
where the subscript GB stands for ``gluon bremsstrahlung'' and where
$\Lambda^2$ is defined in $\overline{\rm MS}$.

Thanks to this NLO universality, in the original DGE paper~\cite{Gardi:2001ny} it was possible to generalize the exponentiation of a single dressed gluon to full next--to--leading logarithmic accuracy, without having to introduce two separate effective charges.
At this level, differences between observables, which affects subleading logarithms as well as power corrections, are encoded only in the different functional form of the characteristic function. Beyond the NLO the `soft' and the `jet' effective charges start differing from the cusp anomalous dimension, which also becomes scheme dependent.

Note that differences with respect to the cusp anomalous dimension appear already in the large--$\beta_0$ limit.  Nevertheless, we expect that similarly to the cusp anomalous dimension, the effective charges $a_{\cal J}^{\Eucl}(\mu^2)$ and $a_{\cal S}^{\Eucl}(\mu^2)$ would be renormalon free\footnote{It should be emphasized that absence of renormalons in $a_{\cal J}^{\Eucl}(\mu^2)$ and $a_{\cal S}^{\Eucl}(\mu^2)$ is obvious in the large--$\beta_0$ limit, where these functions coincide with the one--loop coupling~(\ref{aSJ_one_loop}), but it is far from obvious beyond this limit.}. Thus, the expansions (\ref{Eucl-expansions}) should not develop large subleading corrections. Absence of renormalons in $a_{\cal J}^{\Eucl}(\mu^2)$ and $a_{\cal S}^{\Eucl}(\mu^2)$ implies that also
${\cal J}(\mu^2)$ and  ${\cal S}(\mu^2)$ are renormalon-free, since the small $\mu^2/k^2$ expansion of the characteristic functions in (\ref{J_Sud_anom_dim_conv}) and (\ref{S_Sud_anom_dim_conv}) does not give rise to any non-analytic terms. These relations will be further discussed in Sec.~\ref{sec:Borel_anom_dim} using the Borel formulation.

\vspace*{20pt}
\subsubsection*{Examples}

Let us now use the explicit expressions for the
characteristic functions in Sec.~\ref{sec:calF_results}
together with the known perturbative expansions of the
Sudakov anomalous dimensions, Eqs.~(\ref{A_cusp}),~(\ref{calB})
and (\ref{calD}) with the coefficients in Appendix~\ref{sec:coef},
to derive the corresponding expansions of the effective
charges $a_{\cal J}^{\Eucl}$ and $a_{\cal S}^{\Eucl}$.

\vspace*{10pt}
\noindent
\underline{The jet function}\\
Starting with the jet function case, from (\ref{calF_dot_QD}) we get:
\begin{equation}
\dot{\cal F}_{\cal J}(y)=\theta(y<1) \,\left[1-\frac{y}{2}-\frac{y^2}{2}\right]\qquad \Longrightarrow \qquad \ddot{\cal F}_{\cal J}(y)=\theta(y<1)\, \,y\,\left(\frac{1}{2}-y\right)\,,
\end{equation} so using (\ref{J_Sud_anom_dim_conv1}) we obtain
\begin{align}
\label{J_Sud_anom_dim_conv1_}
{\cal J}(\mu^2)&=\,\frac{1}{\beta_0}\,\int_0^{\infty}\frac{dy}{y}\,
a_{\cal J}^{\Mink}(y\mu^2)\,\ddot{\cal F}_{\cal J}(y)\,
=\,\frac{1}{\beta_0}\,\int_0^{1} dy\,
\,a_{\cal J}^{\Mink}(y\mu^2)\,\left(y+\frac12\right)\,.
\end{align}
The corresponding log-moments defined in (\ref{JkSk_def}) are:
\begin{align}
\begin{split}
J_k&\equiv \int_0^{\infty}\frac{dy}{y}
\,\ln^k(y)\,\ddot{\cal F}_{\cal J}(y)
=\int_0^{1} dy
\,\,\ln^k(y)\,\,\left(y+\frac12\right)=(-1)^k\,\, k!\, (1+2^{-k})/2\,.
\end{split}
\end{align}
Note that these coefficients increase factorially.
The first few numerical values are summarized in Table~\ref{table:Jk_Sk}.

Using now (\ref{j2_to_4_general}) with the $\overline{\rm MS}$
 $b_n$ in Appendix~\ref{sec:coef} we get:
\begin{align}
\label{J_space_likeIn_terms_of_MSbar}
\begin{split}
 a_{\cal J}^{\Eucl}&=\bar{a}
+j_2^E
\bar{a}^2
+j_3^E
\bar{a}^3
+j_4^E
\bar{a}^4
+\cdots\\&=\bar{a} +
a_2\, \bar{a}^2+\Bigg\{
{a_{3}} + {  \frac {101}{18}}  -
{  \frac {3}{2}} \,{a_{2}} +
{  \frac{1}{\beta _{0}}\Bigg[ {\left({  \frac {73}{72}}  - 5\,\zeta
 _3\right)\,C_A + \left(
{  \frac {3}{16}}  - {  \frac {\pi ^{2}}{4}}+ 3\,\zeta _3\right)\,C_F}\bigg]}
\Bigg\}\bar{a}^3\\&+
\Bigg\{
{a_{4}} - {  \frac {9}{4}} \,{a
_{3}} + \left( - {  \frac {3}{8}}  + \pi ^{2}\right)\,{
a_{2}} + {  \frac {901}{216}}  + 2\,\zeta _3 -
{  \frac {5\,\pi ^{2}}{3}}  + \frac{1}{\beta_0}\bigg[\left( - {
\frac {9\,C_F}{8}}  - {  \frac {15\,
C_A}{8}} \right)\,{a_{2}} \\&
\mbox{} + \left({  \frac {1979}{72}}  - {
\frac {37\,\pi ^{2}}{54}}  - {  \frac {50}{3}} \,
\zeta _3 + {  \frac {13\,\pi ^{4}}{120}} \right)\,
C_A + \left({  \frac {993}{32}}  - {  \frac {
45\,\pi ^{2}}{32}}  - {  \frac {23}{4}} \,\zeta _3\right)
\,C_F\bigg]\mbox{} \\&+ \frac{1}{{\beta_0}^2}\bigg[
\left({  \frac {21}{32}} \,{C_A}^{2} +
{  \frac {33}{32}} \,C_F\,C_A\right)\,{
a_{2}} \\ &
\mbox{} + \left( - {  \frac {385}{216}}  - {
\frac {905}{48}} \,\zeta _3 - {  \frac {37\,\pi ^{2}
}{432}}  - {  \frac {41\,\pi ^{4}}{960}}  +
{  \frac {11}{24}} \,\pi ^{2}\,\zeta _3 +
{  \frac {87}{8}} \,\zeta _5\right)\,{C_A}^{2} \\
&\mbox{} + \left( - {  \frac {19723}{768}}  -
{  \frac {3\,\pi ^{2}}{4}}  + {  \frac {
245}{24}} \,\zeta _3 + {  \frac {17\,\pi ^{4}}{240}
}  + {  \frac {1}{8}} \,\pi ^{2}\,\zeta _3 +
{  \frac {45}{8}} \,\zeta _5\right)\,C_F\,
C_A \\
&\mbox{} + \left({  \frac {105}{128}}  - {
\frac {3\,\pi ^{2}}{64}}  + {  \frac {87}{16}} \,
\zeta _3 + {  \frac {3\,\pi ^{4}}{40}}  -
{  \frac {1}{4}} \,\pi ^{2}\,\zeta _3 -
{  \frac {45}{4}} \,\zeta _5\right)\,{C_F}^{2}\bigg]
 \mbox{} \\&+ \frac{1}{{\beta_0}^3}\Bigg[
\left( - {  \frac {511}{1152}}  + {  \frac {35
}{16}} \,\zeta _3\right)\,{C_A}^{3} + \left( - {  \frac {
1795}{2304}}  + {  \frac {17}{8}} \,\zeta _3 +
{  \frac {7\,\pi ^{2}}{64}} \right)\,C_F\,
{C_A}^{2} \\
&\mbox{} + \left( - {  \frac {33}{256}}  - {
\frac {33}{16}} \,\zeta _3 + {  \frac {11\,\pi ^{2}
}{64}} \right)\,{C_F}^{2}\,C_A\bigg]
\Bigg\}\,\bar{a}^4
+
\cdots\,
\end{split}
\end{align}
where the coefficients $a_n$, which are known in full to three--loop order
(see Appendix~\ref{sec:coef}) have not been substituted. As with the anomalous
dimension ${\cal J}$ itself, the only missing ingredient in $a_{\cal J}^{\Eucl}(\mu^2)$
at four--loop  order (N$^3$LO) is the four--loop coefficient of the cusp
anomalous dimension, $a_4$: the combination
${C_R}a_{\cal J}^{\Eucl}/{\beta_0}-{\cal A}$
is known to this order.

The fact that the four--loop coefficient is not known in full is
rather unimportant in practice: the series for the cusp anomalous
dimension ${\cal A}$ in $\overline{\rm MS}$ converges very well,
and $a_4$ can be reliably estimated using Pad\'e approximants~\cite{Ellis:1995jv}
as discussed in~Sec.~4 in~\cite{Moch:2005ba}. For example,
for $N_c=3$ and $N_f=4$ we get:
\begin{align}
\label{Acusp_num}
{\cal A}=\frac{C_R}{\beta_0}
\left[\bar{a} + \bar{a}^2\cdot 0.962\, + \bar{a}^3\cdot
1.159
+\bar{a}^4\cdot \left\{
\begin{array}{l}
 1.341 \qquad [1,2]\,\text{Pad\'e}\\
 1.397 \qquad [2,1]\,\text{Pad\'e}
 \end{array}\right\}+\cdots\right]\,,
\end{align}
where the difference between the two Pad\'e approximant
predictions can be used as a rough measure of uncertainty.

Using these coefficients of ${\cal A}$ in
(\ref{J_space_likeIn_terms_of_MSbar}) we get\footnote{Note in contrast with (\ref{Acusp_num}),
a direct Pad\'e approximant prediction of the N$^3$LO coefficient in
$a_{\cal J}^{\Eucl}$ is unreliable: the [1,2] and [2,1] approximants
differ by much. A direct prediction
for (\ref{aJ_mink_num}) or (\ref{calJ_num}) below is even worse owing to
the large $\pi^2$ terms related to the analytic continuation.}:
\begin{align}
\label{aJ_eucl_num}
a_{\cal J}^{\Eucl}=\bar{a}
+ \bar{a}^2\cdot 0.962\,  - \bar{a}^3\cdot 1.019-\bar{a}^4\cdot
\left\{
\begin{array}{l}
 3.221 \quad \text{using}\, [1,2]\,\text{Pad\'e}\, \text{for}\, a_4\, \text{in}\, (\ref{Acusp_num})\\
 3.165 \quad \text{using}\, [2,1]\,\text{Pad\'e}\, \text{for}\, a_4\, \text{in}\, (\ref{Acusp_num})
 \end{array}\right\}+\cdots\,.
\end{align}
Note that this expansion converges very well,
better than the corresponding Minkowskian coupling:
\begin{align}
\label{aJ_mink_num}
a_{\cal J}^{\Mink}=\bar{a}
+ \bar{a}^2\cdot 0.962\,  - \bar{a}^3\cdot 4.309-\bar{a}^4\cdot
\left\{
\begin{array}{l}
 18.798 \quad \text{using}\, [1,2]\,\text{Pad\'e}\, \text{for}\, a_4\, \text{in}\, (\ref{Acusp_num})\\
 18.742 \quad \text{using}\, [2,1]\,\text{Pad\'e}\, \text{for}\, a_4\, \text{in}\, (\ref{Acusp_num})
 \end{array}\right\}+\cdots\,,
\end{align}
and the physical anomalous dimension ${\cal J}$ itself:
\begin{align}
\label{calJ_num}
\beta_0\,{\cal J} =
\bar{a} + \bar{a}^2\cdot 1.712\,  - \bar{a}^3\cdot
1.061-\bar{a}^4\cdot \left\{
\begin{array}{l}
 17.606 \quad \text{using}\, [1,2]\,\text{Pad\'e}\, \text{for}\, a_4\, \text{in}\, (\ref{Acusp_num})\\
 17.549 \quad \text{using}\, [2,1]\,\text{Pad\'e}\, \text{for}\, a_4\, \text{in}\, (\ref{Acusp_num})
 \end{array}
 \right\}+\cdots\,,
\end{align}
which both contain large $\pi^2$ terms (see (\ref{Mink-expansions}))
owing to the analytic continuation to the timelike region.
It should be emphasized that (at least in the large--$\beta_0$ limit)
\emph{none} of the three has renormalons, \emph{despite} the fact that the log--moments $J_k$ increase
factorially. The most convenient way to see this is by considering
the relations between these quantities in terms of
their Borel representations, see Sec.~\ref{sec:Borel_anom_dim} below.

\vspace*{10pt}
\noindent
\underline{The Quark Distribution function (B decay)}\\
Consider next the soft function associated with the Quark Distribution in an on-shell heavy quark.
From (\ref{calF_dot_QD}) we get:
\begin{equation}
\dot{\cal F}_{{\cal S}_{\QD}}(z)=1-\frac{1}{\left(1+1/z\right)^2}
\qquad \Longrightarrow \qquad\ddot{\cal F}_{{\cal S}_{\QD}}(z)=\frac{2z^2}{(1+z)^3}\,.
\end{equation}
Using (\ref{S_Sud_anom_dim_conv1}) we get
\begin{align}
\label{S_Sud_anom_dim_conv1_}
{{\cal S}_{\QD}}(\mu^2)&=
\,\frac{1}{\beta_0}\,\int_0^{\infty}\frac{dz}{z}
a_{{\cal S}_{\QD}}^{\Mink}(z\mu^2)\,\ddot{\cal F}_{{\cal S}_{\QD}}(z)
\,=\,\frac{1}{\beta_0}\,\int_0^{\infty}dz \,a_{{\cal S}_{\QD}}^{\Mink}(z\mu^2)\,
\frac{2z}{(1+z)^3}\,.
\end{align}
The corresponding log-moments $S_k$ are given by
\begin{align}
\begin{split}
S_k^{\QD}&\equiv \int_0^{\infty}\frac{dz}{z}
\,\ln^k(z)\,\ddot{\cal F}_{{\cal S}_{\QD}}(z)=
\int_0^{\infty}dz \,\ln^k(z)\,
\,\,\frac{2z}{(1+z)^3}.
\end{split}
\end{align}
Their first few values are summarised in Table~\ref{table:Jk_Sk}.

Using (\ref{j2_to_4_general}) we get the following results for coefficients in Eq. (\ref{Eucl-expansions}):
\begin{align}
\label{S_space_likeIn_terms_of_MSbar}
\begin{split}
 a_{{\cal S}_{\QD}}^{\Eucl}(\mu^2)&=\bar{a}
+s_2^E
\bar{a}^2
+s_3^E
\bar{a}^3
+\cdots\\&=\bar{a} +
a_2\, \bar{a}^2+\left[a_3+\frac{10}{3}a_2
-\frac{22}{9}
+\frac{C_A}{\beta_0}\left(\frac{5}{18}\pi^2+\frac{7}{9}
-\frac{9}{2}\zeta_3\right)
\right]\bar{a}^3 +\cdots.
\end{split}
\end{align}
Numerically, we find again very good convergence in the first few orders;
for $N_c=3$ and $N_f=4$ we get:
\begin{align}
a_{{\cal S}_{\QD}}^{\Eucl}=\bar{a}
+ \bar{a}^2\cdot 0.962\,  - \bar{a}^3\cdot 0.799 +\cdots\,.
\end{align}

\vspace*{10pt}
\noindent
\underline{The Drell--Yan soft function}\\
Consider now the soft function associated with Drell--Yan production or Higgs production
through gluon--gluon fusion. The characteristic function is given by
(\ref{Fr-x-scaling-DY1}). Here the second derivative is singular, but using
(\ref{Soft}) one can verify that the relations in
Eq.~(\ref{s2_to_4_general}) hold. Then, using (\ref{B_relation_dotSJ})
to extract the log-moments $S_k$ (see Table~\ref{table:Jk_Sk})
one readily obtains:
\begin{align}
\label{S_DY_space_likeIn_terms_of_MSbar}
\begin{split}
 a_{{\cal S}_{\DY}}^{\Eucl}(\mu^2)&=\bar{a}
+s_2^E
\bar{a}^2
+s_3^E
\bar{a}^3
+s_4^E
\bar{a}^4
+\cdots\\&=\bar{a} +
a_2\, \bar{a}^2
\,+\,
\Bigg\{
{a_{3}} + {  \frac {28}{9}}  +
{  \frac {\,C_A}{{\beta _{
0}}}\left({  \frac {8}{9}}  -
{  \frac {7}{2}} \,\zeta _3\right)}
\Bigg\}\, \bar{a}^3
\\
&+\,\Bigg\{
a_{4}+\left(  - {  \frac {10}{3}}  + 2\,{a_{2}}\right)\,\pi^{2}
+ {  \frac {116}{27}}  + 2\,\zeta _3  \\&
\mbox{} + {  \frac{1}{\beta _{0}} \bigg[{\left( - {  \frac {247\,
\pi ^{2}}{216}}  + {  \frac {23\,\pi ^{4}}{120}}  -
{  \frac {245}{12}} \,\zeta _3 + {
\frac {2905}{144}} \right)\,C_A + \left({  \frac {645}{
32}}  - {  \frac {19}{2}} \,\zeta _3 -
{  \frac {\pi ^{4}}{20}} \right)\,C_F}\bigg]
}   \\&
+\frac{1}{{\beta_{0}}^{2}}\,\bigg[
\left( - {  \frac {11\,\pi ^{4}}{240}}  + \left( -
{  \frac {155}{864}}  + {  \frac {11}{24}
} \,\zeta _3\right)\,\pi ^{2} + 9\,\zeta _5 + {  \frac {
269}{864}}  - {  \frac {581}{48}} \,\zeta _3\right)\,
{C_A}^{2} \\&
\mbox{} + \left( - {  \frac {6839}{384}}  +
{  \frac {73}{12}} \,\zeta _3 + {
\frac {11\,\pi ^{4}}{240}} \right)\,C_F\,C_A \bigg]
\\& + {  \frac{1}{{{\beta _{0}}^{3}}}\bigg[ {\left( - {  \frac {7}{18}
}  + {  \frac {49}{32}} \,\zeta _3\right)\,{C_A}^{3}
 + \left( - {  \frac {11}{18}}  + {  \frac {77
}{32}} \,\zeta _3\right)\,C_F\,{C_A}^{2}}\bigg]}
\Bigg\}\, \bar{a}^4
 \,+\,\cdots\,,
\end{split}
\end{align}
where we used the known coefficients (\ref{d123_DY}).
Using (\ref{a23_MSbar}), Eq.~(\ref{S_DY_space_likeIn_terms_of_MSbar})
yields the following expansion for $N_c=3$ and $N_f=4$:
\begin{align}
a_{{\cal S}_{\DY}}^{\Eucl}=\bar{a}
+ \bar{a}^2\cdot 0.962\,  - \bar{a}^3\cdot 0.508-\bar{a}^4\cdot
\left\{
\begin{array}{l}
 9.07 \quad \text{using}\, [1,2]\,\text{Pad\'e}\, \text{for}\, a_4\, \text{in}\, (\ref{Acusp_num})\\
 9.01 \quad \text{using}\, [2,1]\,\text{Pad\'e}\, \text{for}\, a_4\, \text{in}\, (\ref{Acusp_num})
 \end{array}\right\}+\cdots\,.
\end{align}

\subsection{Renormalization--group evolution of the Sudakov effective charges~\label{sec:ECH_evolution}}

In order to evaluate the anomalous dimensions entering the
Sudakov exponent one needs to perform integrals over the
discontinuity of the running coupling in
(\ref{J_Sud_anom_dim_conv}) and (\ref{S_Sud_anom_dim_conv}).
The scale dependence of the relevant Euclidean couplings can be
computed directly~\cite{Grunberg:1982fw} by solving the appropriate
renormalization--group equations. Upon taking the
derivative of the effective
charges $a_{\cal J}^{\Eucl}(\mu^2)$
 and $a_{\cal S}^{\Eucl}(\mu^2)$ with respect to $\ln\mu^2$ and
expressing the result in terms of the effective charge
itself one arrives at the following renormalization--group equations:
\begin{align}
\begin{split}
\label{J_beta}
\frac{da_{\cal J}^{\Eucl}(\mu^2)}{d\ln\mu^2}&=-
\left(a_{\cal J}^{\Eucl}(\mu^2)\right)^2
\bigg[1+ a_{\cal J}^{\Eucl}(\mu^2)\delta+
\left( a_{\cal J}^{\Eucl}(\mu^2)\right)^2\delta_2^{\cal J}+
\left( a_{\cal J}^{\Eucl}(\mu^2)\right)^3\delta_3^{\cal J}+\cdots\bigg]\,;\\&\qquad
\delta_2^{\cal J}=\delta_2^{\MSbar}+j_3^E-j_2^E\,\delta-\left(j_2^E\right)^2\,
\\
&\qquad
\delta_3^{\cal J}=
\delta_3^{\MSbar}+
2j_4^E+4{\left(j_2^E\right)}^3+\delta {\left(j_2^E\right)}^2
-6 j_2^E j_3^E -2 j_2^E\delta_2^{\MSbar}
\end{split}
\\
\begin{split}
\label{S_beta}
\frac{da_{\cal S}^{\Eucl}(\mu^2)}{d\ln\mu^2}&=-
\left(a_{\cal S}^{\Eucl}(\mu^2)\right)^2
\bigg[1+ a_{\cal S}^{\Eucl}(\mu^2)\delta+
\left( a_{\cal S}^{\Eucl}(\mu^2)\right)^2\delta_2^{\cal S}+
\left( a_{\cal S}^{\Eucl}(\mu^2)\right)^3\delta_3^{\cal S}+\cdots\bigg]\,;\\&\qquad
\delta_2^{\cal S}=\delta_2^{\MSbar}+s_3^E-s_2^E\,\delta-\left(s_2^E\right)^2\\
&\qquad \delta_3^{\cal S}=\delta_3^{\MSbar}+2s_4^E+4
\left(s_2^E\right)^3+\delta \left(s_2^E\right)^2
-6 s_2^E s_3^E -2 s_2^E\delta_2^{\MSbar}
\end{split}
\end{align}

These equations can be truncated at any given order, and integrated.
The first approximation is obtained setting $\delta=\delta_n=0$
which is also the result of the large--$\beta_0$ limit,
\begin{equation}
\label{a_calJS_large_beta0}
\left.a_{\cal J}^{\Eucl}(\mu^2)\right\vert_{\text{large}\,\,\beta_0}=
\left.a_{\cal S}^{\Eucl}(\mu^2)\right\vert_{\text{large}\,\,\beta_0}
=\frac{\bar{a}(\mu^2)}{1+\frac53\bar{a}(\mu^2)}\,.
\end{equation}
Upon using the universal NLO
correction of (\ref{NLO_universality}) to fix the initial condition
one obtains:
\begin{equation}
\label{aSJ_one_loop}
\left.a_{\cal S}^{\Eucl}(\mu^2)\right\vert_{\text{one-loop}}=
\left.a_{\cal J}^{\Eucl}(\mu^2)\right\vert_{\text{one-loop}}=
\frac{1}{\ln\mu^2/\Lambda_{\GB}^2}\,,
\end{equation}
where $\Lambda_{\GB}$ is defined in Eq.~(\ref{Lambda_GB}).
At the next truncation order ($\delta\neq 0$ but
$\delta_{2,\,3\,\ldots}=0$) the effective charges
$a_{\cal J}^{\Eucl}(\mu^2)$ and
$a_{\cal S}^{\Eucl}(\mu^2)$ still coincide. The analytic solution can be written
in terms of the Lambert W function~\cite{Gardi:1998rf,Gardi:1998qr},
facilitating exact analytic continuation to the time--like axis.

Beyond this order $a_{\cal J}^{\Eucl}(\mu^2)$ and
$a_{\cal S}^{\Eucl}(\mu^2)$ start differing from each other. It is
only at this level that the process--dependent nature of soft gluons
radiation reveals itself. Indeed in the large--$\beta_0$ limit
$\delta_n=0$. The evolution equations (\ref{J_beta}) and (\ref{S_beta})
directly reflect~\emph{the non-Abelian nature} of the interaction.
It therefore becomes
particularly important to study the higher--order corrections to
these equations.

Here comes a very interesting observation: in contrast with the
$\overline{\rm MS}$ coupling and with other physical Euclidean
effective charges~\cite{Gardi:1998rf}, the higher--order coefficients $\delta^{\cal J,\,S}_{2,\,3\,\ldots}$ are not all positive. Most
importantly $\delta^{\cal J,\,S}_{2}$ is \emph{negative} in all
the known examples (see below). This opens up a possibility that
an infrared fixed--point would be realized already at the perturbative
level. In this case Landau singularities
may not appear in the entire
first sheet of the complex momentum plane, leading to a causal
analyticity structure~\cite{Gardi:1998rf,Gardi:1998ch}, where the entire
complex momentum plane is mapped into a compact region in the
complex coupling plane.
A Landau singularity does appear in the one--loop coupling
(\ref{aSJ_one_loop}), in the large--$\beta_0$
limit, as well as in the two--loop coupling with any realistic
number of light flavors. Its absence would therefore be a direct consequence of the
non-Abelian nature of soft--gluon interaction, which may be different in different processes.

If this scenario is realized the perturbative solution for the
effective charge can genuinely be a good approximation to the full,
non-perturbative effective charge down to the infrared limit.
In this case the Sudakov factor computed as a dispersive integral
will be in very good control, far
beyond what can be achieved in a fixed--logarithmic--accuracy approach.

Of course, in reality we know, at best, just
$\delta^{\cal J,\,S}_{2}$ and $\delta^{\cal J,\,S}_{3}$ and therefore
it is hard to make any firm conclusions about the size and perturbative
stability of the fixed point. Obviously, a fixed point can always be
washed out by sufficiently large subleading corrections.
Nevertheless, some of the examples considered below
(in particular the Drell--Yan case) are quite suggestive of this
scenario.

It is important to stress
that the infrared finite coupling scenario
discussed here has nothing to do with Analytic Perturbation Theory (APT)
coupling~\cite{Shirkov:1997wi}, which becomes
finite only owing to the ``analytization'' of the Landau
singularities by \emph{imposing} the
dispersion relation~\cite{Shirkov:1997wi,Aglietti_et_al,Aglietti:2006yf}.
This model implicitly assumes the existence of power corrections
that are not of infrared origin. See further discussion of this issue in
Sec.~\ref{sec:PC} and Appendix \ref{sec:Mink_integral}.

To examine the effective--charge beta function in Eqs.~(\ref{J_beta}) and (\ref{S_beta}) it is useful to introduce a decomposition of the coefficients into powers of $\beta_0$
(eliminating the $N_f$ dependence in favor of $\beta_0=\frac{11}{12}C_A-\frac16 N_f$).
To this end we define first the decomposition of the coefficients of the anomalous dimensions $a_i$, $b_i$ and $d_i$ in the $\overline{\rm MS}$ scheme, Eqs.~(\ref{A_cusp}), (\ref{calB}) and (\ref{calD}), respectively:
\begin{align}
\label{beta0_decom_MSbar}
a_i=\frac{1}{{\beta_0}^{i-1}}\sum_{k=0}^{i-1} a_{i,k}{\beta_0}^{k};\qquad
b_i=\frac{1}{{\beta_0}^{i-1}}\sum_{k=0}^{i-1} b_{i,k}{\beta_0}^{k};\qquad
d_i=\frac{1}{{\beta_0}^{i-1}}\sum_{k=0}^{i-1} d_{i,k}{\beta_0}^{k}
\end{align}
and the $\overline{\rm MS}$ beta function
\begin{equation}
\label{beta0_decom_beta_MSbar}
\beta_i^{\MSbar}=\sum_{j=0}^{i} \beta_{i,j}^{\MSbar}\,{\beta_0}^j
\end{equation}
with $\delta_i^{\MSbar}=\beta_i^{\MSbar}/{\beta_0}^{i+1}$. In what follows we omit the superscript
$\overline{\rm MS}$, and simply use $\beta_{i,j}$ when referring to this scheme.
Using similar notation we also define the decomposition of the coefficients of the effective--charge beta function in Eqs.~(\ref{J_beta}) and (\ref{S_beta}):
\begin{align}
\label{beta0_decom_ECH}
\delta_i^{\cal J}=\frac{\beta_i^{\cal J}}{{\beta_0}^{i+1}}=\frac{1}{{\beta_0}^{i+1}}\,
\sum_{j=0}^{i} \delta_{i,j}^{\cal J}\,{\beta_0}^j;\qquad
\delta_i^{\cal S}=\frac{\beta_i^{\cal S}}{{\beta_0}^{i+1}}=\frac{1}{{\beta_0}^{i+1}}\,
\sum_{j=0}^{i} \delta_{i,j}^{\cal S}\,{\beta_0}^j\,,
\end{align}
where the fact that the sum over $j$ goes only up to $j=i$, as in the the $\overline{\rm MS}$ scheme (\ref{beta0_decom_beta_MSbar}),
rather than $j=i+1$ (as occurs in most physical effective charges) is a reflection of the fact that
in the
large--$\beta_0$ limit the Sudakov effective charges are simply related to the $\overline{\rm MS}$ coupling, see (\ref{a_calJS_large_beta0}) above.

Considering the first scheme--dependent coefficient,
\begin{equation}
\label{delta_2_JS}
\delta_2^{\cal J} = \frac{\beta_2^{\cal J}}{{\beta_0}^3} =
\underbrace{{\delta_{2,3}^{\cal J}}}_{\equiv 0}+
\frac{\delta_{2,2}^{\cal J}}{{\beta_0}}
+\frac{\delta_{2,1}^{\cal J}}{{\beta_0}^2}
+\frac{\delta_{2,0}^{\cal J}}{{\beta_0}^3};
\qquad\,\,\,
\delta_2^{\cal S} =\frac{\beta_2^{\cal S}}{{\beta_0}^3} =
\underbrace{{\delta_{2,3}^{\cal S}}}_{\equiv 0}+
\frac{\delta_{2,2}^{\cal S}}{{\beta_0}}
+\frac{\delta_{2,1}^{\cal S}}{{\beta_0}^2}
+\frac{\delta_{2,0}^{\cal S}}{{\beta_0}^3}
\end{equation}
we observe, as anticipated, that $\delta_{2,3}^{\cal J,\, S}=0$. We also find
that $\delta_{2,0}^{\cal J}=\delta_{2,0}^{\cal S}$ and $\delta_{2,1}^{\cal J}=\delta_{2,1}^{\cal S}$ are \emph{universal}, i.e. they depend only on the coefficients of the cusp anomalous dimension and the $\overline{\rm MS}$ beta function (they do not depend on the details of the soft function considered) while  $\delta_{2,2}^{\cal J,\, S}$ are process dependent, depending on
$b_1=J_1$ and $b_2$ or on
$d_1=S_1$ and $d_2$, respectively. The expressions are:
\begin{align}
\label{delta_2_S_ij}
\begin{split}
\delta_{2,0}^{\cal S}&=-a_{2,0} \beta_{1,0}+\beta_{2,0}\\
\delta_{2,1}^{\cal S}&=-a_{2,0}^2+\beta_{2,1}-a_{2,1} \beta_{1,0}-a_{2,0} \beta_{1,1}+a_{3,0}\\
\delta_{2,2}^{\cal S}&=-2 a_{2,0} a_{2,1}+2 S_{1} a_{2,0}+a_{3,1}-a_{2,1} \beta_{1,1}+\beta_{2,2}-2 d_{2,0}\\
\delta_{2,3}^{\cal S}&=-S_{2}+2 S_{1} a_{2,1}+\frac{\pi^2}{3} -2 d_{2,1}-a_{2,1}^2+a_{3,2}=0\,,
\end{split}
\end{align}
and similarly for $\delta_{2,j}^{\cal J}$, with the obvious replacement of the coefficients $d_k$ by the corresponding $b_k$ and $S_k$ by $J_k$.
Note that the vanishing of $\delta_{2,3}^{\cal J,\,S}$ can be verified by taking the large--$\beta_0$ limit of $\delta_{2}^{\cal J\,,S}$ in Eqs.~(\ref{J_beta}) and (\ref{S_beta}) and using
$j_3^E$ and $s_3^E$ of Eqs. (\ref{j2_to_4_general}) and (\ref{s2_to_4_general}), respectively.

At the next order,
\begin{equation}
\delta_3^{\cal J} =
\underbrace{{\delta_{3,4}^{\cal J}}}_{\equiv 0}+
\frac{\delta_{3,3}^{\cal J}}{{\beta_0}}
+\frac{\delta_{3,2}^{\cal J}}{{\beta_0}^2}
+\frac{\delta_{3,1}^{\cal J}}{{\beta_0}^3}
+\frac{\delta_{3,0}^{\cal J}}{{\beta_0}^4};
\qquad
\delta_3^{\cal S} =
\underbrace{{\delta_{3,4}^{\cal S}}}_{\equiv 0}+
\frac{\delta_{3,3}^{\cal S}}{{\beta_0}}
+\frac{\delta_{3,2}^{\cal S}}{{\beta_0}^2}
+\frac{\delta_{3,1}^{\cal S}}{{\beta_0}^3}
+\frac{\delta_{3,0}^{\cal S}}{{\beta_0}^4}\,
\end{equation}
so we find, as expected, that $\delta_{3,4}^{\cal J,\,S}=0$. We also observe that
$\delta_{3,0}^{\cal J}=\delta_{3,0}^{\cal S}$ is universal,  while the other coefficients ($\delta_{3,1}^{\cal J,\,S}$ and $\delta_{3,2}^{\cal J,\,S}$ and $\delta_{3,3}^{\cal J,\,S}$) are process dependent, depending on $d_1=S_1$, $d_2$ and $d_3$ as well as on $S_2$. The general expressions are:
\begin{align}
\label{delta3_ECH}
\begin{split}
\delta_{3,0}^{\cal S}&=-2 \beta_{2,0} a_{2,0}+\beta_{1,0} a_{2,0}^2+\beta_{3,0}
\\
\delta_{3,1}^{\cal S}&=2 a_{4,0}+\beta_{1,1} a_{2,0}^2+4 a_{2,0}^3+\beta_{3,1}-2 \beta_{2,0} a_{2,1}\\&+2 \beta_{1,0} a_{2,0} a_{2,1}+4 a_{2,0} \beta_{1,0} S_{1}-4 \beta_{1,0} d_{2,0}-2 \beta_{2,1} a_{2,0}-6 a_{3,0} a_{2,0}
\\
\delta_{3,2}^{\cal S}&=\left(6 a_{3,0}-12 a_{2,0}^2+4 a_{2,0} \beta_{1,1}+4 a_{2,1} \beta_{1,0}\right) S_{1}+2 a_{4,1}\\&+\beta_{3,2}-6 a_{3,0} a_{2,1}+12 a_{2,0}^2 a_{2,1}+\frac{5\pi^2}{3}  \beta_{1,0}-6 d_{3,0}\\&-2 \beta_{2,1} a_{2,1}+\beta_{1,0} a_{2,1}^2-2 \beta_{2,2} a_{2,0}\\&-6 a_{2,0} a_{3,1}-4 \beta_{1,0} d_{2,1}-4 \beta_{1,1} d_{2,0}+2 \beta_{1,1} a_{2,0} a_{2,1}+12 a_{2,0} d_{2,0}-5 \beta_{1,0} S_{2}
\\
\delta_{3,3}^{\cal S}&=12 S_{1}^2 a_{2,0}+\left(-24 a_{2,0} a_{2,1}+6 a_{3,1}-12 d_{2,0}+4 a_{2,1} \beta_{1,1}\right) S_{1}\\&+\beta_{3,3}+12 a_{2,0} a_{2,1}^2-2 \beta_{2,2} a_{2,1}-6 d_{3,1}+\beta_{1,1} a_{2,1}^2-5 \beta_{1,1} S_{2}-4 \beta_{1,1} d_{2,1}\\&+2 a_{4,2}+12 a_{2,0} d_{2,1}-6 a_{2,0} a_{3,2}+\frac{5\pi^2}{3} \beta_{1,1}+12 a_{2,1} d_{2,0}-6 a_{2,1} a_{3,1}
\\
\delta_{3,4}^{\cal S}&=12 S_{1}^2 a_{2,1}+\left(-12 a_{2,1}^2+6 a_{3,2}-6 S_{2}-12 d_{2,1}\right) S_{1}+2 S_{3}\\&-6 a_{2,1} a_{3,2}-6 d_{3,2}+2 a_{4,3}+4 a_{2,1}^3+12 a_{2,1} d_{2,1}
=0
\end{split}
\end{align}
and similarly for $\delta_{3,j}^{\cal J}$ with the obvious replacements.
Again, the vanishing of $\delta_{3,4}^{\cal J,\,S}$ can be verified by
taking the large--$\beta_0$ limit of
$\delta_{3}^{\cal J\,,S}$ in Eqs.~(\ref{J_beta}) and (\ref{S_beta}) and using
$j_4^E$ and $s_4^E$ in Eqs. (\ref{j2_to_4_general}) and (\ref{s2_to_4_general}), respectively.

We comment that the above universality of $\delta_{2,0}$, $\delta_{2,1}$ and $\delta_{3,0}$
is directly related to the universality of the Banks--Zaks fixed
point~\cite{Grunberg:2006um,Grunberg:2006gd},
which will be proven (to all orders) in Sec.~4.4 below.
This relation follows immediately from Eqs.~(3.3) and (3.6) of
\cite{Grunberg:2001aw}, taking into account the universality of the Banks--Zaks critical exponent.

\subsubsection*{Three--loop examples}

Let us now examine the coefficients of the effective--charge $\beta$
functions for the examples considered in Sec.~\ref{ECH:order_by_order}.
Using Eqs.~(\ref{delta_2_S_ij}) and (\ref{delta_2_S_ij}) with the explicit coefficients in Appendix~\ref{sec:coef} and Table \ref{table:Jk_Sk} (or, alternatively, using in (\ref{J_beta}) and (\ref{S_beta}) the explicit coefficients (\ref{J_space_likeIn_terms_of_MSbar}),
(\ref{S_space_likeIn_terms_of_MSbar}) and (\ref{S_DY_space_likeIn_terms_of_MSbar})) together with the $\overline{\rm MS}$ $\beta$--function coefficients, we obtain the
following results for the three--loop coefficients for the effective--charge $\beta$ functions:
for the jet function (\ref{J_space_likeIn_terms_of_MSbar}) we find
\begin{align}
\label{delta_2_J_decom}
\begin{split}
\delta_2^{\cal J} &= \frac{\delta_{2,2}^{\cal J}}{{\beta_0}}+\frac{\delta_{2,1}^{\cal J}}{{\beta_0}^2}+\frac{\delta_{2,0}^{\cal J}}{{\beta_0}^3}\,,
\end{split}
\end{align}
with
\begin{align}
\label{delta_2_J}
\begin{split}
\delta_{2,2}^{\cal J}&=\left(\frac{53}{32}-\frac{3}{2} \zeta_3+\frac{1}{8} \pi^2\right) C_A+\left(\frac{49}{16}-\frac{1}{4} \pi^2\right) C_F
\\
\delta_{2,1}^{\cal J}&= \left(\frac{1}{120} \pi^4+\frac{5}{48} \pi^2+\frac{73}{96}-\frac{11}{4} \zeta_3\right) {C_A}^2+\left(-\frac{235}{96}+\frac{11}{4} \zeta_3+\frac{1}{16} \pi^2\right) C_F C_A-\frac{3}{32} {C_F}^2
\\
\delta_{2,0}^{\cal J}&=\left(-\frac{301}{512}-\frac{7}{192} \pi^2\right){C_A}^3
+\left(-\frac{11}{64}-\frac{11}{192} \pi^2\right)C_F {C_A}^2
+\frac{11}{128} {C_F}^2 C_A\,.
\end{split}
\end{align}
Another constraint on the jet Sudakov anomalous dimension
is the color structure. In particular, for $N_f=0$ we have:
\begin{equation}
\left.\beta_2^{\cal J}\right\vert_{N_f=0}
= \left(\frac{121}{768}-\frac{121}{576}\pi^2
+\frac{121}{48}\zeta_3\right) {C_A}^2 C_F
+\left(\frac{1729}{1152}-\frac{121}{32}\zeta_3+\frac{21}{128}\pi^2
+\frac{11}{1440}\pi^4\right)\,
{C_A}^3\,,
\end{equation}
where the ${C_F}^3$ and the $C_A{C_F}^2$ components vanishes.
In (\ref{delta_2_J_decom}) the latter implies the
following relation between the coefficients of (\ref{delta_2_J}):
\begin{equation}
\label{color_constraint_jet}
\frac{11}{12}\delta_{2,1}^{{\cal J}({C_F}^2)}+\delta_{2,0}^{{\cal J}({C_F}^2C_A)}=0\,.
\end{equation}
This relation involves only the universal components of $\delta_{2}^{\cal J}$ that coincide with the corresponding ones in $\delta_{2}^{\cal S}$, so it holds for any Sudakov anomalous dimension.

Let us turn now to the soft anomalous dimensions.
For the quark--distribution function (\ref{S_space_likeIn_terms_of_MSbar}) we find
\begin{align}
\begin{split}
\delta_2^{{\cal S}_{\QD}} &= \frac{\delta_{2,2}^{{\cal S}_{\QD}}}{{\beta_0}}
+\frac{\delta_{2,1}^{{\cal S}_{\QD}}}{{\beta_0}^2}+\frac{\delta_{2,0}^{{\cal S}_{\QD}}}{{\beta_0}^3}\,,
\end{split}
\end{align}
with
\begin{align}
\label{delta22_QD}
\begin{split}
\delta_{2,2}^{{\cal S}_{\QD}}&=\left(\frac{23}{8}-3 \zeta_3\right) C_F\,+\,
\left(\frac{97}{32}-\zeta_3\right) C_A
\end{split}
\end{align}
and  $\delta_{2,1}^{{\cal S}_{\QD}}=\delta_{2,1}^{\cal J}$ and
$\delta_{2,0}^{{\cal S}_{\QD}}=\delta_{2,0}^{\cal J}$, given in Eq.~(\ref{delta_2_J}) above.
For the Drell--Yan soft function (\ref{S_DY_space_likeIn_terms_of_MSbar}) we find
\begin{align}
\begin{split}
\delta_2^{{\cal S}_{\DY}} &= \frac{\delta_{2,2}^{{\cal S}_{\DY}}}{{\beta_0}}
+\frac{\delta_{2,1}^{{\cal S}_{\DY}}}{{\beta_0}^2}+\frac{\delta_{2,0}^{{\cal S}_{\DY}}}{{\beta_0}^3}\,,
\end{split}
\end{align}
with
\begin{align}
\label{delta22_DY}
\begin{split}
\delta_{2,2}^{{\cal S}_{\DY}}=\left(\frac{23}{8}-3 \zeta_3\right) C_F\,+\,\frac{65}{32} C_A
\end{split}
\end{align}
and  $\delta_{2,1}^{{\cal S}_{\DY}}=\delta_{2,1}^{\cal J}$ and
$\delta_{2,0}^{{\cal S}_{\DY}}=\delta_{2,0}^{\cal J}$, given in Eq.~(\ref{delta_2_J}) above.

The constraint on the color structure of the soft anomalous
dimensions is more stringent than for the jet:
they are \emph{maximally non-Abelian}.
In particular, for $N_f=0$ we have:
\begin{align}
\begin{split}
\left.\beta_2^{{\cal S}_{\DY}}\right\vert_{N_f=0}&=
\left(\frac{523}{288}+\frac{17}{288}\pi^2-\frac{121}{48}\zeta_3
+\frac{11}{1440}\pi^4\right){C_A}^3\\
\left.\beta_2^{{\cal S}_{\QD}}\right\vert_{N_f=0}&=
\left(\frac{85}{32}+\frac{17}{288}\pi^2-\frac{121}{36}\zeta_3
+\frac{11}{1440}\pi^4\right){C_A}^3\,,
\end{split}
\end{align}
where the ${C_F}^3$, the ${C_F}^2C_A$ and the $C_F{C_A}^2$ components
all vanish.  Thus, beyond (\ref{color_constraint_jet}), which
is obviously realized, there is an additional relation, namely
the $C_F$ component of $\delta_{2,2}^{\cal S}$ is universal:
\begin{equation}
\delta_{2,2}^{{\cal S}(C_F)} = -\frac{144}{121}\delta_{2,0}^{{\cal J,\, S}(C_FC_A^2)}
-\frac{12}{11}\delta_{2,1}^{{\cal J,\, S}(C_AC_F)}=\frac{23}{8}-3\zeta_3\,.
\end{equation}
This general property is of course realized in the examples above,
Eqs. (\ref{delta22_QD}) and (\ref{delta22_DY}). Thus, at this order,
only the $C_A$ component of $\delta_{2,2}^{\cal S}$ distinguishes between
different soft functions.

Incidentally, the numerical values of
$\delta_2^{\cal J}$ \footnote{The corresponding value quoted in \cite{Grunberg:2006gd} is incorrect.} and $\delta_2^{\cal S}$ are
also not far; for example for $N_c=3$ and $N_f=4$
(where $\delta = 0.7392$) one gets:
$\delta_2^{\cal J} \simeq -1.954$,\,\,
$\delta_2^{{\cal S}_{\QD}} \simeq -1.734$,\,\,
and $\delta_2^{{\cal S}_{\DY}} \simeq -1.443$.

As anticipated, the numerical values of $\delta_2^{\cal J}$ and
$\delta_2^{\cal S}$ are all negative, allowing for a perturbative
infrared fixed point. It thus becomes interesting to examine
higher--order corrections to the effective--charge $\beta$ functions.

\subsubsection*{Four--loop examples}

In case of the jet function and the soft function associated with
Drell--Yan production we do have sufficient information to determine
the corresponding \emph{four--loop} coefficients~$\delta_3$ of Eq.~(\ref{delta3_ECH}) using the
Pad\'e approximant predictions for $a_4$ in (\ref{Acusp_num}).
For $N_c=3$ and $N_f=4$ we get the following
effective--charge $\beta$ functions:
\begin{align}
\begin{split}
\label{J_DY_beta_Nf4}
\frac{da_{\cal J}^{\Eucl}(\mu^2)}{d\ln\mu^2}&=-
\left(a_{\cal J}^{\Eucl}(\mu^2)\right)^2\times\bigg[1+ 0.739\cdot a_{\cal J}^{\Eucl}(\mu^2)\\&\,\,
-1.954\cdot\left( a_{\cal J}^{\Eucl}(\mu^2)\right)^2+
\left\{\begin{array}{l}
4.003\quad [1,2]\,\text{Pad\'e}\, \text{for}\, a_4\\
4.116\quad [2,1]\,\text{Pad\'e}\, \text{for}\, a_4
\end{array}\right\}\cdot\left( a_{\cal J}^{\Eucl}(\mu^2)\right)^3+\cdots\bigg]\,
\end{split}
\end{align}
and
\begin{align}
\begin{split}
\label{S_beta_Nf4}
\frac{da_{{\cal S}_{\DY}}^{\Eucl}(\mu^2)}{d\ln\mu^2}&=
-\left(a_{{\cal S}_{\DY}}^{\Eucl}(\mu^2)\right)^2\times
\bigg[1+ 0.739\cdot a_{{\cal S}_{\DY}}^{\Eucl}(\mu^2)\\&\,\,
-1.443\cdot
\left( a_{{\cal S}_{\DY}}^{\Eucl}(\mu^2)\right)^2-
\left\{\begin{array}{l}
10.532\quad [1,2]\,\text{Pad\'e}\, \text{for}\, a_4\\
10.645\quad [2,1]\,\text{Pad\'e}\, \text{for}\, a_4
\end{array}\right\}\cdot
\left( a_{{\cal S}_{\DY}}^{\Eucl}(\mu^2)\right)^3+\cdots\bigg]\,
\end{split}
\end{align}
In case of the jet function effective charge (\ref{J_DY_beta_Nf4})
the positive $\delta_3$ coefficient does not support
the fixed--point scenario; a more definite conclusion
requires higher orders.
However, in the case of the soft Drell--Yan effective
charge $\delta_3$ is negative and sufficiently large to bring the
fixed-point value to the perturbative regime:
at this truncation order we get a zero for the $\beta$ function at
$a_{{\cal S}_{\DY}}\simeq 0.46$ corresponding to
$\alpha_s^{{\cal S}_{\DY}} \simeq 0.7$.  Of course,
since there is no perturbative stability at this order,
higher--order corrections may well change this
infrared fixed--point value.
Further insight on this issue will be gained in the next section.

\subsection{The universal infrared limit of Sudakov effective charges\label{sec:Banks_Zaks}}

If we assume that the Sudakov effective charges do indeed admit a finite infrared limit, it becomes natural to ask what this limit is, and further, to what extent it is universal.
The most natural tool to address these questions is the Banks--Zaks
expansion~\cite{Banks:1981nn,Grunberg:1992mp,Stevenson:1994jd,Caveny:1997yr,Gardi:1998rf,Gardi:1998qr,Gardi:1998ch}.

\subsubsection*{Banks--Zaks fixed point}

Let us briefly recall some terminology. A Banks--Zaks fixed point~\cite{Banks:1981nn} is a non-trivial infrared fixed point, $\beta\left(\alpha(\mu^2\to 0)\right)=0$, occurring in asymptotically free gauge theories with a sufficient amount of matter. In QCD, such a conformal infrared limit characterizes the theory with $N_f/N_c\lsim 33/2$ where the beta function
\begin{equation}
\beta\left(\alpha(\mu^2)\right)=\frac{d\alpha_s(\mu^2)/\pi}{d\ln \mu^2}=
-\beta_0 \left(\alpha_s(\mu^2)/\pi\right)^2
-\beta_1 \left(\alpha_s(\mu^2)/\pi\right)^3
-\beta_2 \left(\alpha_s(\mu^2)/\pi\right)^4
+\cdots
\end{equation}
has $\beta_0>0$ and $\beta_1<0$, so it
is negative for a vanishingly small $\alpha_s(\mu^2)$ admitting asymptotic freedom,  but then changes sign again at $\alpha_s(\mu^2)\simeq -\beta_0/\beta_1$, implying that theory is conformal in the infrared limit, where the coupling saturates at $\alpha_s(\mu^2\to 0 )\simeq -\beta_0/\beta_1$.
In the $\beta_0\to 0$ limit the infrared physics is fully under control of perturbation theory. In real--world QCD both $\beta_0$ and $\beta_1$ are positive, and for most physical effective charges so are the higher--order coefficients $\beta_2$, $\beta_3$ etc.; then there is no infrared fixed point.

The Banks--Zaks
expansion~\cite{Banks:1981nn,Grunberg:1992mp,Stevenson:1994jd,Caveny:1997yr,Gardi:1998rf,Gardi:1998qr,Gardi:1998ch}
amounts to expressing the infrared fixed--point value of the coupling as a systematic expansion in (positive) powers of $\beta_0$. To construct the expansion one eliminates the $N_f$ dependence in favor of $\beta_0=\frac{11}{12}C_A-\frac16 N_f$, and then
solves the equation $\tilde{\beta}\left(\tilde{\alpha}\right)=0$
(the tilde denotes a particular renormalization scheme) order-by-order in $\beta_0$, using the fact that in the formal $\beta_0\to 0$ limit the infrared coupling itself is of the order of $\beta_0$. Using the notation of the previous section for the decomposition of the effective--charge beta function coefficients, the resulting expansion takes the form (see e.g. Sec. 3.1 in \cite{Gardi:1998rf}):
\begin{align}
\label{BZ}
\begin{split}
\frac{\tilde{\alpha}_s\left(\mu^2\to 0\right)}{\pi} &= b_{\BZ}
+\left[{\tilde{\beta}_{1, \,1}} - {  \frac {{\tilde{\beta}_{2, \,0}}}{{\tilde{\beta}_{1, \,0}}}} \right]\,b_{\BZ}^{2}
+ \left[    {\tilde{\beta}_{2, \,1}} + {\tilde{\beta}_{
1, \,1}}^{2} - {  \frac {  {\tilde{\beta}_{3, \,0}} + 3\,{\tilde{\beta}_{1,
\,1}}\,{\tilde{\beta}_{2, \,0}}}{{\tilde{\beta}_{1, \,0}}}}  + {  \frac {2\,{
\tilde{\beta}_{2, \,0}}^{2}}{{\tilde{\beta}_{1, \,0}}^{2}}}     \right] \,b_{\BZ}^{3
}\\
&\hspace*{-60pt}+
\Bigg[   - {\tilde{\beta}_{2, \,2}}
\,{\tilde{\beta}_{1, \,0}} + {\tilde{\beta}_{1, \,1}}^{3} + 3\,{\tilde{\beta}_{2, \,1}}\,{\tilde{\beta}_{1, \,1}}
 + {\tilde{\beta}_{3, \,1}} -{  \frac {  4\,{\tilde{\beta}_{1, \,1}}\,{\tilde{\beta}_{3, \,0}}
  +6\,{\tilde{\beta}_{1, \,1}}^{2}\,{\tilde{\beta}_{2, \,0}} + 4\,{\tilde{\beta}_{2, \,1}}\,{\tilde{\beta}_{2, \,
0}} + {\tilde{\beta}_{4, \,0}}}{{\tilde{\beta}_{1, \,0}}}}  \\
&\hspace*{-60pt} + {  \frac {10\,{
\tilde{\beta}_{1, \,1}}\,{\tilde{\beta}_{2, \,0}}^{2} + 5\,{\tilde{\beta}_{2, \,0}}\,{\tilde{\beta}_{3, \,0}}}{{\tilde{\beta}_{1, \,0}}^{2}}}   - {  \frac {5\,{\tilde{\beta}_{2, \,0}}^{3}}{{\tilde{\beta}_{1, \,0}}
^{3}}} \Bigg]\, b_{\BZ}^{4}\,+\,\cdots
\end{split}
\end{align}
where $b_{\BZ}\equiv -\beta_0/\beta_{1,0}$.
This expansion converges well for an appropriate number of light quark flavors ($N_f/N_c\lsim 33/2$) where a fixed point exists (and is realized perturbatively) while for a realistic number of light flavors, its convergence strongly depends on the observable considered~\cite{Gardi:1998rf}; this is consistent with the expectation that real--world QCD does not have a conformal infrared limit. Physical quantities
may have a finite infrared limit also in the confining phase, which is not driven
by a conformal fixed point and is usually inaccessible to perturbation theory. Yet, it is possible that
certain physical effective charges can be described by perturbation theory down to the
infrared limit and then the Banks--Zaks expansion provides a natural tool to compute it.

\subsubsection*{The universal infrared limit of Sudakov effective charges}

In general the infrared limit of the coupling is scheme dependent. This is true also for physical effective charges and it is reflected in observable--dependent coefficients of the Banks--Zaks expansion starting at NLO \cite{Grunberg:1992mp,Gardi:1998rf}.
Quite remarkably, and in sharp contrast with the general situation, \emph{Sudakov effective charges have a common infrared limit} that is uniquely determined by the cusp anomalous
dimension; as we shall see, \emph{their Banks--Zaks expansion is universal to all
orders in perturbation theory}.

This type of universality has been first observed in Ref.~\cite{Grunberg:2006um} where it was noted
that the Banks--Zaks expansion of the Sudakov effective charges relevant to deep inelastic
structure functions and Drell--Yan production coincide up to the N$^3$LO.
The issue was further developed in Ref.~\cite{Grunberg:2006gd} where a connection was
made with the infrared limit of the cusp anomalous dimension. Here we give a general
proof\footnote{A different argument, which also applies to the more general situation where the infrared fixed points are of a non-perturbative origin and not necessarily related to the cusp anomalous dimension (which may have no fixed point at the non-perturbative level) is given in~\cite{Grunberg:2006gd}: see Eq.~(27) there, where instead a relation to the quark form factor is established.} of the above statements.

To prove the universality of the infrared limit note first that the relations between the physical Sudakov anomalous dimensions and the $\overline{\rm MS}$ ones~(\ref{Sud_anom_dim}) involve a momentum derivative, thus the $\beta$ function itself. Upon assuming that all effective charges involved admit an infrared fixed point, Eq.~(\ref{Sud_anom_dim}) immediately implies that in this limit the term containing the $\beta$ function vanishes, so~\cite{Grunberg:2006gd}:
\begin{equation}
\label{IR_limit_calJS}
C_R\,{\cal J}(\mu^2\to 0) =C_R \,{\cal S} (\mu^2\to 0) = {\cal A} (\mu^2\to 0).
\end{equation}
The next observation is that the dispersion relations (\ref{J_Sud_anom_dim_conv}) and (\ref{S_Sud_anom_dim_conv}) necessarily relate the infrared limit of the Sudakov anomalous dimension with that of the corresponding Sudakov effective charge:
\begin{align}
\label{IR_limit_a}
\begin{split}
{\cal J}(\mu^2\to 0) =\frac{1}{\beta_0}\, a_{{\cal J}}^{\Mink}(\mu^2\to 0)= \frac{1}{\beta_0}\, a_{{\cal J}}^{\Eucl}(\mu^2\to 0); \\
{\cal S}(\mu^2\to 0) =\frac{1}{\beta_0}\, a_{{\cal S}}^{\Mink}(\mu^2\to 0)= \frac{1}{\beta_0}\, a_{{\cal S}}^{\Eucl}(\mu^2\to 0)\,,
\end{split}
\end{align}
where the equality of the Minkowskian and Euclidean effective charges~\cite{Gardi:1998rf} follows directly form~(\ref{Eucl-Mink}). It is worthwhile emphasizing that this equality holds
despite the fact that the $\beta$ function of the Minkowskian effective charge differs from the corresponding Euclidean one by large $\pi^2$ terms at any order (at three--loops and beyond).  The Banks--Zaks expansion of a Minkowskian effective charge is identical to the corresponding Euclidean one (Sec. 3.3 in~\cite{Gardi:1998rf}).

Using in (\ref{BZ}) the results for the effective--charge beta function of the Sudakov effective charges, Eqs. (\ref{delta_2_S_ij}) and (\ref{delta3_ECH}), we find, as implied by
(\ref{IR_limit_calJS}) and (\ref{IR_limit_a}), that all the process--dependent coefficients drop out: at each order in the Banks--Zaks expansion only the cusp anomalous dimension and the QCD beta function coefficients appear.
The result for the first few orders is:
\begin{align}
\label{BZ_Sud}
\begin{split}
\frac{1}{\beta_{0}} \,a_{\cal J,\,S}^{\Eucl}\left(\mu^2\to 0\right) &
=  b_{\BZ}
+ \left[{a_{2, \,0}} + {\beta_{1, \,1}} -
{  \frac {{\beta_{2, \,0}}}{{\beta_{1, \,0}}}} \right]\,
b_{\BZ}^{2}
+
\Bigg[ \left(2\,{\beta_{1, \,1}} - {  \frac {2\,{\beta_{2,
\,0}}}{{\beta_{1, \,0}}}} \right)\,{a_{2, \,0}} - {a_{2, \,1}}\,{\beta
 _{1, \,0}} \\
&\hspace*{-40pt} + {a_{3, \,0}} + {\beta_{2, \,1}} + {\beta_{1, \,1}}^{2
} + {  \frac { - {\beta_{3, \,0}} - 3\,{\beta_{1,
\,1}}\,{\beta_{2, \,0}}}{{\beta_{1, \,0}}}}  + {
\frac {2\,{\beta_{2, \,0}}^{2}}{{\beta_{1, \,0}}^{2}}}
 \Bigg]
b_{\BZ}^{3}\\
&\hspace*{-40pt}+
\Bigg[  \bigg(   2\,
{\beta_{2, \,1}} + 3\,{\beta_{1, \,1}}^{2} + {
\frac { - 8\,{\beta_{1, \,1}}\,{\beta_{2, \,0}} - 2\,{\beta_{3
, \,0}}}{{\beta_{1, \,0}}}}  + {  \frac {5\,{\beta_{2, \,0}}^{2}}{{\beta_{1, \,0}}^{2}}}  \bigg) \,{a_{2, \,0
}} \\
&\hspace*{-40pt} + \left(2\,{\beta_{2, \,0}} - 2\,{\beta_{1, \,0}}\,{\beta_{
1, \,1}}\right)\,{a_{2, \,1}} + \left( - {  \frac {3\,{\beta_{2
, \,0}}}{{\beta_{1, \,0}}}}  + 3\,{\beta_{1, \,1}}\right)\,{a_{3, \,0
}} - {\beta_{1, \,0}}\,{a_{3, \,1}} - {\beta_{2, \,2}}\,{\beta_{1, \,0}}  \\
&\hspace*{-40pt} + {a_{4, \,0}}+ {\beta_{1, \,1}}^{3} + {\beta_{3, \,1}} + 3\,{\beta_{2, \,1}}\,{\beta_{1, \,1}} - {  \frac {  4\,{
\beta_{2, \,1}}\,{\beta_{2, \,0}} + 6\,{\beta_{1, \,1}}^{2}\,{
\beta_{2, \,0}} + 4\,{\beta_{1, \,1}}\,{\beta_{3, \,0}} + {\beta_{4, \,0}}}{{\beta_{1, \,0}}}}  \\
&\hspace*{-40pt} + {  \frac {10\,{\beta_{1, \,1}}\,{\beta_{2
, \,0}}^{2} + 5\,{\beta_{2, \,0}}\,{\beta_{3, \,0}}}{{\beta_{1
, \,0}}^{2}}}  - {  \frac {5\,{\beta_{2, \,0}}^{3}}{
{\beta_{1, \,0}}^{3}}}  \Bigg] \,b_{\BZ}^{4}\,+\cdots \,,
\end{split}
\end{align}
where the coefficients of the cusp anomalous dimension $a_{i, j}$ and the QCD beta function $\beta_{i, j}$ are defined in~(\ref{beta0_decom_MSbar}) and (\ref{beta0_decom_beta_MSbar}), respectively. Both these quantities are defined in the $\overline{\rm MS}$ scheme, but this scheme dependence cancels at each order in the expansion, as both the l.h.s and the expansion parameter are physical quantities.

Finally, for $N_c=3$ QCD  we find the following numerical values:
\begin{align}
\label{BZ_Sud_Nc3}
\begin{split}
\frac{1}{\beta_{0}} \,a_{\cal J,\,S}^{\Eucl}\left(\mu^2\to 0\right) \,\,&=\,\,
 b_{\BZ} +\left[\frac{21947}{10272}-\frac{1}{4} \pi^2\right]\, b_{\BZ}^2
\\& +\left[\frac{98425175}{17585664}-\frac{21947}{20544} \pi^2-
\frac{275}{214} \zeta_3+\frac{11}{80} \pi^4\right]\, b_{\BZ}^3
\\&
+\bigg[
a_{4,0}+\frac{16}{107} \beta_{4,0}+\frac{25685}{54784} \pi^4
+\left(-\frac{4955688493}{422055936}-\frac{5335}{856}\zeta_3\right)\pi^2
\\&+\frac{4750031643031}{1083839643648}-\frac{75857683}{3297312}\zeta_3
 \bigg]
\, b_{\BZ}^4+\cdots \\
 &\hspace*{-20pt} \simeq
 b_{\BZ}-0.3308 \, b_{\BZ}^2\, +\,6.902\, b_{\BZ}^3+
\left[a_{4,0}+0.1495\,\beta_{4,0}-167.43\right]\,b_{\BZ}^4+\cdots \,.
\end{split}
\end{align}
For $N_f=4$ we obtain the following numerical values at the first three truncation orders:
$a_{\cal J,\,S}^{\Eucl}\left(\mu^2\to 0\right)/\beta_0=0.312$,\, $0.279$ and $0.488$, corresponding to $a_{\cal J,\,S}^{\Eucl}\left(\mu^2\to 0\right)=0.649$,\, $0.582$ and $1.017$, respectively.
The higher--order corrections are certainly significant, in particular the NNLO ones.
The results suggest a fixed point value that is marginally perturbative.
This is consistent with the conclusion we reached in Sec.~\ref{sec:ECH_evolution} by considering directly the fixed--point solution of the effective--charge beta function.

\section{A dispersive representation of the moment--space Sudakov exponent~\label{sec:dispersive_exponent}}
\setcounter{footnote}{1}

In Sec.~\ref{sec:general_disparsive_formulae} we have written a dispersive
representation for the physical Sudakov anomalous dimensions.
Its ingredients have been studied in detail,
the characteristic functions in Sec.~\ref{sec:kernel} and the
corresponding effective charges in Sec.~\ref{sec:general_dispersive}.
We are now ready to use these tools to construct a dispersive
representation of the Sudakov exponent based on the definitions in Sec.~\ref{sec:setup}.
We proceed as follows: in Sec.~\ref{sec:dispersive_exponent_derivation} we present a short derivation of the main result, a general dispersive representation of the Suakov exponent. Then, in Sec.~\ref{sec:mom_space_chars_and_expansions} we present explicit results for the moment--space characteristic functions entering the Sudakov exponent in various examples and analyze their limits. Finally, in Sec.~\ref{sec:virt} we return to the general discussion by considering in detail the role of virtual corrections in the large--$N$ limit. This discussion elucidates the interpretation of the dispersive formula.

\subsection{The Sudakov exponent: a general dispersive representation~\label{sec:dispersive_exponent_derivation}}

The crucial step here is going from momentum space, where the
real--emission contribution is finite and can be considered
separately, to moment space, where an infrared singularity is
generated, which requires cancellation with virtual corrections.
This cancellation is incorporated in the Sudakov exponent of Eq.~(\ref{Exponent}), or
equivalently
in (\ref{Exponent_Laplace_with_Mellin_const}) or (\ref{Exponent_Laplace}).

It is convenient to consider first the Sudakov evolution equation
obtained through the logarithmic derivative of (\ref{Exponent_Laplace}):
\begin{align}
\label{diff_Exponent_Laplace}
\begin{split}
\frac{d\ln \overline{\rm Sud}(m^2,N)}{d \ln m^2} & =
C_R\int_0^{\infty}\,\frac{dr}{r}
\Big[{\rm e}^{-Nr}-{\rm e}^{-r}\Big] \,\frac{d R(m^2,r)}{d\ln m^2}\\
& = C_R\int_0^{\infty}\,\frac{dr}{r}
\Big[{\rm e}^{-Nr}-{\rm e}^{-r}\Big] \,
\Big\{\,{\cal J}(r m^2)-{\cal S}(r^2 m^2)\Big\}\\
& = C_R\int_0^{\infty}\,\frac{dr}{r}
\Big[{\rm e}^{-Nr}-{\rm e}^{-r}\Big] \,{\cal J}(r m^2)
-C_R\int_0^{\infty}\,\frac{dr}{r}
\Big[{\rm e}^{-Nr}-{\rm e}^{-r}\Big] \,{\cal S}(r^2 m^2)\,,
\end{split}
\end{align}
where in the second line we have inserted the
derivative of the momentum--space kernel (\ref{dR}) that admits
infrared factorization.
Owing to the cancellation between the real and virtual terms
for $r\to 0$, the `jet' and `soft' contributions to
(\ref{diff_Exponent_Laplace}) are
\emph{separately} infrared finite. This explains the
third line.
The purely real Laplace integral,
\[
\int_0^{\infty}\,\frac{dr}{r} {\rm e}^{-Nr}\,{\cal J}(r m^2),
\]
however, is infrared divergent. Here, the dispersive integral
becomes handy. Substituting into (\ref{diff_Exponent_Laplace})
the dispersive representation of the Sudakov anomalous dimensions
according to (\ref{J_Sud_anom_dim_conv}) and
(\ref{S_Sud_anom_dim_conv}), respectively, we obtain:
\begin{align}
\label{diff_Exponent_Laplace1}
\begin{split}
\frac{d\ln \overline{\rm Sud}(m^2,N)}{d \ln m^2}
& = \frac{C_R}{\beta_0}\,\Bigg\{
\int_0^{\infty}\,\frac{dr}{r}\,\Big[{\rm e}^{-Nr}-{\rm e}^{-r}\Big]
\,\int_0^{\infty}\frac{d\epsilon}{\epsilon}
\rho_{\cal J}(\epsilon m^2)
\left[\dot{\cal F}_{\cal J}\left(\epsilon/r\right)
-\dot{\cal F}_{\cal J}(0)\right]
\\&\hspace*{30pt}
-
\int_0^{\infty}\,\frac{dr}{r}\,\Big[{\rm e}^{-Nr}-{\rm e}^{-r}\Big]
\,\int_0^{\infty}\frac{d\epsilon}{\epsilon}
\rho_{\cal S}(\epsilon m^2) \left[
\dot{\cal F}_{\cal S}\left(\epsilon/r^2\right)
-\dot{\cal F}_{\cal S}\left(0\right)\right]
\Bigg\}\,,
\end{split}
\end{align}
where, as usual, $\epsilon=\mu^2/m^2$.
Next, changing the order of integration and
defining the moment--space characteristic functions by
the Laplace transform of the momentum--space ones:
\begin{align}
\label{G_def}
\begin{split}
\dot{\cal G}_{\cal J}\left(N\epsilon\right)
&=\int_0^{\infty}\,\frac{dr}{r}\,{\rm e}^{-Nr}\,
\dot{\cal F}_{\cal J}\left(\epsilon/r\right)\\
\dot{\cal G}_{\cal S}\left(N^2\epsilon\right)
&=\int_0^{\infty}\,\frac{dr}{r}\,{\rm e}^{-Nr}\,
\dot{\cal F}_{\cal S}\left(\epsilon/r^2\right)\,,
\end{split}
\end{align}
we obtain an elegant dispersive representation of the Sudakov
evolution equation:
\begin{align}
\label{diff_Exponent_Laplace2}
\begin{split}
\frac{d\ln \overline{\rm Sud}(m^2,N)}{d \ln m^2} &=
\frac{C_R}{\beta_0}\,\Bigg\{\,
\int_0^{\infty}\frac{d\epsilon}{\epsilon} \,\,
\rho_{\cal J}(\epsilon m^2)
\left(\dot{\cal G}_{\cal J}\left(\epsilon N\right)-
\dot{\cal G}_{\cal J}\left(\epsilon\right)+\ln N\right)
\\&\hspace*{75pt}-
\int_0^{\infty}\frac{d\epsilon}{\epsilon} \,\,
\,\rho_{\cal S}(\epsilon m^2)
\left(\dot{\cal G}_{\cal S}\left(\epsilon N^2\right)-
\dot{\cal G}_{\cal S}\left(\epsilon\right)+\ln N\right)
\Bigg\}\,,
\end{split}
\end{align}
where the $\ln N$ terms originates in the terms
proportional to $\dot{\cal F}_{\cal J}(0)$ and
$\dot{\cal F}_{\cal S}(0)$ in (\ref{diff_Exponent_Laplace1}),
where we have evaluated the trivial integral
\[
\int_0^\infty\frac{dr}{r}\left[{\rm e}^{-Nr}-{\rm e}^{-r}\right]=-\ln N
\]
and used Eq.~(\ref{dotF_equality}) setting
$\dot{\cal F}_{\cal J}(0)=\dot{\cal F}_{\cal S}(0)=1$.
We have explicitly written the
`jet' and `soft' contributions to the r.h.s of
(\ref{diff_Exponent_Laplace2}) as two separate integrals, in order to emphasize the fact that they are separately finite:
the combinations
\begin{align}
\label{power_like_fall_off}
\begin{split}
\dot{\cal G}_{\cal J}\left(\epsilon N\right)-
\dot{\cal G}_{\cal J}\left(\epsilon\right) + \ln N\qquad {\text{and}}\qquad
\dot{\cal G}_{\cal S}\left(\epsilon N^2\right)-
\dot{\cal G}_{\cal S}\left(\epsilon\right) + \ln N
\end{split}
\end{align}
are each falling as a power of $\epsilon$; indeed
the expansion of the characteristic functions at small arguments starts as
\begin{align}
\label{small_epsilon_behaviour}
\begin{split}
\dot{\cal G}_{\cal J}\left(\epsilon N\right)&=
-\ln(\epsilon N)+J_1-\gamma_E+{\cal O}(\epsilon N)\\
\dot{\cal G}_{\cal S}\left(\epsilon N^2\right)&=
-\frac12\ln(\epsilon N^2)+\frac12 S_1-\gamma_E+{\cal O}\left((\epsilon N^2)^{1/2}\right)\,,
\end{split}
\end{align}
where $S_1$ and $J_1$ are the log-moments defined in (\ref{JkSk_def}). These expansions are most easily derived using Eq.~(\ref{inv_G_laplace_B_JS}) below. Explicit examples will be given in Sec.~\ref{sec:mom_space_chars_and_expansions}.

Because `jet' and `soft' contributions to the r.h.s of
(\ref{diff_Exponent_Laplace2}) are separately finite, evolution equations similar to (\ref{diff_Exponent_Laplace2}) can be written separately for the `jet' and `soft'
Sudakov factors (see e.g. Eqs. (8) and (22) in Ref.~\cite{Gardi:2006jc}) whose product is $\overline{\rm Sud}(m^2,N)$. The definition of such factors, however, requires to introduce a
factorization scale. This will not be necessary in the derivation we present below, where we shall combine the two dispersive integrals in (\ref{diff_Exponent_Laplace2}) into one.

Note that integrating Eq.~(\ref{diff_Exponent_Laplace2}) by parts we obtain an alternative expression in terms of the Sudakov effective charges:
\begin{align}
\label{diff_Exponent_Laplace2_by_parts}
\begin{split}
\frac{d\ln \overline{\rm Sud}(m^2,N)}{d \ln m^2} &=
\frac{C_R}{\beta_0}\,\Bigg\{\,
\int_0^{\infty}\frac{d\epsilon}{\epsilon} \,\,
a_{\cal J}^{\Mink}(\epsilon m^2)
\left(\ddot{\cal G}_{\cal J}\left(\epsilon N\right)-
\ddot{\cal G}_{\cal J}\left(\epsilon\right)\right)
\\&\hspace*{75pt}-
\int_0^{\infty}\frac{d\epsilon}{\epsilon} \,\,
\,a_{\cal S}^{\Mink}(\epsilon m^2)
\left(\ddot{\cal G}_{\cal S}\left(\epsilon N^2\right)-
\ddot{\cal G}_{\cal S}\left(\epsilon\right)\right)
\Bigg\}\,,
\end{split}
\end{align}

The final step in deriving a dispersive representation of the exponent
is to integrate the evolution equation (\ref{diff_Exponent_Laplace2}).
Quite conveniently, the entire $m^2$ dependence of the r.h.s in
(\ref{diff_Exponent_Laplace2})
is  through $\rho_{\cal J}(\epsilon m^2)$
and $\rho_{\cal S}(\epsilon m^2)$. Using
(\ref{eq:int_discontinuity_JS}) we can therefore readily integrate
(\ref{diff_Exponent_Laplace2}) getting:
\begin{align}
\label{Dispersive_Exponent_Laplace}
\begin{split}
\overline{\rm Sud}(m^2,N) &=
\exp\Bigg\{\frac{C_R}{\beta_0}\,\,\int_0^{\infty}\frac{d\epsilon}{\epsilon} \,\,
\bigg[a_{\cal J}^{\Mink}(\epsilon m^2)
\left(\dot{\cal G}_{\cal J}\left(\epsilon N\right)-
\dot{\cal G}_{\cal J}\left(\epsilon\right)+\ln N\right)
\\&\hspace*{75pt}-\,a_{\cal S}^{\Mink}(\epsilon m^2)
\left(\dot{\cal G}_{\cal S}\left(\epsilon N^2\right)-
\dot{\cal G}_{\cal S}\left(\epsilon\right)+\ln N\right)
\bigg]\Bigg\}\,,
\end{split}
\end{align}
which is the main result of this section.
Explicit results for the moment--space characteristic functions
$\dot{\cal G}_{\cal J,\,S}$ in a few examples will be computed and analyzed in the next section; the final expressions are compiled in Table~\ref{table:dcalG}.

The exponent (\ref{Dispersive_Exponent_Laplace}) is
finite owing to the exact
cancellation of logarithmic singularities to any order
in perturbation theory; there are two levels
of cancellation: first between real and virtual terms
within the separate `jet' and `soft' parts,
and second between these terms. The latter cancellation mixes the infrared and
ultraviolet in an interesting way. To see this note that
for $\epsilon\to 0$ the convergence of the separate `jet' and
`soft' parts in (\ref{Dispersive_Exponent_Laplace}) follows from
the convergence of (\ref{diff_Exponent_Laplace2})
(the power--like falloff of the combinations in~(\ref{power_like_fall_off}) above).
On the other hand, in contrast with (\ref{diff_Exponent_Laplace2}),
for $\epsilon\to \infty$ the integrals corresponding to the
`jet' and the `soft' parts in
(\ref{Dispersive_Exponent_Laplace}) are \emph{not} separately finite.
For the $\dot{\cal G}_{\cal J,\,S}$ terms convergence for
$\epsilon\to \infty$ is guaranteed by the
Laplace weight in (\ref{G_def}).
The ultraviolet divergence is therefore entirely due to
the $\ln N$ term, identifying its origin as the cusp singularity.
In conclusion, the $\ln N$ terms are required for the
separate `jet' and `soft' integrals to converge
for $\epsilon\to 0$, but render each of them divergent for
\hbox{$\epsilon\to \infty$}.
Nevertheless, the exponent as a whole
is finite owing to the fact that
$a_{\cal S}^{\Mink}(\epsilon m^2)$ and
$a_{\cal J}^{\Mink}(\epsilon m^2)$ are equal at leading and at next-to-leading order (see Sec.~(\ref{sec:general_disparsive_formulae})) and therefore their difference behaves as $\alpha_s^3(\mu^2)$ at large $\mu^2$.

\begin{table}[htb]
  \centering
  \begin{tabular}{|l|l|}
    \hline
    &\vspace*{-13pt}\\
     process        & moment--space characteristic function, \,\,
$\dot{\cal G}_{\cal J}(\xi)^{\qquad}$
    \\ \hline\hline
    &\vspace*{-6pt}\\
    \begin{tabular}{l}
    jet function\\
    (e.g. DIS)
    \end{tabular}
    & $
    {\displaystyle \left(1+\frac{\xi}{2}-\frac{\xi^2}{4}\right)\,{\rm Ei}(1,\xi)
+\left(\frac{\xi}{4}-\frac{3}{4}\right)\,{\rm e}^{-\xi}}$
 \\
    &\vspace*{-6pt}\\
    \hline\hline
    &\vspace*{-13pt}\\
     process        & moment--space characteristic function, \,\,
$\dot{\cal G}_{\cal S}(\nu^2)^{\qquad}$
    \\ \hline
    &\vspace*{-6pt}\\
    \begin{tabular}{l}
    B decay; HQ  \\
    Fragmentation
    \end{tabular}
    & $
    {\displaystyle \frac12+
\left[\frac{\nu}{2} \cos(\nu)-\sin(\nu)\right]\left({\rm Si}(\nu)-\frac{\pi}{2}\right)
-\left[\frac{\nu}{2} \sin(\nu)+\cos(\nu)\right]{\rm Ci}(\nu)}$
  \\
    &\vspace*{-6pt}\\
    \hline
    &\vspace*{-6pt}\\
    \begin{tabular}{l}
    Drell--Yan\,; \\
    $gg \to$ Higgs
    \end{tabular}
    & $
    {\displaystyle K_0(2\nu)}$
 \\
    &\vspace*{-6pt}\\
    \hline
    &\vspace*{-12pt}\\
    \begin{tabular}{l}
    $e^+e^-\to \text{jets}$\\
    C parameter\\ ($c=C/6$)
    \end{tabular}
    & $
{\scriptsize \displaystyle \frac{\pi\,\nu}{4}-\gamma_E-
\ln(\nu)\,-\,
\frac{\nu^2}{8}\,\, _3F_2
\left(
\left.\begin{array}{l}
1, 1\\
\frac{3}{2}, \frac{3}{2}, 2
\end{array}\right\vert
\frac{\nu^2}{16}\right)
\,+\,\frac{\pi\,\nu^3}{192}\,\, _3F_2
\left(
\left.\begin{array}{l}
1, \frac{3}{2}\\
2, 2, \frac{5}{2}
\end{array}\right\vert
\frac{\nu^2}{16}
\right)
}$
 \\
    &\vspace*{-12pt}\\\hline
    &\vspace*{-12pt}\\
    \begin{tabular}{l}
    $e^+e^-\to \text{jets}$\\
    Thrust\\ ($t=1-T$)
    \end{tabular}
     & $
     {\displaystyle {\rm Ei}(1,\nu)}$\vspace*{-12pt} \\
    &\\\hline
  \end{tabular}
  \caption{Summary of results for the moment--space characteristic
  functions $\dot{\cal G}_{{\cal J}}(\xi=N\epsilon)$ and
$\dot{\cal G}_{{\cal S}}(\nu^2=N^2\epsilon)$
of some inclusive distributions. }
  \label{table:dcalG}
\end{table}

Note that a somewhat different picture emerges in the
approximation where
$a_{\cal J}^{\Mink}(\epsilon m^2)=a_{\cal J}^{\Mink}(\epsilon m^2)
=a_{\GB}^{\Mink}(\epsilon m^2)$, which is valid to NLL accuracy (see
(\ref{aSJ_one_loop})) or in the large--$\beta_0$ limit.
In this case, the $\ln N$ terms cancel, and we obtain
\begin{align}
\label{Dispersive_Exponent_Laplace_NLL}
\begin{split}
\left.\overline{\rm Sud}(m^2,N)\right\vert_{\rm NLL}
&=
\exp\Bigg\{\frac{C_R}{\beta_0}\,\,\int_0^{\infty}
\frac{d\epsilon}{\epsilon} \,\,a_{\GB}^{\Mink}(\epsilon m^2)\,\,
\\& \times
\bigg[
\left(\dot{\cal G}_{\cal J}\left(\epsilon N\right)-
\dot{\cal G}_{\cal J}\left(\epsilon\right)\right)
-
\left(\dot{\cal G}_{\cal S}\left(\epsilon N^2\right)-
\dot{\cal G}_{\cal S}\left(\epsilon\right)\right)
\bigg]\Bigg\}\,.
\end{split}
\end{align}
Here each of the terms is separately finite for $\epsilon\to \infty$,
but the $\epsilon\to 0$ singularities cancel in two levels: first
between real and virtual, and second between
the `jet' and `soft' terms.

\subsection{Moment--space characteristic functions and their expansions\label{sec:mom_space_chars_and_expansions}}

Let us examine now some explicit examples. Having determined
the functional form of the momentum--space characteristic
functions (Table~\ref{table:calF_B})
we can readily compute the Laplace integrals (\ref{G_def})
to determine the moment--space ones.
The final results are collected in Table~\ref{table:dcalG}.

In what follows we give further details on the calculation of the characteristic functions and address the
convergence of the dispersive integral in the exponent
(\ref{Dispersive_Exponent_Laplace}) considering
the asymptotic behavior of these functions at small and large
$\epsilon$. In the Drell--Yan case we extend the discussion and compare the dispersive formalism developed here with the joint resummation formalism or Ref.~\cite{Laenen:2000ij}.
For Drell--Yan we also derive an alternative representation of the exponent in terms of a Euclidean integral~\cite{Grunberg:2006gd}. Such a representation does not exist in the other examples.

\vspace*{10pt}
\noindent
\underline{The jet function}\\
Let us begin by constructing an explicit expression for the
moment--space characteristic function of the `jet',
$\dot{\cal G}_{\cal J}(N\mu^2/m^2)$, based on the momentum--space
one given in (\ref{calF_dot_QD}). According to (\ref{G_def}) we have:
\begin{align}
\label{WJ}
\begin{split}
\dot{\cal G}_{\cal J}(\xi)&= \int_0^{\infty}\frac{dr}{r}\,{\rm e}^{-Nr} \,\,\dot{\cal F}_{\cal J}\left(\frac{\mu^2}{m^2 r}\right)\\
&= \int_0^{\infty}\frac{dr}{r}\,{\rm e}^{-Nr} \,\,
\,\theta(r>\mu^2/m^2)\,\left[1-\frac12\frac{\mu^2}{m^2 r}-\frac12\left(\frac{\mu^2}{m^2 r}\right)^2\right]\\
&=\int_1^{\infty}\frac{dw}{w}\,{\rm e}^{-\xi w}\left[1-\frac12 \frac{1}{w}-\frac12 \frac{1}{w^2}\right]\\
&=\left(1+\frac{\xi}{2}-\frac{\xi^2}{4}\right)\,{\rm Ei}(1,\xi)
+\left(\frac{\xi}{4}-\frac{3}{4}\right)\,{\rm e}^{-\xi}
\end{split}
\end{align}
where $\xi\equiv N\mu^2/m^2$ and where the
Exponential--integral function is defined as usual by
\begin{equation}
\label{Ei}
{\rm Ei}(p,\xi) \equiv
\int_1^{\infty}\,\frac{d\alpha}{\alpha^p}
{\rm e}^{-\alpha \xi}.
\end{equation}

Let us now examine the convergence of the $\epsilon$
integral in (\ref{Dispersive_Exponent_Laplace}).
First, for $\epsilon\to \infty$ there is convergence
for each of the separate $\dot{\cal G}_{\cal J}(\xi)$ terms
owing to the behavior of this function for large $\xi$:
\begin{align}
\begin{split}
\dot{\cal G}_{\cal J}(\xi)&\simeq \left[\frac32\frac{1}{\xi^2}+{\cal O}\left(\frac{1}{\xi^3}\right)\right]\,{\rm e}^{-\xi}.
\end{split}
\end{align}

In the $\epsilon\to 0$ limit there is convergence in
(\ref{Dispersive_Exponent_Laplace}) only owing to cancellations between the separate terms.
We obtain
\begin{align}
\label{WJ_IR}
\begin{split}
&\dot{\cal G}_{\cal J}(\epsilon N)=
\left(-\gamma_E-\ln(N)-\ln(\epsilon)-\frac{3}{4}\right)
+\left(2 -\frac12   \Big(\gamma_E+\ln(N)+\ln(\epsilon)\Big)\right)
N \epsilon+{\cal O}(\epsilon^2)\,,
\end{split}
\end{align}
making the combination
$\dot{\cal G}_{\cal J}\left(\epsilon N\right)-
\dot{\cal G}_{\cal J}\left(\epsilon\right)+\ln N$ power suppressed
as required for the integral to converge.

\vspace*{10pt}
\noindent
\underline{The Quark Distribution function (B decay)}\\
Let us consider now the `soft' function associated
with the Quark Distribution in an on-shell heavy quark.
Using in (\ref{G_def}) the explicit expression for the momentum--space
characteristic function in (\ref{calF_dot_QD}) we obtain :
\begin{align}
\label{WS}
\begin{split}
\dot{\cal G}_{{\cal S}_{\QD}}(\nu^2)&=\int_0^{\infty}\frac{dr}{r}\,{\rm e}^{-Nr}\,
\dot{\cal F}_{{\cal S}_{\QD}}\left(\frac{\mu^2}{m^2 r^2}\right)\\
&=\int_0^{\infty}\frac{d\eta}{\eta}\,{\rm e}^{-\nu\,\eta}\left[\frac{\eta^2}{1+\eta^2}+\frac{\eta^2}{(1+\eta^2)^2}
\right]\\
&=\frac12
+\frac{1}{2} \Big({\rm e}^{-i \nu}{\rm Ei}(1,-i \nu)
+ {\rm e}^{i \nu} {\rm Ei}(1,i \nu)\Big)
+\frac{i \nu}{4} \Big( {\rm e}^{-i \nu} {\rm Ei}(1,-i\nu)
-  {\rm e}^{i \nu}  {\rm Ei}(1,i \nu)\Big)\\
&=\frac12+
\left[\frac{\nu}{2} \cos(\nu)-\sin(\nu)\right]\left({\rm Si}(\nu)-\frac{\pi}{2}\right)
-\left[\frac{\nu}{2} \sin(\nu)+\cos(\nu)\right]{\rm Ci}(\nu)\,,
\end{split}
\end{align}
where we have defined $\nu=N\mu/m=N\sqrt{\epsilon}$
and changed variables
to $\eta^2=r^2/\epsilon$. The
Exponential--integral function is defined in (\ref{Ei}) above,
and
\begin{align}
\begin{split}
{\rm Si}(\nu)   &\equiv \int_0^\nu\frac{dt}{t}\sin(t)\,,\\
{\rm Ci}(\nu)   &\equiv \gamma_E + \ln(\nu) +
\int_0^\nu \frac{dt}{t}\,(\cos(t)-1)\,.
\end{split}
\end{align}

Let us examine the convergence of the $\epsilon$ integral in
(\ref{Dispersive_Exponent_Laplace}). For
$\epsilon\to \infty$ we find
\begin{align}
\begin{split}
\dot{\cal G}_{{\cal S}_{\QD}}(\nu)&\simeq \frac{2}{\nu^2}+
{\cal O}\left(\frac{1}{\nu^3}\right)\,,
\end{split}
\end{align}
which guarantees convergence separately for each of the
$\dot{\cal G}_{{\cal S}_{\QD}}$ terms.
As expected, in the $\epsilon\to 0$ limit
there is convergence in (\ref{Dispersive_Exponent_Laplace}) only owing to
cancellations between the terms.
We obtain
\begin{align}
\label{WS_IR}
\begin{split}
\dot{\cal G}_{{\cal S}_{\QD}}(N^2{\epsilon})=\left(-\gamma_E-\ln(N)+\frac12 -\frac12   \ln(\epsilon)\right)
+\frac{\pi}{4}   N    \epsilon^{1/2}
-\frac14   N^2  \epsilon+
\frac{\pi}{24}  N^3    \epsilon^{3/2}
+{\cal O}(\epsilon^2)
\end{split}
\end{align}
making the combination
$\dot{\cal G}_{{\cal S}_{\QD}}\left(N^2 \epsilon\right)
-
\dot{\cal G}_{{\cal S}_{\QD}}\left(\epsilon\right)+\ln N$ power suppressed.

\vspace*{10pt}
\noindent
\underline{The Drell--Yan soft function}\\
Next, let us consider the `soft' function associated with
Drell--Yan production. Using in (\ref{G_def}) the expression for the
momentum--space characteristic function (\ref{Fr-x-scaling-DY1}),
we obtain:
\begin{align}
\label{WS_DY}
\begin{split}
\dot{\cal G}_{{\cal S}_{\DY}}(\nu^2)\,=\,
\int_0^{\infty}\frac{dr}{r}\,{\rm e}^{-Nr}\,
\dot{\cal F}_{{\cal S}_{\DY}}\left(\frac{\mu^2}{Q^2 r^2}\right)
\,=\,\frac{1}{2}\int_0^{1}\frac{d\zeta}{\zeta\sqrt{1-\zeta}}\,{\rm e}^{-2\nu/\,\sqrt{\zeta}}\,\,
\,=\,K_0(2\nu)\,,
\end{split}
\end{align}
where we have defined $\nu=N\mu/m=N\sqrt{\epsilon}$
and changed the integration variable from $r$ to $\zeta=4\epsilon/r^2$. Here $K_0$
is the modified Bessel function of the second kind.

To examine the convergence of the dispersive integral in the exponent
(\ref{Dispersive_Exponent_Laplace}) for $\epsilon\to \infty$, let us
expand (\ref{WS_DY}) at large $\nu=N\sqrt{\epsilon}$. The result is:
\begin{align}
\dot{\cal G}_{{\cal S}_{\DY}}(\nu^2)&=\,\frac{1}{2}\,{\rm e}^{-2\nu}\,
\sqrt{\frac{\pi}{\nu}}\,\left(
1-\frac{1}{16}\frac{1}{\nu}+{\cal O}\left(\frac{1}{\nu^2}\right)\right).
\end{align}
Obviously, the integrals over the $\dot{\cal G}_{{\cal S}_{\DY}}$ terms converge for large $\epsilon$.

Next, consider the $\epsilon\to 0$ limit. Here the expansion yields
\begin{align}
\label{exp_G_DY}
\dot{\cal G}_{{\cal S}_{\DY}}(N^2\epsilon)&=
\Big(-\ln\left(N\sqrt{\epsilon}\right)-\gamma_E\Big)
+\Big(-\ln\left(N\sqrt{\epsilon}\right)-\gamma_E+1\Big) N^2\epsilon
+{\cal O}\left(\left(N^2\epsilon\right)^2\right)\,
\end{align}
making the combination $\dot{\cal G}_{{\cal S}_{\DY}}(N^2\epsilon)
-
\dot{\cal G}_{{\cal S}_{\DY}}(\epsilon)+\ln N$ power suppressed at small
$\epsilon$, as required for convergence. Note that only {\em even} powers of $\nu$ appear in this expansion.

Let us return now to the physical Drell--Yan kernel. The Laplace equivalent of (\ref{dG_N}), which differs from it only by ${\cal O}(1/N)$ terms for $N\to \infty$, is
\begin{equation}\label{DY_kernel_Mink1}\frac{d\ln G_{q\bar{q}}(N,Q^2,\mu_F^2)}{d\ln Q^2}\simeq\frac{d\ln \widetilde{\rm Sud}_{\DY}(Q^2,N)}{d \ln Q^2}+\frac{d\ln H_{\DY}(Q^2,\mu^2_F)}{d\ln Q^2}+{\cal O}(1/N)\,,\end{equation}
with
\begin{align}
\label{DY_kernel_Mink}
\begin{split}
&\frac{d\ln \widetilde{\rm Sud}_{\DY}(Q^2,N)}{d \ln Q^2}=
2C_F\int_0^{\infty}\frac{dr}{r} \left({\rm e}^{-Nr}-\theta(r<1)\right) {\cal S}_{\DY}(Q^2r^2)\\
&\hspace*{3pt}=\frac{2C_F}{\beta_0}\int_0^{\infty}\frac{dr}{r}
\left({\rm e}^{-Nr}-\theta(r<1)\right) \int_0^{\infty}
\frac{dp^2}{p^2}\rho_{{\cal S}_{\DY}}(p^2)\left[\dot{\cal F}_{{\cal S}_{\DY}}\left(\frac{p^2}{Q^2r^2}\right)-
\dot{\cal F}_{{\cal S}_{\DY}}(0)\right]
\\
&\hspace*{3pt}=\frac{2C_F}{\beta_0}
\int_0^{\infty}\frac{dp^2}{p^2}\,\rho_{{\cal S}_{\DY}}(p^2)
\left[\dot{\cal G}_{{\cal S}_{\DY}}\left(\frac{N^2p^2}{Q^2}\right)-\frac12 {\cal F}_{{\cal S}_{\DY}}\left(\frac{p^2}{Q^2}\right)+\gamma_E+\ln(N)
\right]
\\
&\hspace*{3pt}=\frac{2C_F}{\beta_0}
\int_0^{\infty}\,\frac{dp^2}{p^2}\,a_{{\cal S}_{\DY}}^{\Mink}(p^2)
\frac{d}{d\ln Q^2}\left[\dot{\cal G}_{{\cal S}_{\DY}}\left(\frac{N^2p^2}{Q^2}\right)-\frac12
{\cal F}_{{\cal S}_{\DY}}\left(\frac{p^2}{Q^2}\right)\right]\\
&\hspace*{3pt}=\frac{2C_F}{\beta_0}
\int_0^{\infty}\,\frac{dp^2}{p^2}\,a_{{\cal S}_{\DY}}^{\Mink}(p^2)
\frac{d}{d\ln Q^2}\bigg\{K_0\left(2\sqrt{{N^2p^2}/{Q^2}}\right)
-\frac12\theta(p^2/Q^2<1/4)\times
\\&\hspace*{150pt}\,
\Big[2\ln\frac{1+\sqrt{1-4p^2/Q^2}}{2}-\ln(p^2/Q^2)
\Big]\,\bigg\}
\\
&\hspace*{3pt}\simeq \frac{2C_F}{\beta_0}
\int_0^{Q^2/4}\,\frac{dp^2}{p^2}\,a_{{\cal S}_{\DY}}^{\Mink}(p^2)
\frac{d}{d\ln Q^2}\bigg\{K_0\left(2\sqrt{{N^2p^2}/{Q^2}}\right)
-\ln(\sqrt{Q^2/p^2})
\Big]\,\bigg\}\,,
\end{split}
\end{align}
where in the second line we used the dispersive representation of ${\cal S}_{\DY}$ from (\ref{S_Sud_anom_dim_conv}), in the third (see derivation in Sec.~\ref{sec:virt})
we performed the integration over $r$ and substituted the Laplace transform of $\dot{\cal F}_{\DY}$ according to (\ref{G_def}) (and used $\dot{\cal F}_{\DY}(0)=1$), in the fourth we performed integration-by-parts and in the fifth we inserted the explicit expressions from (\ref{WS_DY}) and (\ref{calF_DY_z}). Finally, in the last line we have changed the upper limit of integration over the $K_0$ term, which modifies the result by an exponentially small correction at large $N$, and neglected the first term in the square brackets,
$2\ln\Big(({1+\sqrt{1-4p^2/Q^2}})/{2}\Big)$, which generates ${\cal O}(\Lambda^2/Q^2)$
power--suppressed corrections.

Note that the logarithmic derivative with respect to $Q^2$ cannot be pulled out of the $p^2$ integral in (\ref{DY_kernel_Mink}), which would then diverge. This divergence is the usual collinear divergence associated with the evolution of the quark distribution ---
in the Sudakov exponent of infrared and collinear safe distributions, Eq.~(\ref{widetilde_Exponent_Laplace_with_Mellin_const}), it cancels with the jet subprocess. In order to obtain $\widetilde{\rm Sud}_{\DY}(Q^2,N,\mu^2_F)$ one would obviously need to introduce a factorization scale.

It is interesting to compare our final result for the large--$N$ Drell--Yan kernel (\ref{DY_kernel_Mink}) with the equivalent expression derived in the joint resummation formalism~\cite{Laenen:2000ij}. To this end consider the result for the Eikonal cross section, Eq. (48) in \cite{Laenen:2000ij}, where dimensional regularization ($D=4-2\varepsilon$) is used to regularize the collinear divergence in the exponent:
\begin{align}
\label{joint_resummation}
\begin{split}
\bar{\sigma}_{q\bar{q}}(N,bQ,\varepsilon)&=\exp\Bigg\{
2\int_0^{Q^2}  dp^2 \int {dk^2} \int\frac{d^{2-2\varepsilon}k_{\perp}}{{\Omega}_{1-2\varepsilon}}\,
\omega_{q\bar{q}}(k^2,k^2+k_{\perp}^2,\mu^2,\varepsilon)\,\delta\Big(k^2+k_{\perp}^2-p^2\Big)\\
&\hspace*{100pt}\times \left[{\rm e}^{-ibk_{\perp}}\,K_{0}\left(2\sqrt{N^2p^2/Q^2}\right)-\ln\sqrt{Q^2/p^2}\right]
\Bigg\}\,
\end{split}
\end{align}
where $b$ is the Fourier conjugate to the transverse momentum of the system, and $\omega_{q\bar{q}}$
is the ``web''~\cite{Gatheral:1983cz} corresponding to radiation from the annihilating lightlike quark and antiquark, defined at fixed transverse momentum.
One immediately recognizes that upon considering the small $b$ limit, corresponding to the ``total'' cross section (or to a case where the resummation of transverse momentum logarithms is unimportant compared to threshold logarithms, $bm\ll N$),  Eq.~(\ref{joint_resummation}) reproduces Eq.~(\ref{DY_kernel_Mink}) up to\footnote{
Finite terms are generated by a different upper limit on the momentum integration, $Q^2$ in (\ref{joint_resummation}) versus $Q^2/4$ in (\ref{DY_kernel_Mink}), which amounts to a different split of the virtual corrections between the Sudakov factor and the hard function that multiplies it.} terms that are finite at $N\to \infty$,
 provided one makes the following identification\footnote{In the $\varepsilon\to 0$ limit, the result is independent on the dimensional regularization scale $\mu$. For example, in case of a single gluon emission (Eq. (45) in~\cite{Laenen:2000ij}), the r.h.s in (\ref{Web_identification}) is
$\lim_{\varepsilon\to 0} C_F(\alpha_s/\pi) (4\pi{\rm e}^{-\gamma_E}\mu^2/p^2)^{\varepsilon} \,/p^2$. }:
\begin{align}
\label{Web_identification_}
a_{{\cal S}_{\DY}}^{\Mink}(p^2)\,=\,
\lim_{\varepsilon\to 0} \,a_{{\cal S}_{\DY}}^{\Mink}(p^2,\mu^2,\varepsilon)
\end{align}
where
\begin{align}
\label{Web_identification}
\frac{C_F}{\beta_0} \,\frac{a_{{\cal S}_{\DY}}^{\Mink}(p^2,\mu^2,\varepsilon)}{p^2}\,\equiv\,
\int dk^2 \int\frac{d^{2-2\varepsilon}k_{\perp}}{{\Omega}_{1-2\varepsilon}}\,
\omega_{q\bar{q}}(k^2,k^2+k_{\perp}^2,\mu^2,\varepsilon) \,\delta\Big(k^2+k_{\perp}^2-p^2\Big)\,.
\end{align}
This gives a direct diagrammatic interpretation to the Minkowskian effective charge in terms of webs.

Eq.~(\ref{DY_kernel_Mink}) is the standard dispersive representation of the kernel that involves the \emph{Minkowskian} effective charge. In the Drell--Yan case, the exists an alternative \emph{Euclidean} representation~\cite{Grunberg:2006gd}. To derive it one defines a Euclidean characteristic function by the following dispersion relation:
\begin{equation}
\label{G_Eucl}
\ddot{\cal G}_{{\cal S}_{\DY}}(\epsilon)=\epsilon \int_0^{\infty}\frac{dy}{(\epsilon+y)^2}\,\ddot{\cal G}_{{\cal S}_{\DY}}^{\Eucl}(y)
\end{equation}
Substituting this expression for $\ddot{\cal G}_{{\cal S}_{\DY}}$ in
the soft part of
(\ref{diff_Exponent_Laplace2}) (after integration by parts) one gets:
\begin{align}
\label{DY_kernel_Eucl}
\begin{split}
\frac{d\ln \overline{\rm Sud}_{\DY}(Q^2,N)}{d \ln Q^2} &=
\frac{2C_F}{\beta_0}
\int_0^{\infty}\,\frac{d\mu^2}{\mu^2}\,a_{{\cal S}_{\DY}}^{\Mink}(\mu^2)
\,\int_0^{\infty}\frac{dy}{y}
{\displaystyle{\frac{\frac{\mu^2}{yQ^2}}{\left(1+\frac{\mu^2}{yQ^2}\right)^2}}}
\left[\ddot{\cal G}_{{\cal S}_{\DY}}^{\Eucl}\left({yN^2}\right)-
\ddot{\cal G}_{{\cal S}_{\DY}}^{\Eucl}\left({y}\right)\right]\\
&=
\frac{2C_F}{\beta_0}
\int_0^{\infty}\frac{dy}{y}\,a_{{\cal S}_{\DY}}^{\Eucl}(yQ^2)
\,
\left[\ddot{\cal G}_{{\cal S}_{\DY}}^{\Eucl}\left({yN^2}\right)-
\ddot{\cal G}_{{\cal S}_{\DY}}^{\Eucl}\left({y}\right)\right]
\end{split}
\end{align}
where in the second line we used the relation between the Minkowskian and the Euclidean effective charges, Eq.~(\ref{Eucl-Mink}).

According to (\ref{WS_DY}), the explicit result for $\ddot{\cal G}_{{\cal S}_{\DY}}^{\Eucl}(y)$, defined in (\ref{G_Eucl}), is~\cite{Grunberg:2006jx,Grunberg:2006gd}
\begin{equation}
\label{WS_DY_Eucl}
\ddot{\cal G}_{{\cal S}_{\DY}}^{\Eucl}(y)=\frac12\,J_0(2\sqrt{y})\,,
\end{equation}
where $J_0$ is the Bessel function of the first kind.

Finally, note that when using the Euclidean representation (\ref{DY_kernel_Eucl}), power--like infrared sensitivity (i.e. infrared renormalons) is directly reflected in the expansion of the Euclidean characteristic function near zero. This should be contrasted with the Minkowskian representation where only non-analytic terms in the expansion are associated with renormalons (see Sec.~\ref{sec:PC} below). In the case of Drell--Yan production, only even powers of $N\Lambda/m$ appear~\cite{Beneke:1995pq}, as can be seen~\cite{Grunberg:2006gd} from the fact that $J_0$ is an even function of its argument. Equivalently in the Minkowskian representation the non-analytic terms in the expansion of $K_0$ appear with even powers (see Eq.~(\ref{exp_G_DY})) --- this has been pointed out in \cite{Laenen:2000ij,Laenen:2000hs}.

\vspace*{10pt}
\noindent
\underline{The soft functions in event--shape distributions}\\
Let us compute the moment--space characteristic
functions of the thrust and the $C$ parameter.

The case of the thrust is simple:
$\dot{\cal F}_{{\cal S}_t}(z)=\theta(z<1)$.  Using (\ref{G_def})
we obtain
\begin{align}
\dot{\cal G}_{{\cal S}_{t}}(\nu^2)=
\frac{1}{2}\int_0^{\infty}\frac{dz}{z}\,{\rm e}^{-\nu/\sqrt{z}}\,
\dot{\cal F}_{{\cal S}_t}(z)=\frac{1}{2}\int_0^1\,
\frac{dz}{z}\,{\rm e}^{-\nu/\sqrt{z}}={\rm Ei}(1,\nu).
\end{align}

The case of the $C$ parameter is slightly more complicated.
Let us write the momentum--space characteristic function as
\[\dot{\cal F}_{{\cal S}_c}(z)=\theta(z<4)+\theta(z>4)\left[1-\frac{1}
{\sqrt{1-4/z}}\right]
\]
and split the integral accordingly:
\begin{align}
\begin{split}
\dot{\cal G}_{{\cal S}_{c}}(\nu^2)&=
\frac{1}{2}\int_0^{\infty}\frac{dz}{z}\,{\rm e}^{-\nu/\sqrt{z}}\,
\dot{\cal F}_{{\cal S}_c}(z)=\frac{1}{2}\left\{
\int_0^4\,
\frac{dz}{z}\,{\rm e}^{-\nu/\sqrt{z}}
+\int_4^{\infty}\,
\frac{dz}{z}\,{\rm e}^{-\nu/\sqrt{z}}\left[1-\frac{1}
{\sqrt{1-4/z}}\right]\right\}\\
&={\rm Ei}(1,\nu/2)+
\int_0^1\frac{dx}{x}\,{\rm e}^{-\nu\,x/2}\left[1-\frac{1}
{\sqrt{1-x^2}}\right]\\
&=
{\displaystyle \frac{\pi\,\nu}{4}-\gamma_E-
\ln(\nu)\,-\,
\frac{\nu^2}{8}\,\, _3F_2
\left(
\left.\begin{array}{l}
1, 1\\
\frac{3}{2}, \frac{3}{2}, 2
\end{array}\right\vert
\frac{\nu^2}{16}\right)
\,+\,\frac{\pi\,\nu^3}{192}\,\, _3F_2
\left(
\left.\begin{array}{l}
1, \frac{3}{2}\\
2, 2, \frac{5}{2}
\end{array}\right\vert
\frac{\nu^2}{16}
\right)}\,.
\end{split}
\end{align}
Let us examine the convergence of the $\epsilon$ integral in
(\ref{Dispersive_Exponent_Laplace}). For
$\epsilon\to \infty$ we find exponential suppression for the thrust and power suppression
for the $C$ parameter:
\begin{align}
\begin{split}
\dot{\cal G}_{{\cal S}_{t}}(\nu^2)&\simeq {\rm e}^{-\nu}\,\left[\frac{1}{\nu}+
{\cal O}\left(\frac{1}{\nu^2}\right)\right]\,;
\\
\dot{\cal G}_{{\cal S}_{c}}(\nu^2)&\simeq -\frac{2}{\nu^2}+{\cal O}
\left(\frac{1}{\nu^4}\right)\,.
\end{split}
\end{align}
In either case each of the $\dot{\cal G}_{{\cal S}}$ terms in (\ref{Dispersive_Exponent_Laplace})
converge.

For $\epsilon\to 0$ we find
\begin{align}
\begin{split}
\dot{\cal G}_{{\cal S}_{t}}(\nu^2)&\simeq
-\ln(\nu)-\gamma_E+\nu-\frac{1}{4}\nu^2+\frac{1}{18}\nu^3+{\cal O}(\nu^4)\,;
\\
\dot{\cal G}_{{\cal S}_{c}}(\nu^2)&\simeq
-\ln(\nu)-\gamma_E+\frac{\pi}{4}\nu-\frac{1}{8}\nu^2+\frac{\pi}{192}\nu^3+{\cal O}(\nu^4)\,,
\end{split}
\end{align}
making the combinations $\dot{\cal G}_{{\cal S}}(N^2\epsilon)
- \dot{\cal G}_{{\cal S}}(\epsilon)+\ln N$ power suppressed at small
$\epsilon$, as required for convergence.

\subsection{The large--$N$ limit, strict factorization and finite terms~\label{sec:virt}}

As discussed in Sec.~\ref{sec:setup}, the Sudakov exponent in a given process
is unique as far as the terms that diverge at large--$N$ are concerned, but it is
subject to convention in what concerns finite terms for $N\to \infty$ as well as
any ${\cal O}(1/N)$ corrections. This has been illustrated by introducing three definitions for the Sudakoiv factor in Eqs.~(\ref{Exponent}), (\ref{Exponent_Laplace_with_Mellin_const}) and (\ref{Exponent_Laplace}). Our purpose here is to identify, in the framework of the dispersive approach, the unique large--$N$ limit of the Sudakov exponent, which in contrast with Eq.~(\ref{Dispersive_Exponent_Laplace_NLL}) above, involves two scales only: $m^2/N$ and $m^2/N^2$ (such a strict scale separation is a must in an effective field theory framework). We will also show that a convention--dependent part that violates this strict scaling is in fact necessary to render the Sudakov exponent finite.

\subsubsection*{The large--$N$ limit: disentangling $N$--dependent and constant terms}

Finite terms for $N\to \infty$ are generated by purely virtual corrections, which  are proportional to $\delta(r)$ in momentum space. Purely virtual corrections are partially accounted for within the exponent and partially in the hard function $H(m^2)$ in (\ref{fact_Mellin}) or (\ref{fact_Laplace_with_Mellin_const}).
This split depend on the convention adopted, except for the requirement that all infinities are assigned to the exponent where they cancel against the integral over the real--emission part.  Let us make this dependence parametric by generalizing Eq.~(\ref{Exponent_Laplace_with_Mellin_const}) though a modification of the cutoff value:
\begin{align}
\label{Exponent_Laplace_with_Mellin_const_r0}
\widetilde{\rm Sud}(m^2,N,r_0)  =\exp\left\{C_R\int_0^{\infty}\,\frac{dr}{r} \Big[{\rm e}^{-Nr}-{\theta(r<r_0)}\Big] R(m^2,r)\right\}\,,
\end{align}
with
\begin{equation}
\label{fact_Laplace_with_Mellin_const_r0}
\Gamma_N(m^2)=H(m^2,r_0)\,\times\,\widetilde{\rm Sud}(m^2,N,r_0)+\widetilde{\Delta \Gamma}_N(m^2).
\end{equation}
For $r_0=1$ the constant term ${\cal O}(N^0)$ in the exponent matches the original Mellin--transform definition (\ref{Exponent}).

It is straightforward to repeat the derivation of the dispersive representation with this definition. Proceeding as in (\ref{diff_Exponent_Laplace}) and (\ref{diff_Exponent_Laplace1})
we obtain:
\begin{align}
\label{diff_Exponent_Laplace_with_Mellin_const}
\begin{split}
\frac{d\ln \widetilde{\rm Sud}(m^2,N,r_0)}{d \ln m^2} & =
C_R\int_0^{\infty}\,\frac{dr}{r}
\Big[{\rm e}^{-Nr}-\theta(r<r_0)\Big] \,\frac{d R(m^2,r)}{d\ln m^2}\\
& \hspace*{-40pt}= \frac{C_R}{\beta_0}\int_0^{\infty}\,
\frac{d\epsilon}{\epsilon}\int_0^{\infty}\,\frac{dr}{r}
\Big[{\rm e}^{-Nr}-\theta(r<r_0)\Big] \,\times\\&
\bigg[\,\,
\rho_{\cal J}(\epsilon m^2)
\left(\dot{\cal F}_{\cal J}\left(\epsilon/r\right)
-\dot{\cal F}_{\cal J}(0)\right)
-
\rho_{\cal S}(\epsilon m^2) \left(
\dot{\cal F}_{\cal S}\left(\epsilon/r^2\right)
-\dot{\cal F}_{\cal S}\left(0\right)\right)\bigg]
\\
& \hspace*{-40pt}= \frac{C_R}{\beta_0} \int_0^{\infty}\,
\frac{d\epsilon}{\epsilon}
\bigg[
\rho_{\cal J}(\epsilon m^2){\cal H}_{\cal J}(N,\epsilon,r_0)
-\rho_{\cal S}(\epsilon m^2){\cal H}_{\cal S}(N,\epsilon,r_0)
\bigg]\,,
\end{split}
\end{align}
with
\begin{align}
\label{H_def}
\begin{split}
{\cal H}_{\cal J}(N,\epsilon,r_0)&\equiv
\int_0^{\infty}\,\frac{dr}{r}
\Big[{\rm e}^{-Nr}-\theta(r<r_0)\Big]\,\left(\dot{\cal F}_{\cal J}\left(\epsilon/r\right)
-\dot{\cal F}_{\cal J}(0)\right)\\
&=\dot{\cal G}_{\cal J}(N\epsilon)-{\cal F}_{\cal J}(\epsilon/r_0)+\ln(r_0) +\gamma_E+\ln(N)
\\
{\cal H}_{\cal S}(N,\epsilon,r_0)&\equiv
\int_0^{\infty}\,\frac{dr}{r}
\Big[{\rm e}^{-Nr}-\theta(r<r_0)\Big]\,\left(
\dot{\cal F}_{\cal S}\left(\epsilon/r^2\right)
-\dot{\cal F}_{\cal S}\left(0\right)\right)\\
&= \dot{\cal G}_{\cal S}(N^2\epsilon)-\frac12 {\cal F}_{\cal S}(\epsilon/r_0^2)+\ln(r_0)+\gamma_E+\ln(N)
\end{split}
\end{align}
where in the last step in (\ref{diff_Exponent_Laplace_with_Mellin_const}) we performed the integration over $r$ substituting $\dot{\cal G}_{\cal J}(N\epsilon)$ and
$\dot{\cal G}_{\cal J}(N^2\epsilon)$ according to
(\ref{G_def}) and used (\ref{cal_F_from_d_cal_F}) to write
\[
\int_0^{r_0}\frac{dr}{r}
\dot{\cal F}_{\cal J}(\epsilon/r)={\cal F}_{\cal J}(\epsilon/r_0)\,;\qquad \quad \int_0^{r_0}
\frac{dr}{r}
\dot{\cal F}_{\cal S}(\epsilon/r^2)=\frac12 {\cal F}_{\cal S}(\epsilon/r_0^2)\,.
\]
Note that ${\cal H}_{\cal J,\,S}$ in Eq.~(\ref{H_def}) are infrared finite owing to the subtraction of $\theta(r<r_0)$ in the square brackets; they are ultraviolet finite because of the differences
$\dot{\cal F}_{\cal J}\left(\epsilon/r\right)
-\dot{\cal F}_{\cal J}\left(0\right)$
 and
 $\dot{\cal F}_{\cal S}\left(\epsilon/r^2\right)
-\dot{\cal F}_{\cal S}\left(0\right)$, respectively.
Finally, performing the $\ln m^2$ integration in (\ref{diff_Exponent_Laplace_with_Mellin_const})
we obtain the result for the Sudakov factor:
\begin{align}
\label{widetilde_Exponent_Laplace_with_Mellin_const}
\begin{split}
{\widetilde{\rm Sud}(m^2,N,r_0)} & = \exp\Bigg\{\frac{C_R}{\beta_0}\int_0^{\infty}\,
\frac{d\epsilon}{\epsilon}
\bigg[
a^{\Mink}_{\cal J}(\epsilon m^2) {\cal H}_{\cal J}(N,\epsilon,r_0)
-a^{\Mink}_{\cal S}(\epsilon m^2) {\cal H}_{\cal S}(N,\epsilon,r_0)
\bigg]\Bigg\}\,.
\end{split}
\end{align}
Explicit results for ${\cal H}_{\cal J,\,S}(N,\epsilon,r_0)$
can be readily obtained by substituting $\dot{\cal G}_{\cal J,\,S}$ and ${\cal F}_{\cal J,\,S}$
of Tables~\ref{table:dcalG} and~\ref{table:calF}, respectively into Eq.~(\ref{H_def}).

Let us now check the convergence of the integral over $\epsilon$ in Eq.~(\ref{widetilde_Exponent_Laplace_with_Mellin_const}).
Consider first the $\epsilon\to 0$ limit.
Using the small--$\epsilon$ expansion of $\dot{\cal G}_{\cal J\,,S}$ in (\ref{small_epsilon_behaviour}) and that of
${\cal F}_{\cal J\,,S}$ in (\ref{calF_small_epsilon1}) we observe that the combinations
in Eq.~(\ref{H_def}) are power--suppressed at $\epsilon\to 0$.
Next consider the $\epsilon\to \infty$ limit. $\dot{\cal G}_{\cal J\,,S}$ and
${\cal F}_{\cal J\,,S}$ are (at least) power suppressed at large
$\epsilon$. The remaining terms $\ln(r_0)+\gamma_E+\ln(N)$ are common to the `jet' and `soft' part, and therefore the integrand is proportional to the difference of the two effective charges,
$a_{\cal J}^{\Mink}(\epsilon m^2)-a_{\cal S}^{\Mink}(\epsilon m^2)$, which behaves as $\alpha_s^3(\mu^2)$ at large $\mu^2$. Thus, the integral (\ref{widetilde_Exponent_Laplace_with_Mellin_const}) is well-defined.

Let us now isolate the $N$--dependent terms in the exponent. This can be elegantly done by considering the limit $r_0\to \infty$. Expanding ${\cal H}_{\cal J,\,S}(N,\epsilon,r_0)$ at large $r_0$ we obtain:
\begin{align}
\label{H_limit}
\begin{split}
{\cal H}_{\cal J}(N,\epsilon,r_0\gg 1)&=\Delta \dot{\cal G}_{\cal J}^{(r+v)}(N\epsilon)\,\,+\,\,{\cal O}(\epsilon/r_0)
\\
{\cal H}_{\cal S}(N,\epsilon,r_0\gg 1)&= \Delta \dot{\cal G}_{\cal S}^{(r+v)}(N^2\epsilon) \,\,+\,\,{\cal O}(\epsilon/r_0^2)
\end{split}
\end{align}
where we used the expansion of ${\cal F}_{\cal J\,,S}$ at small arguments, Eq.~(\ref{calF_small_epsilon1}), and defined
\begin{align}
\label{G_r+v}
\begin{split}
\Delta \dot{\cal G}_{\cal J}^{(r+v)}(N\epsilon)&\equiv \int_0^{\infty}\,\frac{dr}{r}
\Big[{\rm e}^{-Nr}-1\Big]\,\left(\dot{\cal F}_{\cal J}\left(\epsilon/r\right)
-\dot{\cal F}_{\cal J}(0)\right)\\
&=\dot{\cal G}_{\cal J}(N\epsilon)+\ln(N\epsilon)-J_1 +\gamma_E
\\
\Delta \dot{\cal G}_{\cal S}^{(r+v)}(N^2\epsilon)&\equiv\int_0^{\infty}\,\frac{dr}{r}
\Big[{\rm e}^{-Nr}-1\Big]\,\left(
\dot{\cal F}_{\cal S}\left(\epsilon/r^2\right)-\dot{\cal F}_{\cal S}(0)\right)\\
&=\dot{\cal G}_{\cal S}(N^2\epsilon)+\frac12\ln(N^2\epsilon)-\frac12 S_1 +\gamma_E\,,
\end{split}
\end{align}
where the superscript $(r+v)$ indicates that in contrast with $\dot{\cal G}_{\cal J,\,S}$ of (\ref{G_def}), which consists solely of real emission contributions, through the subtraction of $1$ in the square brackets (\ref{G_r+v})
includes also virtual corrections. This guarantees convergence for $r\to 0$.
Note that $\Delta \dot{\cal G}_{\cal J}^{(r+v)}(N\epsilon)$ and $\Delta \dot{\cal G}_{\cal S}^{(r+v)}(N^2\epsilon)$ converge for $r\to \infty$ owing to the subtraction of $\dot{\cal F}_{\cal J}(0)$ and $\dot{\cal F}_{\cal S}(0)$, respectively. Thus, similarly to ${\cal H}_{\cal J,\,S}(N,\epsilon,r_0)$ these objects are well--defined. Based on these definitions and (\ref{G_def}) it is straightforward to recover the expressions in terms of
$\dot{\cal G}_{\cal J}(N\epsilon)$ and $\dot{\cal G}_{\cal S}(N^2\epsilon)$, respectively.

In contrast to ${\cal H}_{\cal J,\,S}(N,\epsilon,r_0)$ the functions $\Delta \dot{\cal G}_{\cal J}^{(r+v)}(N\epsilon)$
and $\Delta \dot{\cal G}_{\cal S}^{(r+v)}(N^2\epsilon)$ depend on a single argument.
Using (\ref{H_limit}) we therefore find that the $N$--dependence of the jet function in (\ref{widetilde_Exponent_Laplace_with_Mellin_const}) involves \emph{only} the jet mass scale $m^2/N$ while that of the soft function \emph{only} the soft scale $m^2/N^2$. This is the infrared factorization property we have already encountered in Sec.~\ref{sec:kernel} when working in momentum space.
We also observe, however, that despite the fact that the integral in (\ref{widetilde_Exponent_Laplace_with_Mellin_const}) is well defined for any finite $r_0$, the $r_0\to \infty$ limit does not exist:
by taking $r_0\to \infty$ inside the $\epsilon$ integral a divergence is generated for $\epsilon\to\infty$. Thus, the ${\cal O}(\epsilon/r_0)$ and ${\cal O}(\epsilon/r_0^2)$ contributions that distinguish ${\cal H}_{\cal J,\, S}(N,\epsilon,r_0)$ from their simple $r_0\to \infty$ limits (thus violating strict factorization) are in fact necessary for convergence.

Having separated the $N$--dependent terms from the cutoff--dependent constant terms,
Eq.~(\ref{widetilde_Exponent_Laplace_with_Mellin_const}) can be elegantly written as follows:
\begin{align}
\label{widetilde_Exponent_Laplace_with_Mellin_const1}
\begin{split}
 \widetilde{\rm Sud}(m^2,N,r_0) &=
\exp\,\Bigg\{\,
\frac{C_R}{\beta_0} \int_0^{\infty}\frac{d\epsilon}{\epsilon} \,\,
\Bigg[a^{\Mink}_{\cal J}(\epsilon m^2)
\Bigg(\Delta \dot{\cal G}_{\cal J}^{(r+v)}(N\epsilon)-\Delta{\cal F}_{\cal J}(\epsilon/r_0)\Bigg)
\\&\hspace*{45pt}-
\,\,
a^{\Mink}_{\cal S}(\epsilon m^2)
\Bigg(\Delta \dot{\cal G}_{\cal S}^{(r+v)}(N^2\epsilon)
-\frac{1}{2}\Delta{\cal F}_{\cal S}(\epsilon/r_0^2)
\Bigg)\Bigg]
\Bigg\}\,,
\end{split}
\end{align}
where for the $r_0$ dependent terms we used the following definitions
\begin{align}
\label{int_F_J}
\begin{split}
\Delta{\cal F}_{\cal J}(\epsilon/r_0)\equiv -\int_0^{\epsilon/r_0}\,\frac{dy}{y}\,\left[\dot{\cal F}_{\cal J}\left(y\right)-\dot{\cal F}_{\cal J}\left(0\right)\right]\
&={\cal F}_{\cal J}\left(\epsilon/r_0\right)+\ln\left(\epsilon/r_0\right)-J_1\\
\Delta{\cal F}_{\cal S}(\epsilon/r_0^2)\equiv-\int_0^{\epsilon/r_0^2}\,\frac{dy}{y}\,\left[\dot{\cal F}_{\cal S}\left(y\right)-\dot{\cal F}_{\cal S}\left(0\right)\right]\
&={\cal F}_{\cal S}\left(\epsilon/r_0^2\right)+\ln\left(\epsilon/r_0^2\right)-S_1\,.
\end{split}
\end{align}
Note that $\Delta \dot{\cal G}_{\cal J}^{(r+v)}(N\epsilon)$ and $\Delta \dot{\cal G}_{\cal S}^{(r+v)}(N^2\epsilon)$ in (\ref{G_r+v}) are essentially $\dot{\cal G}_{\cal J}(N\epsilon)$ and
$\dot{\cal G}_{\cal S}(N^2\epsilon)$, respectively, minus the leading terms in their expansions,
Eq.~(\ref{small_epsilon_behaviour}). Similarly $\Delta{\cal F}_{\cal J}(\epsilon/r_0)$  and
$\Delta{\cal F}_{\cal S}(\epsilon/r_0^2)$ in (\ref{int_F_J}) are
${\cal F}_{\cal J}\left(\epsilon/r_0\right)$ and ${\cal F}_{\cal S}(\epsilon/r_0^2)$ minus the leading terms in their expansions, Eq.~(\ref{calF_small_epsilon1}). Consequently, the convergence of (\ref{widetilde_Exponent_Laplace_with_Mellin_const1}) for $\epsilon\to 0$ is guaranteed for each of the separate terms, which all fall as $\epsilon^1$ in this limit. In contrast, the convergence for
$\epsilon\to \infty$ involves cancellation between \emph{all} the terms.

\subsubsection*{Re-derivation of the exponent using strictly factorized components}

We saw that upon removing the regulator $r_0$ one recovers strictly factorized functions that fully capture the $N$--dependent terms of the Sudakov exponent at large $N$. We can now repeat the derivation of the dispersive representation of the exponent with no constant terms, Eq.~(\ref{Dispersive_Exponent_Laplace}), using strictly factorized components.
To this end let us define the formal `jet' and `soft' Sudakov exponents:
\begin{equation}
\label{limit_Sud_exp_J}
E_{\cal J}(m^2/N)=C_R\, \int_0^{\infty}\,\frac{dr}{r} \left[{\rm e}^{-Nr}-1\right]\,{\cal J}(r m^2)\end{equation}
and
\begin{equation}
\label{limit_Sud_exp_S}
E_{\cal S}(m^2/N^2)=C_R\, \int_0^{\infty}\,\frac{dr}{r} \left[{\rm e}^{-Nr}-1\right]\,{\cal S}(r^2 m^2)\,,
\end{equation}
which include both real and virtual corrections. Owing to the virtual term $(-1)$ these expressions are infrared finite, but since the integration extends to infinity \emph{they are ultraviolet divergent}.
In Appendix \ref{sec:taking_large_N} we show that these formal definitions correspond to taking the
large $N$--limit of the finite--$N$ physical evolution kernel, where the limit is taken with fixed $N\mu^2/m^2$ for the `jet' and fixed $N^2\mu^2/m^2$ for the soft function.  The characteristic functions of Eq.~(\ref{G_r+v}) are then identified with the corresponding large--$N$ limits of the full finite--$N$ characteristic functions computed in Ref.~\cite{DMW}.

The Sudakov evolution equation, Eq.~(\ref{diff_Exponent_Laplace}), can be expressed in terms of \emph{finite combinations} of these quantities, where the ultraviolet divergences cancel out. We have
\begin{equation}
 \label{blocks}
 \frac{d\ln \overline{\rm Sud}(m^2,N)}{d \ln m^2}=\left[E_{\cal J}(m^2/N)-E_{\cal J}(m^2)\right]-\left[E_{\cal S}(m^2/N^2)-E_{\cal S}(m^2)\right]\,.
\end{equation}
Substituting into (\ref{limit_Sud_exp_J}) and (\ref{limit_Sud_exp_S})
the dispersive representation of the Sudakov anomalous dimensions
according to (\ref{J_Sud_anom_dim_conv}) and
(\ref{S_Sud_anom_dim_conv}), respectively, we obtain the moment-space representations:
\begin{align}
\label{limit_Sud_exp_JS_disp}
\begin{split}
E_{\cal J}(m^2/N)
&=\frac{C_R}{\beta_0}\,
\int_0^{\infty}\frac{d\epsilon}{\epsilon} \,\,
\rho_{\cal J}(\epsilon m^2)\Delta\dot{\cal G}_{\cal J}^{(r+v)}\left(\epsilon N\right)\\
E_{\cal S}(m^2/N^2)
&=\frac{C_R}{\beta_0}\,
\int_0^{\infty}\frac{d\epsilon}{\epsilon} \,\,
\rho_{\cal S}(\epsilon m^2)
\Delta\dot{\cal G}_{\cal S}^{(r+v)}\left(\epsilon N^2\right)\,,
\end{split}
\end{align}
where, as usual, $\epsilon=\mu^2/m^2$, and $\Delta\dot{\cal G}_{\cal J}^{(r+v)}\left(\epsilon N\right)$ and
$\Delta\dot{\cal G}_{\cal S}^{(r+v)}\left(\epsilon N^2\right)$ have been defined in
(\ref{G_r+v}). Note that here the integrands are well--defined and integrable near $\epsilon\to 0$ but, as expected based on the definitions (\ref{limit_Sud_exp_J}), the integrals diverge for $\epsilon\to \infty$.

Integrating (\ref{limit_Sud_exp_JS_disp}) by parts we obtain an alternative form in terms of the Sudakov effective charges:
\begin{align}
\label{limit_Sud_exp_JS_disp1}
\begin{split}
E_{\cal J}(m^2/N)
&=\frac{C_R}{\beta_0}\,
\int_0^{\infty}\frac{d\epsilon}{\epsilon} \,\,
a_{\cal J}^{\Mink}(\epsilon m^2)\Delta\ddot{\cal G}_{\cal J}^{(r+v)}\left(\epsilon N\right)\\
E_{\cal S}(m^2/N^2)
&=\frac{C_R}{\beta_0}\,
\int_0^{\infty}\frac{d\epsilon}{\epsilon} \,\,
a_{\cal S}^{\Mink}(\epsilon m^2)
\Delta\ddot{\cal G}_{\cal S}^{(r+v)}\left(\epsilon N^2\right)\,.
\end{split}
\end{align}

Next, substituting in Eq.~(\ref{blocks}) the dispersive representations~(\ref{limit_Sud_exp_JS_disp})
of the jet and soft Sudakov exponents, we get the dispersive representation of the Sudakov
evolution equation:
\begin{align}
\label{diff_Exponent_Laplace3}
\begin{split}
\frac{d\ln \overline{\rm Sud}(m^2,N)}{d \ln m^2} &=
\frac{C_R}{\beta_0}\,\Bigg\{\,
\int_0^{\infty}\frac{d\epsilon}{\epsilon} \,\,
\rho_{\cal J}(\epsilon m^2)
\left[\Delta\dot{\cal G}_{\cal J}^{(r+v)}\left(\epsilon N\right)-
\Delta\dot{\cal G}_{\cal J}^{(r+v)}\left(\epsilon\right)\right]
\\&\hspace*{45pt}-
\int_0^{\infty}\frac{d\epsilon}{\epsilon} \,\,
\rho_{\cal S}(\epsilon m^2)
\left[\Delta\dot{\cal G}_{\cal S}^{(r+v)}\left(\epsilon N^2\right)-
\Delta\dot{\cal G}_{\cal S}^{(r+v)}\left(\epsilon\right)\right]
\Bigg\}\,.
\end{split}
\end{align}
Using the expressions for $\Delta\dot{\cal G}_{\cal J,\,S}^{(r+v)}$ in Eq.~(\ref{G_r+v}) this yields Eq.~(\ref{diff_Exponent_Laplace2}). Finally,
integration over $\ln m^2$ gives:
\begin{align}
\label{Dispersive_Exponent_Laplace1}
\begin{split}
\overline{\rm Sud}(m^2,N) &=
\exp\Bigg\{\frac{C_R}{\beta_0}\,\,\,
\int_0^{\infty}\frac{d\epsilon}{\epsilon} \,\,
\Bigg[a_{\cal J}^{\Mink}(\epsilon m^2)
\Bigg(\Delta \dot{\cal G}_{\cal J}^{(r+v)}\left(\epsilon N\right)-
\Delta\dot{\cal G}_{\cal J}^{(r+v)}\left(\epsilon\right)\Bigg)\\&\hspace*{75pt}-
\,\,
a_{\cal S}^{\Mink}(\epsilon m^2)
\Bigg(\Delta\dot{\cal G}_{\cal S}^{(r+v)}\left(\epsilon N^2\right)-\Delta
\dot{\cal G}_{\cal S}^{(r+v)}\left(\epsilon\right)\Bigg)\Bigg]
\Bigg\}\,,
\end{split}
\end{align}
which yields Eq.~(\ref{Dispersive_Exponent_Laplace}). Thus, we have recovered the final result of Sec.~(\ref{sec:dispersive_exponent_derivation}) in terms of the characteristic functions
$\Delta\dot{\cal G}_{\cal J,\,S}^{(r+v)}$ that include virtual corrections. This derivation elucidates
the origin of the $\ln (N)$ terms in Eq.~(\ref{Dispersive_Exponent_Laplace}). Eq.~(\ref{Dispersive_Exponent_Laplace1}) also exhibits explicitly the fact that the integral converges for $\epsilon\to 0$: here each term is separately power suppressed in this limit. Convergence in the $\epsilon\to \infty$ limit relies on cancellation between the terms.

\section{Relation with the Borel representation of the Sudakov
exponent~\label{sec:Borel}}

The Borel formulation of Sudakov resummation has been the subject of much theoretical work over the past few
years~\cite{Gardi:2001ny,Gardi:2002bg,Gardi:2002xm,Gardi:2005yi,Andersen:2005bj,Gardi:2006jc}. It led to
 a successful phenomenology  in a range of applications. The purpose of the present section is to study in depth the relation between this formulation and the dispersive one, which was presented in the previous sections. We will show that the two provide different ways of summing up \emph{the same set of radiative corrections}. In the next section we shall address potential differences at the power level.

\subsection{Borel representation of the anomalous
dimensions~\label{sec:Borel_anom_dim}}

Let us begin by introducing the scheme--invariant Borel representations of the Euclidean  effective charges~(\ref{Eucl-expansions}):
\begin{align}
\begin{split}
\label{eq:int_discontinuity_SJ_space_like}
a_{\cal S}^{\Eucl}(\mu^2)&=
\, \int_0^{\infty} du\, T(u)\,\left(\frac{\Lambda^2}{\mu^2}\right)^{u}
 \,
B\big[a_{\cal S}^{\Eucl}\big](u)\\
a_{\cal J}^{\Eucl}(\mu^2)&=
\, \int_0^{\infty} du\, T(u)\, \,
\left(\frac{\Lambda^2}{\mu^2}\right)^{u} \,
B\left[a_{\cal J}^{\Eucl}\right](u)\,,
\end{split}
\end{align}
where $\Lambda^2$ is defined in $\overline{\rm MS}$.
We emphasize that the Borel integrals do not necessarily converge for
arbitrary $\mu^2$: all that is required is that they exist for sufficiently large $\mu^2$.
As usual, while the integrand in (\ref{eq:int_discontinuity_SJ_space_like}) is
scheme invariant~\cite{Grunberg:1992hf}, the separation between $T(u)$ and the Borel function
$B\big[a_{\cal J,\,S}^{\Eucl}\big](u)$ itself is scheme
dependent. In practice (see e.g.~Eq.~(2.8) in \cite{Andersen:2006hr})
it is convenient to use the 't Hooft scheme where $T(u)$ absorbs
the dependence on the two--loop coupling $\delta=\beta_1/{\beta_0}^2$,
\begin{align}
T(u)=\frac{|u\delta|^{u\delta}{\rm e}^{-u\delta}}{\Gamma(1+u\delta)}.
\end{align}
Note that in the large--$\beta_0$ limit
$\left.T(u)\right\vert_{\text{large}\, \beta_0}=1$
and
\begin{equation}
\label{a_JS_large_beta0}
\left.
B\left[a_{\cal J}^{\Eucl}\right](u)\right\vert_{\text{large}\, \beta_0}
=
\left.
B\big[a_{\cal S}^{\Eucl}\big](u)\right\vert_{\text{large}\, \beta_0}
={\rm e}^{\frac53\,u}.
\end{equation}

According to (\ref{Eucl-Mink}) the Borel representations of the timelike couplings corresponding to
(\ref{eq:int_discontinuity_SJ_space_like}) are obtained by analytic
continuation:
\begin{align}
\begin{split}
\label{eq:int_discontinuity_SJ}
a_{\cal S}^{\Mink}(\mu^2)&=
\, \int_0^{\infty} du\, T(u)\,
\left(\frac{\Lambda^2}{\mu^2}\right)^{u} \, \frac{\sin \pi u}{\pi u} \,
B\big[a_{\cal S}^{\Eucl}\big](u)\\
a_{\cal J}^{\Mink}(\mu^2)&=
\, \int_0^{\infty} du\, T(u)
\,\left(\frac{\Lambda^2}{\mu^2}\right)^{u} \, \frac{\sin \pi u}{\pi u} \,
B\left[a_{\cal J}^{\Eucl}\right](u).
\end{split}
\end{align}
Substituting these expressions into (\ref{J_Sud_anom_dim_conv}) and
(\ref{S_Sud_anom_dim_conv}), respectively, we obtain the
Borel representation of the physical Sudakov anomalous
dimensions of (\ref{Sud_anom_dim})~\cite{Gardi:2002xm,Gardi:2006jc}:
\begin{align}
\begin{split}
\label{Sud_anom_dim_Borel}
{\cal J}(\mu^2)&=\frac{1}{\beta_0}\int_0^{\infty}du \, T(u)
\left(\frac{\Lambda^2}{\mu^2}\right)^u B_{\cal J}(u)\\
{\cal S}(\mu^2)&=\frac{1}{\beta_0}\int_0^{\infty}du \, T(u)
\left(\frac{\Lambda^2}{\mu^2}\right)^u B_{\cal S}(u).
\end{split}
\end{align}
where
\begin{align}
\label{B_JS_mathbb}
\begin{split}
B_{\cal J}(u)&=\frac{\sin \pi u}{\pi u} \,
B\left[a_{\cal J}^{\Eucl}\right](u)\,\,\times\,\,{\mathbb B}_{\cal J}(u)\\
B_{\cal S}(u)&=\frac{\sin \pi u}{\pi u} \,
B\big[a_{\cal S}^{\Eucl}\big](u)\,\,\times\,\,{\mathbb B}_{\cal S}(u)\,,
\end{split}
\end{align}
with
\begin{align}
\label{B_relation_dotSJ}
\begin{split}
\,{\mathbb B}_{\cal J}(u)
&=\,-\,u\,\int_0^{\infty}\frac{dy}{y} \, y^{-u}\,\dot{\cal F}_{\cal J}(y)
\,= \,\int_0^{\infty}\frac{dy}{y} \, y^{-u}\,\ddot{\cal F}_{\cal J}(y)
\,= \,\sum_{k=0}^{\infty}\frac{(-1)^k\,u^k}{k!}\,J_k\,,\\
{\mathbb B}_{\cal S}(u)
&=\,-\,u\,\int_0^{\infty}\frac{dz}{z} \, z^{-u}\,\dot{\cal F}_{\cal S}(z)
\,= \,\int_0^{\infty}\frac{dz}{z} \, z^{-u}\,\ddot{\cal F}_{\cal S}(z)
\,= \,\sum_{k=0}^{\infty}\frac{(-1)^k\, u^k}{k!}\,S_k,
\end{split}
\end{align}
where $\dot{\cal F}_{\cal J,\,S}(z)$ are the characteristic functions defined
in the large--$\beta_0$ limit~(\ref{R_int_by_parts}) and $J_k$ and
$S_k$ are the corresponding log-moments defined in~(\ref{JkSk_def}). Thus, the expansion of ${\mathbb B}_{\cal J,\,S}(u)$ also provides a convenient way to extract the
log-moments. Obviously, the same information about higher orders and infrared sensitivity
(power corrections) contained in the characteristic functions $\dot{\cal F}_{\cal J,\,S}$
is encapsulated by the corresponding Borel functions
${\mathbb B}_{\cal J,\,S}(u)$. Table~\ref{table:calF_B} summarizes the
explicit results for ${\mathbb B}_{\cal J,\,S}(u)$ in
various inclusive distributions.

Note that 
\begin{equation}
\label{B_JS_mathbb_large_beta0}
\left.
B_{\cal J,\cal S}(u)\right\vert_{\text{large}\, \beta_0}
=\frac{\sin \pi u}{\pi u} \,\,{\rm e}^{\frac53\,u}\,\,
\times\,\,{\mathbb B}_{\cal J,\cal S}(u).
\end{equation} 
Although ${\mathbb B}_{\cal J,\cal S}(u)$ have
singularities at certain integer values of the Borel variable, these \emph{do not generate renormalons} at large $\beta_0$  in the
anomalous dimensions ${\cal J}(k^2)$ and ${\cal S}(k^2)$ because
of the $\sin (\pi u)$ factor associated with the timelike nature of these quantities. 

We further expect that $B\big[a_{\cal J,\,S}^{\Eucl}\big](u)$ and thus, in virtue of (\ref{B_JS_mathbb}), also $B_{\cal J,\,S}(u)$ are renormalon free \emph{at finite $\beta_0$}. Absence of renormalons in anomalous dimensions is expected on general grounds\footnote{Note however we deal here with {\em physical} anomalous dimensions, for which this expectation might not be realized.}, but a formal proof in the non-Abelian case is still absent and would be very important.

When using the Borel formulation, the \emph{deviation} of
$B_{\cal J,\,S}(u)$  from their large--$\beta_0$ limits is computed
order-by-order in perturbation theory. It amounts to a \emph{multiplicative}
modification of the large--$\beta_0$ result~\cite{Gardi:2002xm} by a
factor $V_{\cal J,\,S}(u)$, see e.g. Eq.~(19) and (26)
in~\cite{Gardi:2006jc}, which we can now readily identify as
\begin{equation}
V_{\cal J,\,S}(u)=\frac{B\big[a_{\cal J,\,S}^{\Eucl}\big](u)}{\hspace*{20pt}
\left.
B\big[a_{\cal J,\,S}^{\Eucl}\big](u)\right\vert_{\text{large}\, \beta_0}}=
B\big[a_{\cal J,\,S}^{\Eucl}\big](u)\,\,{\rm e}^{-\frac53 u}\,.
\end{equation}
This gives a new perspective on the physical meaning of this function.

\subsection{The Sudakov exponent~\label{sec:Borel_Sud}}

To demonstrate the correspondence between the dispersive
representation of the Sudakov exponent and the Borel one, let us
now re-derive the Borel formula for the exponent
({\it cf.}~Eq.~(27) in \cite{Gardi:2006jc}) following the steps
of Sec.~\ref{sec:dispersive_exponent} above. We will then present
a second derivation starting from the dispersive formula
(\ref{Dispersive_Exponent_Laplace}).

\subsubsection*{Direct derivation}

Inserting the Borel representation of the Sudakov anomalous dimensions
in (\ref{Sud_anom_dim_Borel}) into (\ref{diff_Exponent_Laplace})
we obtain:
\begin{align}
\label{diff_Exponent_Laplace_Borel}
\begin{split}
\frac{d\ln \overline{\rm Sud}(m^2,N)}{d \ln m^2}
& = C_R\int_0^{\infty}\,\frac{dr}{r}
\Big[{\rm e}^{-Nr}-{\rm e}^{-r}\Big] \,{\cal J}(r m^2)
-C_R\int_0^{\infty}\,\frac{dr}{r}
\Big[{\rm e}^{-Nr}-{\rm e}^{-r}\Big] \,{\cal S}(r^2 m^2)\\
& = \frac{C_R}{\beta_0}\Bigg\{\int_0^{\infty}\,\frac{dr}{r}
\Big[{\rm e}^{-Nr}-{\rm e}^{-r}\Big] \,
\int_0^{\infty}du \, T(u)
\left(\frac{\Lambda^2}{r m^2}\right)^u B_{\cal J}(u)\\
&\hspace*{30pt}-\int_0^{\infty}\,\frac{dr}{r}
\Big[{\rm e}^{-Nr}-{\rm e}^{-r}\Big]
\int_0^{\infty}du \, T(u)\left(\frac{\Lambda^2}{r^2 m^2}\right)^u B_{\cal S}(u)
\Bigg\}\,,
\end{split}
\end{align}
where the two terms corresponding to the `jet' and `soft'
contributions to the r.h.s are separately finite. Here one can change
the order of integration using $u$ as an infrared regulator. Using
\[
\int_0^{\infty}\frac{dr}{r}\,{\rm e}^{-Nr}\, r^{-u}=N^u\,\Gamma(-u)\,,
\]
we obtain an elegant Borel representation of the evolution equation:
\begin{align}
\label{diff_Exponent_Laplace_Borel1}
\begin{split}
\frac{d\ln \overline{\rm Sud}(m^2,N)}{d \ln m^2}
& = \frac{C_R}{\beta_0}\Bigg\{
\int_0^{\infty}du \, T(u)
\left(\frac{\Lambda^2}{m^2}\right)^u B_{\cal J}(u)\,\Gamma(-u)(N^u-1)\\
&\hspace*{30pt}-
\int_0^{\infty}du \, T(u)\left(\frac{\Lambda^2}{m^2}\right)^u
B_{\cal S}(u)\,\Gamma(-2u)(N^{2u}-1)
\Bigg\}\,,
\end{split}
\end{align}
where, as before, the two Borel integrals are separately finite,
for $u\to 0$. We note that the terms proportional to $N^u$ and $N^{2u}$ in Eq.~(\ref{diff_Exponent_Laplace_Borel1}), respectively, are the Borel representations of the formal `jet' and `soft' Sudakov exponents (\ref{limit_Sud_exp_J}) and (\ref{limit_Sud_exp_S}), where $u$ serves as a regulator.

It is straightforward to see the relation with
Eq.~(\ref{diff_Exponent_Laplace2})\footnote{Note that upon using
(\ref{eq:int_discontinuity_SJ}), Eqs.~(\ref{comp_disp_Borel_J}) and
(\ref{comp_disp_Borel_S})  can be derived from (\ref{G_int_Borel}) below.}:
\begin{align}
\label{comp_disp_Borel_J}
\begin{split}
&\int_0^{\infty}\frac{d\epsilon}{\epsilon} \,\,
\rho_{\cal J}(\epsilon m^2)
\left(\dot{\cal G}_{\cal J}\left(\epsilon N\right)-
\dot{\cal G}_{\cal J}\left(\epsilon\right)+\ln N\right)
= \int_0^{\infty}\frac{d\epsilon}{\epsilon} \,\,
a_{\cal J}^{\Mink}(\epsilon m^2)
\left(\ddot{\cal G}_{\cal J}\left(\epsilon N\right)-
\ddot{\cal G}_{\cal J}\left(\epsilon\right)\right)\\
&\hspace*{100pt}=
\int_0^{\infty}du \, T(u)
\left(\frac{\Lambda^2}{m^2}\right)^u B_{\cal J}(u)\,\Gamma(-u)(N^u-1)
\end{split}
\end{align}
and
\begin{align}
\label{comp_disp_Borel_S}
\begin{split}
&\int_0^{\infty}\frac{d\epsilon}{\epsilon} \,\,
\,\rho_{\cal S}(\epsilon m^2)
\left(\dot{\cal G}_{\cal S}\left(\epsilon N^2\right)-
\dot{\cal G}_{\cal S}\left(\epsilon\right)+\ln N\right)
=\int_0^{\infty}\frac{d\epsilon}{\epsilon} \,\,
\,a_{\cal S}^{\Mink}(\epsilon m^2)
\left(\ddot{\cal G}_{\cal S}\left(\epsilon N^2\right)-
\ddot{\cal G}_{\cal S}\left(\epsilon\right)\right)\\
&\hspace*{100pt}=
\int_0^{\infty}du \, T(u)\left(\frac{\Lambda^2}{m^2}\right)^u
B_{\cal S}(u)\,\Gamma(-2u)(N^{2u}-1)
\,.
\end{split}
\end{align}
Note that in (\ref{diff_Exponent_Laplace_Borel1}), (\ref{comp_disp_Borel_J}) and
(\ref{comp_disp_Borel_S}) one must assume some prescription for
the Borel singularities at positive integer and half integer values of $u$.

The final step in the derivation of the Sudakov factor
is to integrate (\ref{diff_Exponent_Laplace_Borel1})
over $\ln m^2$. As in (\ref{Dispersive_Exponent_Laplace}) this
requires to combine the two integrals over $u$ in
(\ref{diff_Exponent_Laplace_Borel1}). The result ({\it cf.}~Eq.~(27) in
\cite{Gardi:2006jc}) is:
\begin{align}
\label{Exponent_Laplace_Borel2}
\begin{split}
\overline{\rm Sud}(m^2,N)
& = \exp\Bigg\{-
\frac{C_R}{\beta_0}
\int_0^{\infty}\frac{du}{u} \, T(u)\left(\frac{\Lambda^2}{m^2}\right)^u
\bigg[B_{\cal J}(u)\,\Gamma(-u)(N^u-1)\\
&\hspace*{180pt}-
B_{\cal S}(u)\,\Gamma(-2u)(N^{2u}-1)\bigg]
\Bigg\}\,.
\end{split}
\end{align}
This is the Borel equivalent of (\ref{Dispersive_Exponent_Laplace}).

\subsubsection*{Derivation based on the dispersive formula}

An alternative route is to insert the Borel representation of the
Minkowskian couplings (\ref{eq:int_discontinuity_SJ}) directly into
Eq.~(\ref{Dispersive_Exponent_Laplace}). This yields:
\begin{align}
\label{Disp_Borel_Exponent_Laplace}
\begin{split}
\overline{\rm Sud}(m^2,N) &=
\exp\Bigg\{\frac{C_R}{\beta_0}\,\,\int_0^{\infty}
\frac{d\epsilon}{\epsilon} \\
&
\bigg[\int_0^{\infty} du\, T(u)
\,\left(\frac{\Lambda^2}{\epsilon m^2}\right)^{u} \, \frac{\sin \pi u}{\pi u} \,
B\left[a_{\cal J}^{\Eucl}\right](u)
\left(\dot{\cal G}_{\cal J}\left(\epsilon N\right)-
\dot{\cal G}_{\cal J}\left(\epsilon\right)+\ln N\right)
\\&\hspace*{5pt}-\,\int_0^{\infty} du\, T(u)\,
\left(\frac{\Lambda^2}{\epsilon m^2}\right)^{u} \, \frac{\sin \pi u}{\pi u} \,
B\big[a_{\cal S}^{\Eucl}\big](u)
\left(\dot{\cal G}_{\cal S}\left(\epsilon N^2\right)-
\dot{\cal G}_{\cal S}\left(\epsilon\right)+\ln N\right)
\bigg]\Bigg\}\,.
\end{split}
\end{align}
Upon combining the two Borel integrals and
changing the order of integration the $\ln N$ terms cancel and
the $\epsilon$
integrals over the $\dot{\cal G}_{\cal J,\,S}$ terms take the form:
\begin{align}
\label{G_int_Borel}
\begin{split}
\int_0^{\infty}\frac{d\epsilon}{\epsilon} \epsilon^{-u}
\Big(\dot{\cal G}_{\cal J}\left(\epsilon N\right)-
\dot{\cal G}_{\cal J}\left(\epsilon \right)\Big)
&=-\frac{1}{u}(N^{u}-1)\Gamma(-u)\,{\mathbb{B}}_{\cal J}(u)\\
\int_0^{\infty}\frac{d\epsilon}{\epsilon} \epsilon^{-u}
\Big(\dot{\cal G}_{\cal S}\left(\epsilon N^2\right)-
\dot{\cal G}_{\cal S}\left(\epsilon \right)\Big)
&=-\frac{1}{u}(N^{2u}-1)\Gamma(-2u)\,{\mathbb{B}}_{\cal S}(u)
\,
\end{split}
\end{align}
where we have computed the integrals using the relations (\ref{G_def})
and (\ref{B_relation_dotSJ}), which together yield
a direct relation between the moment--space
characteristic functions $\dot{\cal G}_{\cal J,\,S}\left(\epsilon\right)$
and the Borel functions
${\mathbb{B}}_{\cal J,\,S}(u)$ (all functions are defined
in the large--$\beta_0$ limit):
\begin{align}
\label{G_laplace_B_J}
\begin{split}
\left.\begin{array}{l}{\displaystyle
\dot{\cal G}_{\cal J}(\epsilon)=\int_0^{\infty}\,\frac{dy}{y} \,{\rm e}^{-\epsilon/y}\,
\dot{\cal F}_{\cal J}(y)}\\
\\
{\displaystyle \mathbb{B}_{\cal J}(u)=-u\int_0^{\infty}\frac{dy}{y}\,y^{-u}\,
\dot{\cal F}_{\cal J}(y)}
\end{array}\right\}\Longrightarrow \,\,\,
\int_0^{\infty}\frac{d\epsilon}{\epsilon} \,\,\epsilon^{-u}
\dot{\cal G}_{\cal J}\left(\epsilon\right)
&=-\frac{1}{u}\Gamma(-u)\,{\mathbb{B}}_{\cal J}(u)
\end{split}
\end{align}
and
\begin{align}
\label{G_laplace_B_S}
\begin{split}
\left.\begin{array}{l}{\displaystyle
\dot{\cal G}_{\cal S}(\epsilon)=\frac12
\int_0^{\infty}\,\frac{dz}{z}\, {\rm e}^{-\sqrt{\epsilon}/\sqrt{z}}\,
\dot{\cal F}_{\cal S}(z)}\\
\\
{\displaystyle \mathbb{B}_{\cal S}(u)=-u\int_0^{\infty}
\frac{dz}{z}\,z^{-u}\,
\dot{\cal F}_{\cal S}(z)}
\end{array}\right\}\Longrightarrow \,\,\,
\int_0^{\infty}\frac{d\epsilon}{\epsilon} \,\,\epsilon^{-u}
\dot{\cal G}_{\cal S}\left(\epsilon\right)
&=-\frac{1}{u}\Gamma(-2u)\,{\mathbb{B}}_{\cal S}(u)
\end{split}
\end{align}
It is straightforward to check this correspondence
in the explicit examples considered above.
Using (\ref{G_int_Borel}) in (\ref{Disp_Borel_Exponent_Laplace})
we immediately recover the known form of the exponent in
(\ref{Exponent_Laplace_Borel2}).

Finally, it is useful to write the inverse relations to (\ref{G_laplace_B_J}) and
(\ref{G_laplace_B_S}):
\begin{align}
\label{inv_G_laplace_B_JS}
\begin{split}
\dot{\cal G}_{\cal J}(\epsilon)=-\frac{1}{2\pi i}\int_{-i\infty}^{i\infty}\frac{du}{u}\, \epsilon^{u}
\,\Gamma(-u)\,\mathbb{B}_{\cal J}(u)
\\
\dot{\cal G}_{\cal S}(\epsilon)=-\frac{1}{2\pi i}\int_{-i\infty}^{i\infty}\frac{du}{u}\, \epsilon^{u}
\,\Gamma(-2u)\,\mathbb{B}_{\cal S}(u)\,
.
\end{split}
\end{align}
Using these relations with the $u\to 0$ expansion of $\mathbb{B}_{\cal J,\,S}(u)$ in
(\ref{B_relation_dotSJ}) we can easily derive the ${\cal O}(\epsilon^0)$ term in the
expansion of $\dot{\cal G}_{\cal J,\,S}(\epsilon)$ in (\ref{small_epsilon_behaviour}) by applying
the Cauchy theorem.

\section{Power corrections in the Sudakov exponent~\label{sec:PC}}
\setcounter{footnote}{1}

As discussed in the introduction infrared renormalons are present in the Sudakov exponent. This is evident in Eq.~(\ref{Exponent_Laplace_Borel2}).
As always, infrared renormalons reflect sensitivity to
large--distance dynamics and indicate the presence of non-perturbative
power corrections. From (\ref{Exponent_Laplace_Borel2}) one can immediately deduce that
these power corrections (1) exponentiate together with the logarithms; and (2)
are enhanced at large $N$: the `jet' and `soft' contributions scale as integer
powers of $N\Lambda^2/m^2$ and $N\Lambda/m$, respectively.
Such parametrically--enhanced power corrections appear
exclusively through the Sudakov factor and have a significant
impact on the distribution in the threshold region.

The renormalon technique offers here a unique window into the non-perturbative
side of the problem. Renormalon analysis allows us to find the specific
pattern of power corrections~\cite{Gardi:2006jc}: in a given process
only certain power corrections appear while others are absent.
This pattern of power corrections is linked with the symmetry properties of the source of soft gluon radiation, and it is very much process dependent.
The absence of specific power terms can be seen as an extension of
the notion of infrared finiteness beyond the logarithmic level, i.e. ``infrared safety at the power level''.

In the Borel formulation the presence of infrared renormalon ambiguities is
transparent: the Borel integral in (\ref{Exponent_Laplace_Borel2}) is obstructed by simple poles at integer and half--integer values of $u$, except where the Borel transforms of the Sudakov anomalous dimensions $B_{\cal S}(u)$ and $B_{\cal J}(u)$ of (\ref{B_JS_mathbb}) vanish.
It is natural then~\cite{Gardi:2001ny,Gardi:2006jc} to take the Principal Value of the integral in (\ref{Exponent_Laplace_Borel2}) as a definition of the perturbative sum, and use the renormalon ambiguities as a basis for parametrization of power corrections.

In the dispersive approach, power--like sensitivity is
somewhat less obvious. Consider for example the \emph{soft} (${\cal S}$) Sudakov factor as can be obtained from (\ref{diff_Exponent_Laplace2}) through integration--by--parts (an example is provided by Eq.~(\ref{DY_kernel_Mink}) for the Drell--Yan case):
\begin{equation}
\label{E_Mink}
\left.
\frac{d\ln \overline{\rm Sud}(m^2,N)}{d \ln m^2}\right\vert_{\cal S}= -\frac{C_R}{\beta_0}
\int_0^{\infty}\frac{d\epsilon}{\epsilon}\,a_{{\cal S}}^{\Mink}(\epsilon\,m^2)
\,
\left[\ddot{\cal G}_{{\cal S}}\left({N^2\epsilon}\right)-
\ddot{\cal G}_{{\cal S}}\left({\epsilon}\right)\right]\equiv
E_{\Mink}(m^2,N)\,.
\end{equation}
Only {\em non-analytic} terms in the small--$\epsilon$ behavior of the Minkowskian characteristic function $\ddot{\cal G}_{{\cal S}}\left({\epsilon}\right)$
are related with renormalons~\cite{DMW,Grunberg:1998ix,Gardi:1999dq}, indicating corresponding power corrections.
In a Euclidean formulation,
\begin{equation}
\label{E_Eucl}
\left.\frac{d\ln \overline{\rm Sud}(m^2,N)}{d \ln m^2}
\right\vert_{\cal S}=-\frac{C_R}{\beta_0}
\int_0^{\infty}\frac{dy}{y}\,a_{{\cal S}}^{\Eucl}(y\,m^2)
\,
\left[\ddot{\cal G}_{{\cal S}}^{\Eucl}\left({N^2y}\right)-
\ddot{\cal G}_{{\cal S}}^{\Eucl}\left({y}\right)\right]\,,
\end{equation}
such as the one of Eq.~(\ref{DY_kernel_Eucl}) in the Drell--Yan case, the relation is simpler: all the terms in the small--$y$ expansion of $a_{{\cal S}}^{\Eucl}$ are related with renormalons. Although a Euclidean formulation (\ref{E_Eucl}) does not exist in general --- amongst the examples considered in this paper it exists \emph{only} in the Drell--Yan case ---
it is instructive to consider it first since, in contrast with the Minkowskian one, it facilitates separating between the infrared and ultraviolet contributions in a straightforward manner by splitting the momentum integral:
\begin{equation}
\label{kernel_Eucl_split}
\left.\frac{d\ln \overline{\rm Sud}(m^2,N)}{d \ln m^2}
\right\vert_{\cal S} =
E_{\Eucl}^{\IR}(m^2,N;\mu^2_I)+E_{\Eucl}^{\UV}(m^2,N;\mu^2_I)
\equiv
E_{\Eucl}(m^2,N)\,
\end{equation}
where
\begin{align}
\label{E_IR_DY}
E_{\Eucl}^{\IR}(m^2,N;\mu^2_I)&\equiv-\frac{C_R}{\beta_0}
\int_0^{y_I}\frac{dy}{y}\,a_{{\cal S}}^{\Eucl}(y\,m^2)
\,
\left[\ddot{\cal G}_{{\cal S}}^{\Eucl}\left({N^2y}\right)-
\ddot{\cal G}_{{\cal S}}^{\Eucl}\left({y}\right)\right]
\end{align}
where $\Lambda \ll\mu_I\ll Q/N$ is a momentum cutoff and $y_I=\mu^2_I/m^2$.
Using (\ref{E_Eucl}) and (\ref{kernel_Eucl_split}) the remainder is given by
\begin{align}
\label{E_UV_DY}
E_{\Eucl}^{\UV}(m^2,N;\mu^2_I)&=-\frac{C_R}{\beta_0}
\int_{y_I}^{\infty}\frac{dy}{y}\,a_{{\cal S}}^{\Eucl}(y\, m^2)
\,
\left[\ddot{\cal G}_{{\cal S}}^{\Eucl}\left({N^2y}\right)-
\ddot{\cal G}_{{\cal S}}^{\Eucl}\left({y}\right)\right]\,.
\end{align}
One can now expand the Euclidean characteristic function at small $y$ under the integral (\ref{E_IR_DY}), getting the power corrections. According to
Eq.~(\ref{G_Eucl}) the Euclidean characteristic function $\ddot{\cal G}_{{\cal S}}^{\Eucl}\left({y}\right)$ is the integral over the discontinuity of the Minkowskian one:
\begin{align}
\label{G_discont}
 \frac{d\ddot{\cal G}_{\cal S}^{\Eucl}(y)}{d\ln 1/y} =
 -\frac{1}{\pi}\,{\rm Im}\left\{
 \ddot{\cal G}_{\cal S}(-y-i0)\right\}\,;\qquad
 \ddot{\cal G}_{\cal S}^{\Eucl}(y)= -\int_{y}^{\infty}\frac{d\epsilon}{\epsilon}\,\frac{1}{\pi}\,{\rm Im}\left\{
 \ddot{\cal G}_{\cal S}(-\epsilon-i0)\right\}\,.
\end{align}
This is why only non-analytic terms in
the expansion of the Minkowskian characteristic function $\ddot{\cal G}_{{\cal S}}\left({\epsilon}\right)$ are relevant.

The extension to the general case, where (\ref{E_Eucl}) does not exists, is
based on (\ref{kernel_Eucl_split})
where $E_{\Eucl}^{\IR}(m^2,N;\mu^2_I)$ is still defined by (\ref{E_IR_DY}) \emph{but} $E_{\Eucl}^{\UV}(m^2,N;\mu^2_I)$ \emph{does not have a representation of the form (\ref{E_UV_DY})}; it must instead be defined by first reverting to a Borel representation \cite{Grunberg:1998ix,Gardi:1999dq}.
Since in this section we are focussing on power corrections, we do not write down the explicit form of the latter, but instead refer the reader to \cite{Grunberg:1998ix,Gardi:1999dq}.
There is one point that needs to be stressed, however: Eq.~(\ref{kernel_Eucl_split}) will {\em not} coincide with the Minkowskian dispersive integral (\ref{E_Mink})
\emph{unless the Euclidean Sudakov coupling  is causal}.

A causal coupling admits the dispersion relation (\ref{Eucl-Mink}): it is analytic except for the cut on the timelike axis.
In this case the entire first sheet of the complex momentum plane is mapped onto a compact domain in the complex coupling plane whose boundary is the image of the time--like axis (the simplest example where this is realized is a two--loop coupling with negative $\beta_1/\beta_0$, see Fig. 2 in Ref.~\cite{Gardi:1998ch}).
The three-- and four--loop coefficients computed in Sec.~(\ref{sec:ECH_evolution}) in a few examples, have been found to have opposite signs to the one-- and two--loop ones, opening the possibility that a non-trivial infrared fixed-point appears in these effective charges within perturbation theory. While the stability of this fixed point and the presence or absence of Landau singularities obviously depends on higher orders, a causal structure
becomes possible.
 In this scenario the dispersion representation (\ref{Eucl-Mink})
holds for the perturbative coupling itself, which does not need to be
 modified by any power terms, and then
\[
E_{\Mink}(m^2,N)=E_{\Eucl}(m^2,N).
\]
$E_{\Mink}(m^2,N)$ still differs from the Principal Value of the Borel sum. But this difference is then entirely due to power corrections that \emph{are} related to infrared renormalons~\cite{Grunberg:1995vx,Dokshitzer:1995af,Grunberg:1996hu,Grunberg:1998ix}, and genuinely reflect sensitivity to long-distance dynamics. The physical distribution is expected to differ from this perturbative result by power corrections whose pattern follows the renormalons.

In contrast \emph{if} $a_{\cal S}^{\Eucl}(k^2)$ has Landau singularities in the first sheet of the complex
$k^2$ plane, as indeed occurs in the large--$\beta_0$ limit (see
Appendix~\ref{sec:Renormalon_sum_techniques}), then
\[
E_{\Mink}(m^2,N)\neq E_{\Eucl}(m^2,N),
\]
and, although $E_{\Mink}(m^2,N)$ in (\ref{E_Mink}) is still finite and yields a \emph{unique result},  the dispersion integral (\ref{E_Mink}) differs from
the corresponding Borel sum by power corrections that are \emph{not related with infrared renormalons}~\cite{Ball:1995ni,DMW,Beneke:1994qe,Gardi:1999dq,Grunberg:1998ix}.

In the presence of Landau singularities in the Euclidean couplings,
the Minkowskian couplings $a_{\cal J, S}^{\Mink}(k^2)$ are still finite in the infrared
limit. One may therefore be tempted to consider the
result of (\ref{Dispersive_Exponent_Laplace}) or even the approximation of (\ref{Dispersive_Exponent_Laplace_NLL})
as a ``non-perturbative model''\footnote{A related model has been proposed in
\cite{Aglietti:2006yb},
where the fixed--logarithmic--accuracy formula for the exponent has been rewritten in terms of
the analytic coupling (\ref{APT}).}.
One should be aware, however, that this result involves additional power corrections that are not associated with long--distance dynamics.
As we recall in Appendix \ref{sec:Renormalon_sum_techniques}, using the dispersive integral amounts to ``analytization'' of the coupling that removes the Landau singularity. This is why the dispersive integral differs from the Borel sum by power terms that are not all related to renormalons. In the large--$\beta_0$ limit this difference is given by the second term in the curly brackets in~(\ref{RQ}). An explicit example is given in Appendix~\ref{sec:Mink_integral}.

In the following we derive the small--$\epsilon$ expansion of the Minkowkian characteristic function, and then deduce from it the corresponding expansion of its discontinuity, namely of the Euclidean characteristic function.


The relations (\ref{inv_G_laplace_B_JS}) can readily be used to see that the singularities of
$\Gamma(-u)\,\mathbb{B}_{\cal J}(u)$ and $\Gamma(-2u)\,\mathbb{B}_{\cal S}(u)$ at integer
or half integer values of $u$, $u=k/2$ ($k$ is a positive integer),
are associated with terms of order ${\cal O}(\epsilon^{k/2})$ in the expansion of the
characteristic functions $\dot{\cal G}_{\cal J,\,S}(\epsilon)$ at small~$\epsilon$.
Non-analytic terms scaling as $\epsilon^{k/2}$
in this expansion are related to infrared renormalons at $u=k/2$
and therefore indicate ${\cal O}((\Lambda/m)^k)$ power corrections in the exponent. Owing to the $N$--dependence of the exponent (\ref{Dispersive_Exponent_Laplace}) through $\dot{\cal G}_{\cal J}(N\epsilon)$ and $\dot{\cal G}_{\cal S}(N^2\epsilon)$, power corrections associated with the `jet' and `soft' subprocesses will scale as
$(N\epsilon)^{k/2}$ and $(N^2\epsilon)^{k/2}$, respectively.

To see the precise relation between Borel singularities in (\ref{Exponent_Laplace_Borel2}) and non-analytic terms in (\ref{Dispersive_Exponent_Laplace}), let us first recall the Borel singularity structure and then translate it into
the terminology of the dispersive approach.
The universal $\Gamma(-u)$ and $\Gamma(-2u)$ factors give rise to poles in (\ref{Exponent_Laplace_Borel2}).
The functions $B_{{\cal J}\,,{\cal S}}(u)$ multiplying them may nevertheless vanish, cancelling the
 potential pole. According to (\ref{B_JS_mathbb}) this depends on the properties of $\mathbb{B}_{\cal J,\,S}(u)$
as well as the Borel transform of the corresponding Sudakov effective charge $B\left[a_{\cal J,\,S}^{\Eucl}\right](u)$. 
We assume, as in previous work~\cite{Gardi:2006jc}, that the latter is innocuous: in the large--$\beta_0$ limit it is given by (\ref{a_JS_large_beta0}), so it has no poles nor zeros at integer or half--integer locations, and we assume that singularities or zeros will not develop there beyond this limit. 
In contrast $\mathbb{B}_{\cal J,\,S}(u)$ may have poles at integer values of $u$,  as well as zeros at integer and half integer values of $u$.
Note that the assumption above implies that the zeros of  $B_{\cal J,\,S}(u)$ present at large--$\beta_0$, which cancel some potential renormalon singularities in the exponent, are still present in the full theory.

Thus, according to (\ref{inv_G_laplace_B_JS}) non-analytic terms in the characteristic function emerge in the following cases:
\begin{itemize}
\item{}
In $\dot{\cal G}_{\cal J}(\epsilon)$, when $\mathbb{B}_{\cal J}(u)$ has a pole, i.e. when
$\Gamma(-u)\,\mathbb{B}_{\cal J}(u)$ has a double
pole. This occurs at $u=1$ and $u=2$ (see Table~\ref{table:calF_B}). Indeed, upon expanding the
explicit result for $\dot{\cal G}_{\cal J}(\epsilon)$ we get:
\begin{align}
\begin{split}
\dot{\cal G}_{\cal J}(\epsilon)&\simeq
\left(-\gamma_E-\ln(\epsilon)-\frac{3}{4}\right)
+\left(2-\frac{1}{2}\gamma-\frac{1}{2}\ln(\epsilon)\right)\epsilon
+\left(-\frac{3}{8}+\frac{1}{4}\gamma_E+\frac{1}{4}\ln(\epsilon)\right)\epsilon^2\\
&-\frac{5}{72}\epsilon^3+{\cal O}(\epsilon^4)\,.
\end{split}
\end{align}
The non-analytic terms $\epsilon\ln(\epsilon)$ and $\epsilon^2\ln(\epsilon)$ in this expansion
indicates, respectively, $N\Lambda^2/m^2$ and $N^2\Lambda^4/m^4$ power
corrections in the exponent (\ref{Dispersive_Exponent_Laplace});
\,see e.g.~\cite{Gardi:2002bk,DGE,Gardi:2002xm,Grunberg:2006gd}.
\item{} In $\dot{\cal G}_{\cal S}(\epsilon)$, at all half integer values of $u$, $u=k/2$
where $k$ is an odd number, except when $\mathbb{B}_{\cal S}(u)$ vanishes. In this case
$\dot{\cal G}_{\cal S}(\epsilon)$ develops a square-root singularity
$\left(\sqrt{\epsilon}\right)^{k}$.
Thus, $\dot{\cal G}_{\cal S}(N^2\epsilon)$ in~(\ref{Dispersive_Exponent_Laplace}) gives rise to ${\cal O}((N\Lambda/m)^k)$ corrections.
\item{} In $\dot{\cal G}_{\cal S}(\epsilon)$, at integer values of $u$ ($u=k/2$
with even $k$) where
$\mathbb{B}_{\cal S}(u)$ has a pole (so $\Gamma(-2u)\,\mathbb{B}_{\cal S}(u)$ has a double pole).
In this case $\dot{\cal G}_{\cal S}(\epsilon)$ develops a logarithmic singularity:
$\epsilon^{k/2}\ln (\epsilon)$, leading to ${\cal O}((N\Lambda/m)^k)$ corrections in~(\ref{Dispersive_Exponent_Laplace}).
\end{itemize}
The three cases above  are precisely the ones where $\left. B_{{\cal J},{\cal S}}(u)\right\vert_{\text{large}\, \beta_0}$ in Eq.~(\ref{B_JS_mathbb_large_beta0}) does not vanish, i.e. where the singularities in $\Gamma(-u)$ and $\Gamma(-2u)$ are not cancelled. 

Altogether the soft characteristic function has the following generic structure at small~$\epsilon$:
\begin{align}
\begin{split}
\label{small_epsilon}
\dot{\cal G}_{\cal S}( \epsilon)&=
-\frac12\ln( \epsilon)+\frac12 S_1-\gamma_E+
\sum_{k\, \text{odd}} C_k \left(\sqrt{\epsilon}\right)^{k}\,+\,
\sum_{k\, \text{even}}C_k \, \epsilon^{k/2}\,\ln (\epsilon)\,
+\,{\text{analytic\, terms}}\,.
\end{split}
\end{align}
or
\begin{align}
\begin{split}
\label{small_epsilon_d}
\ddot{\cal G}_{\cal S}(\epsilon)&=
\frac12+
\sum_{k\, \text{odd}} \frac{k}{2}\, C_k \left(\sqrt{\epsilon}\right)^{k}\,+\,
\sum_{k\, \text{even}}\frac{k}{2}\,C_k \, \epsilon^{k/2}\,\ln (\epsilon)\,+\,{\text{analytic\, terms}}\,.
\end{split}
\end{align}
Thus, upon extracting the discontinuity according to (\ref{G_discont}) one obtains:
\begin{align}
\begin{split}
\label{small_epsilon_d_Im_d}
 \frac{d\ddot{\cal G}_{\cal S}^{\Eucl}(y)}{d\ln 1/y}=-\frac{1}{\pi}{\rm Im}\left\{\ddot{\cal G}_{\cal S}(-y)\right\}&=-
\sum_{k\, \text{odd}} \frac{k}{2}\, C_k \,\frac{\sin(k\pi/2)}{\pi}\,\left(\sqrt{y}\right)^{k}\,-\,
\sum_{k\, \text{even}}\frac{k}{2}\,C_k \, y^{k/2}\,.
\end{split}
\end{align}
or
\begin{align}
\begin{split}
\label{small_epsilon_d_Im}
 \ddot{\cal G}_{\cal S}^{\Eucl}(y)=
\sum_{k\, \text{odd}} \, C_k \,\frac{(-1)^{(k-1)/2}}{\pi}\,\left(\sqrt{y}\right)^{k}\,+\,
\sum_{k\, \text{even}}\,C_k \, y^{k/2}\,\equiv \sum_{k=1}^{\infty} c_k \, y^{k/2}.
\end{split}
\end{align}
The coefficients $C_k$ of the first few non-analytic terms in specific examples
are summarized in Table \ref{table:non_analytic_terms}.
 The table clearly demonstrates that
the renormalon singularity structure is process dependent.
\begin{table}[h]
  \centering
  \begin{tabular}{|l|l|c|c|c|c|c|c|}
    \hline
     process         & & $\xi\ln(\xi)$  & $\xi^2\ln(\xi)$ & $\xi^3\ln(\xi)$ &
$\xi^4\ln(\xi)$ & $\xi^5\ln(\xi)$&
$\xi^6\ln(\xi)$
    \\\hline\hline
    \begin{tabular}{l}
    jet function\\
    (e.g. DIS)
    \end{tabular}
    & ${\cal J}$           & $-1/2$& $1/4$& 0&0 & 0& $0$
    \\
    \hline
\end{tabular}\\\vspace*{15pt}
\begin{tabular}{|l|l|c|c|c|c|c|c|}
\hline
     process         & & $\nu$  & $\nu^2\ln(\nu)$ &$\nu^3$ & $\nu^4\ln(\nu)$ & $\nu^5$&
$\nu^6\ln(\nu)$
    \\ \hline\hline
    \begin{tabular}{l}
    B decay; HQ  \\
    Fragmentation
    \end{tabular}
    & ${\cal S}_{\QD}$     &$\pi/4$& 0& $\pi/24$& $1/24$& $-\pi/160$ &
    $-1/360$
      \\
    \hline
    \begin{tabular}{l}
    Drell--Yan\,; \\
    $gg \to$ Higgs
    \end{tabular}
    & ${\cal S}_{\DY}$ &0& $-1$&
    $0 $& $-1/4$& $0$ &
    $-1/36$
    \\
    \hline
    \begin{tabular}{l}
    $e^+e^-\to \text{jets}$\\
    C parameter
    \end{tabular}
    & ${\cal S}_{c}$      & $\pi/4$& 0& $\pi/192 $& $0$ & $\pi/20480$
    & $0$
    \\\hline
    \begin{tabular}{l}
    $e^+e^-\to \text{jets}$\\
    Thrust
    \end{tabular}
     & ${\cal S}_{t}$       &1 & 0& 1/18& 0& 1/600& 0
     \vspace*{-12pt}\\
    &&&&&&&\\\hline
  \end{tabular}
  \caption{Summary of results for the first few coefficients of non-analytic
terms in the small--$\epsilon$ expansion of the moment--space characteristic functions
$\dot{\cal G}_{\cal J}(\xi=N\epsilon)$ and $\dot{\cal G}_{\cal S}(\nu^2=N^2\epsilon)$
  of some inclusive distributions ({\it cf.} Tables \ref{table:dcalG} and \ref{table:calF_B}).
  }\label{table:non_analytic_terms}
\end{table}

The small--momentum expansion of the Euclidean characteristic function (\ref{small_epsilon_d_Im})
can be readily used inside the integral in Eq.~(\ref{E_IR_DY}) to obtain:
\begin{align}
\label{E_IR_DY_expanded}
E_{\Eucl}^{\IR}(m^2,N;\mu^2_I)&\simeq -\frac{C_R}{\beta_0}\,\sum_{k=1}^{\infty} c_k \,\times\,(N^k-1)\,\times\,
\underbrace{\int_0^{y_I}\frac{dy}{y}\,y^{k/2}\,a_{{\cal S}}^{\Eucl}(y\,m^2)}_{\equiv M_k} \,.
\end{align}
Obviously, in the infrared region the coupling may differ from its perturbative part. This is encapsulated
in the moments. In this way the moment $M_k=\,{\cal O}
\left((\mu_I/m)^{\,k}\right)$ can be used to parametrize non-perturbative power corrections of order ${\cal O}\left((N\Lambda/m)^k\right)$ with any $k$ in the Sudakov exponent.
Recall that the $k=1$ moment (with a choice of $\mu_I=1$ or $2$ GeV) has been used in
\cite{DMW} and following work (see e.g.~\cite{Dokshitzer:1997iz,Dokshitzer:1998kz,Dokshitzer:1998pt,Dasgupta:2003iq})
to parametrize power corrections to event--shape distributions. The framework presented here allows one to identify this coupling to higher orders in perturbation theory as well as to parametrize in a similar manner the corrections that scale as ${\cal O}\left((N\Lambda/m)^k\right)$ using the higher moments.

Assuming that the couplings $a_{\cal J, S}^{\Eucl}$ have a causal analyticity structure one can directly parametrize these functions in the infrared consistently with the dispersion relation (\ref{Eucl-Mink}) as well as their known ultraviolet evolution of Eqs. (\ref{J_beta}) and (\ref{S_beta}); a simple one-loop model of this kind has been considered in
Ref.~\cite{Webber:1998um}. 
Upon taking the time--like discontinuity to obtain $a_{\cal J, S}^{\Mink}$ one can readily evaluate the Minkowskian integral in the exponent,  Eq.~(\ref{Dispersive_Exponent_Laplace}), getting the all--order sum already including power corrections of the renormalon type.
An even simpler possibility would be to parametrize directly the Minkowskian couplings  $a_{\cal J, S}^{\Mink}$ in the infrared, consistently with their ultraviolet evolution\footnote{In this case one should make sure that the resulting Euclidean coupling is causal at the \emph{perturbative} level, and does not involve extra unwelcome power corrections, as in the APT example!}.
In this way one can set up a simple yet fully consistent  ``power--correction phenomenology'' without ever dealing with individual powers and without introducing any momentum cutoff.

\section{Conclusions~\label{sec:conc}}

We have presented a general formalism for Sudakov resummation based on dispersion integrals.
The expression for the Sudakov factor in infrared and collinear safe distributions involving a jet function (${\cal J}$) and a soft function (${\cal S}$) is summarized by Eqs.~(\ref{Dispersive_Exponent_Laplace1}) and (\ref{Dispersive_Exponent_Laplace}) or (\ref{widetilde_Exponent_Laplace_with_Mellin_const1}).
This formulation consists of two
ingredients~\cite{Grunberg:2006gd,Grunberg:2006ky}: the first, $\dot{\cal G}_{\cal J,\,S}$,  are \emph{characteristic functions} that are computed analytically in the large--$\beta_0$ limit and encode information on power corrections; the second, $a_{\cal J,\,S}$, are \emph{Sudakov effective charges} that are defined order--by--order in perturbation theory and encapsulate the non-Abelian nature of the interaction. In what concerns the soft function (${\cal S}$) both these ingredients are \emph{process--dependent} owing to the fact that large--angle soft gluon radiation depends on the space--time geometry of the hard partons that radiate. In contrast, there is a \emph{unique} jet function (${\cal J}$) characterizing soft and collinear radiation from a jet with a fixed invariant mass. Explicit results for the characteristic functions in a variety of processes have been compiled in Table~\ref{table:dcalG}, and the corresponding Sudakov effective charges have been computed and analyzed in Sec.~\ref{sec:general_dispersive}.

The Sudakov effective charges are directly related to the physical anomalous dimensions.
They can be computed to any order based on the conventional anomalous dimensions defined by dimensional regularization. Furthermore, comparison with the joint resummation formalism~\cite{Laenen:2000ij} leads to a direct diagrammatic interpretation of the Sudakov effective charges in terms of `webs', see (\ref{Web_identification}) above.

The dispersive approach presented here provides a realization of DGE~\cite{Gardi:2001ny,Gardi:2002bg,DGE,Gardi:2006jc}: it goes beyond the resummation of Sudakov logarithms per se by incorporating an internal all--orders resummation of running--coupling corrections. This resummation guarantees renormalization--group invariance. Owing to the enhancement of subleading logarithms that are associated with the running of the coupling~\cite{Gardi:2001ny,DGE}, this additional resummation leads to a significant improvement over the conventional approach to Sudakov resummation where a renormalization--scheme--dependent truncation is performed, guided solely by the logarithmic accuracy criterion.
Beyond the perturbative level, the present approach facilitates a systematic analysis of power corrections based on renormalon ambiguities which reveal themselves through the discontinuities of
the characteristic function $\dot{\cal G}_{\cal J,\,S}$.

We have shown that there is a direct correspondence between the scheme--invariant Borel formulation and the dispersive one, and derived all--order relations between their ingredients. The two formulations capture \emph{the same} set of radiative corrections. Yet, at the power level they provide different regularizations of the sum. As far as renormalon singularities are concerned, this difference does not pose any difficulty:
the two regularizations are in principal equivalent, as they differ by power corrections of the same parametric form that need be introduced to account for genuine non-perturbative effects.
In contrast, if the Sudakov effective charges have Landau singularities, as occurs for example in the large--$\beta_0$ limit, the dispersion relation is violated. Then, the dispersion integral differs from the Borel sum by additional power corrections that are not related with renormalons.
This complication does not arise if the Sudakov effective charges have a causal analyticity structure, making them consistent with the dispersion relation.

The dispersive approach offers a convenient way to parametrize the
non-perturbative power corrections exposed by the renormalons by means of integrals over the coupling in the infrared
region~\cite{DMW}. We find that in the context of Sudakov resummation, the infrared--finite--coupling approach by Dokshitzer, Marchesini and Webber is of special interest: in contrast with the general situation here the coupling --- the Sudakov effective charge --- can be systematically identified to any order. Moreover, as discussed in Sec.~\ref{sec:ECH_evolution}, there are indications that these effective charges may reach a finite limit in the infrared already within perturbation theory. Remarkably,  despite the fact that the evolution of the Sudakov effective charges becomes process--dependent at three loops, the infrared limit itself turns out to be universal: it depends only on the cusp anomalous dimension.

A particularly attractive example from the point of view of the dispersive approach is that of Drell--Yan or Higgs production. This has two aspects: first, there exists a Euclidean representation (\ref{DY_kernel_Eucl}) where the identification of large--distance effects is transparent, and second the relevant Sudakov effective charge admits an infrared fixed point already at the perturbative level as both the three--loop and four--loop coefficients of the effective--charge beta function are negative, see Eq.~(\ref{S_beta_Nf4}).

A significant effort has been put in the past decade in examining the hypothesis of a \emph{universal}~\cite{DMW,Dokshitzer:1997iz,Dokshitzer:1998kz,Dokshitzer:1998pt} infrared--finite coupling in the context of event--shape distributions. Experimental data were primarily used to test the assumed universality of the \emph{first} power moment of this ``effective coupling''. Indeed, overall this assumption is supported by data~\cite{Dasgupta:2003iq}. Despite the different resummation formalism utilized in these studies compared to the one presented here, the notion of the infrared--finite coupling, and the way it relates to power corrections, are the same. So far this 
``effective coupling'' has only been identified to NLO, and its universality has not been established theoretically\footnote{See however Refs.~\cite{Lee:2006fn,Lee:2007jr} (as well as earlier work, especially~\cite{Korchemsky:1999kt,Korchemsky:2000kp,Gardi:2001ny})
where universality of the \emph{leading} power correction (the shift) in a class of event--shape distribution is  established independently of any infrared--finite--coupling hypothesis.}.
This has now changed: the present formalism uniquely identifies the effective charges relevant for Sudakov resummation to all orders in perturbation theory, and shows that their universality does not extend beyond the NLO. It nevertheless shows that their \emph{infrared limit} is universal. \emph{Approximate universality} of power corrections, determined by the first few power moments of the coupling, is expected as a by-product.

DGE, in its Borel formulation, has been successfully applied to phenomenology in a range of processes~\cite{Gardi:2006jc}. Most importantly, it extends the range of applicability of resummed perturbation theory closer to threshold, far beyond what can be achieved by a conventional, fixed--logarithmic--accuracy approach.   This should be attributed to two main factors: first the additional resummation performed, and second the possibility to identify correctly the pattern of power corrections, especially the first one or two renormalon singularities.
The main challenge has been to find an effective parametrization of power corrections that captures the dependence on $N\Lambda/m$ for $N\gg m/\Lambda$, where the power expansion completely breaks down.
Even if one assumes that non-perturbative corrections are directly proportional to the corresponding renormalon residues, it remains difficult to control them in practice since the residues vary going beyond the large--$\beta_0$ limit. Indeed, little is known about the Borel transform $B[a_{\cal S}^{\Eucl}](u)$ in (\ref{Exponent_Laplace_Borel2}) beyond the large--$\beta_0$ limit and away from the vicinity of the origin.
When using the dispersive formulation instead, the problem translates into the parametrization of the Sudakov effective charge $a_{\cal S}^{\Eucl}(\mu^2)$ itself in the infrared region. This should be an easier function to constrain, especially if $a_{\cal S}^{\Eucl}(\mu^2)$ tends to a finite limit in the infrared already within perturbation theory, making it a slowly varying function. Constraining the coupling over the infrared region may be further helped by the universality of its infrared limit.
It should be nevertheless stressed that the dispersive approach is advantageous only upon assuming that the relevant couplings are causal, free of Landau singularities. In the opposite case using the Borel formulation~\cite{Gardi:2006jc} is more straightforward.

Finally, parametrization of the Sudakov effective charges over the infrared region provides an alternative to the shape--function approach, one that matches smoothly onto the perturbative description, does not require the introduction of an explicit infrared cutoff, and most importantly, is better constrained.


\vskip .3cm

\section*{Acknowledgements}

E.G. wishes to thank Bryan Webber for useful discussions. E.G. is grateful to the \'Ecole Polytechnique for hospitality when this work was initiated, the University of Edinburgh particle physics group for hospitality in its final stages, and finally the HEP group in Cavendish Laboratory for a very enjoyable time over the past four years.
G.G. benefited from early discussions with Yuri Dokshitzer, Pino Marchesini and George Sterman.

\vskip .4cm

\appendix

\section{Coefficients of the Sudakov anomalous dimensions in $\overline{\rm MS}$\label{sec:coef}}

The coefficients of the cusp anomalous dimension, corresponding to (\ref{A_cusp}) are known to three--loop order~\cite{Moch:2004pa}:
\begin{eqnarray}
\label{a23_MSbar}
 a_1&=&1 \nonumber\\
 a_2&=& {\displaystyle \frac {5}{3}}  +
{\displaystyle {\frac{\mathit{C_A}}{\beta_{0}} \left({\displaystyle
\frac {1}{3}}  - {\displaystyle \frac {\pi ^{2}}{12}} \right)\,}},
\\ \nonumber
 a_3&=&  - {\displaystyle \frac {1}{3}}  + \frac{1}{\beta_0}\left[
{\displaystyle  {\left({\displaystyle \frac {55}{16}}  - 3\,\zeta_3
 \right)\,\mathit{C_F} + \left({\displaystyle \frac {253}{72}}  -
{\displaystyle \frac {5\,\pi ^{2}}{18}}  + {\displaystyle \frac {
7}{2}} \,\zeta_3\right)\,\mathit{C_A}}}  \right]\nonumber \\
\nonumber \mbox{}&&\hspace*{30pt} +  \frac{1}{{\beta_0}^2}
\left[\left( - \frac {605}{192}  +  \frac {11}{4} \,\zeta_3\right)
\,\mathit{C_A}\,\mathit{C_F}\,+\, \left( - {\displaystyle \frac
{7}{18}}  - {\displaystyle \frac {\pi ^{2}}{18}} - {\displaystyle
\frac {11}{4}} \,\zeta_3 +  \frac {11\,\pi ^{4}}{720}
\right)\,\mathit{C_A}^{2} \right].
\end{eqnarray}

The coefficients of the jet--function anomalous dimension ${\cal B}$ (associated with an unresolved jet with a constrained mass) are known to three--loop order~\cite{Moch:2005ba}:
\begin{align}
\begin{split}
b_1 &= -\frac34\\
b_2 &= \frac{\pi^2}{6}-\frac {247}{72}  -
\frac{C_A}{\beta_0}\left(\frac{73}{144}-\frac{5}{2}\zeta_3\right)
+\frac{C_F}{\beta_0}\left(\frac{\pi^2}{8}-\frac{3}{32}
-\frac{3}{2}\zeta_3\right)\\
b_3 &= -\frac{4357}{648}+\frac{29}{36}\pi^2-\frac{2}{3}\zeta_3 \\& +
\frac{1}{\beta_0}\Bigg[\left(-\frac{5501}{576}+\frac{25}{32}\pi^2-\frac{1}{3}\zeta_3\right)C_F
+\left(-\frac{1807}{216}+\frac{283}{648}\pi^2+\frac{335}{36}\zeta_3-\frac{13}{360}\pi^4\right)
C_A\Bigg]\\
&+\frac{1}{{\beta_0}^2}
\Bigg[\bigg(-\frac{29}{128}-\frac{3}{64}\pi^2-\frac{17}{16}\zeta_3-\frac{\pi^4}{40}+\frac{1}{12}\pi^2
\zeta_3+\frac{15}{4}\zeta_5\bigg){C_F}^2\\
&+\bigg(\frac{55543}{6912}+\frac{1}{32}\pi^2-\frac{245}{72}\zeta_3-\frac{17}{720}\pi^4-\frac{1}{24}
\pi^2\zeta_3-\frac{15}{8}\zeta_5\bigg)C_A C_F\\
&+\bigg(\frac{4891}{10368}-\frac{115}{2592}\pi^2+\frac{605}{144}\zeta_3+\frac{41}{2880}\pi^4
-\frac{11}{72}\pi^2\zeta_3-\frac{29}{8}\zeta_5\bigg){C_A}^2\Bigg]\,.
\end{split}
\end{align}

The coefficients of the Drell--Yan anomalous dimension ${\cal D}_{\DY}$ (associated with large--angle soft radiation from two lightlike partons that annihilate to produce a non-colored heavy object) are known to three--loop order~\cite{Moch:2005ky,Laenen:2005uz}:
\begin{align}
\label{d123_DY}
\begin{split}
d_1^{\DY} &= 0\\
d_2^{\DY} &= -\frac{14}{9}+\frac{1}{3}\pi^2+\frac{1}{2}
\frac{C_A}{\beta_0}\left[-\frac{8}{9}+\frac{7}{2}\zeta_3\right]\\
d_3^{\DY} &= -\frac{116}{81}+{10}{9}\pi^2+\frac{10}{3}\zeta_3+\frac{1}{2\beta_0}
\Bigg[\left(-\frac{23}{180}\pi^4+\frac{517}{324}\pi^2+\frac{245}{18}\zeta_3-\frac{2345}{216}\right)C_A\\
&+\left(-\frac{1711}{144}+\frac{19}{3}\zeta_3+\frac{\pi^4}{30}+\frac{\pi^2}{2}\right)C_F\Bigg]\\
&+\frac{1}{2{\beta_0}^2}\Bigg[\left(-\frac{223}{1296}\pi^2+\frac{11}{360}\pi^4-\frac{485}{1296}-6\zeta_5-
\frac{11}{36}\pi^2\zeta_3+\frac{371}{72}\zeta_3\right){C_A}^2\\
&+\left(-\frac{11}{24}\pi^2+\frac{18821}{1728}-\frac{209}{36}\zeta_3-\frac{11}{360}\pi^4\right)C_F C_A\Bigg].
\end{split}
\end{align}

The coefficients of the anomalous dimension ${\cal D}_{\QD}$  corresponding to the heavy--quark distribution function~\cite{Korchemsky:1992xv} as well as the heavy--quark fragmentation function are known to two--loop order~\cite{Gardi:2005yi}:
\begin{align}
\begin{split}
d_1^{\QD} &= 1\\
d_2^{\QD} &= \frac {1}{9}  -\frac{C_A}{\beta_0}
\left[\frac{\pi^2}{12}+\frac{11}{18}
-\frac{9}{4}\zeta_3\right]\,.
\end{split}
\end{align}

\section{Renormalon sum: Borel and dispersive representations\label{sec:Renormalon_sum_techniques}}

There exist two convenient representations of a dressed gluon which result in two different formulations of resummation formulae:
\begin{itemize}
\item{} {\bf The Borel method:} The running coupling, which include the effect of dressing, can be written as
\begin{equation}
\label{eq:alpha_V_Borel}
 \frac{\alpha_s^V(-k^2)}{\pi} = \frac{\alpha_s(\mu^2)}{\pi}
\,\frac{1}{1+\Pi(k^2)}\,=\,
\frac{1}{\beta_0}\, \int_0^{\infty}du\, T(u)\, \left(\frac{\Lambda_V^2}{-k^2}\right)^{u}
\ ,
\end{equation}
where $k$ is the gluon momentum and $\Pi(k^2)$ is the vacuum polarization function, renormalized at $\mu^2$. The coupling is defined in the spacelike region $-k^2>0$ and then analytically continued to the complex momentum plane. The superscript $V$ stands for the $V$ scheme (defined by the potential between two heavy quarks)
where $\Lambda_V^2=\Lambda^2{\rm e}^{\frac53u}$
({\it cf.} (\ref{Lambda_GB})).
For one--loop running coupling,  $T(u)=1$. In this case one recovers
\begin{equation}
\label{one_loop}
\left.\frac{\alpha_s^V(-k^2)}{\pi}\right\vert_{\rm one-loop}
= \frac{1}{\beta_0}\,\frac{1}{\ln\left(
-{k^2}/{\Lambda_V^2}\right)}
\ .
\end{equation}
An all--order
resummation of running--coupling corrections in a given quantity $R(Q^2)$ with a single dressed gluon can thus be achieved by performing the
momentum integration with the modified propagator
\begin{equation}
\label{eq:Borel_replacement}
 \frac{1}{-k^2-i0} \,\,\to\,\,
\frac{1}{(-k^2-i0)^{1+u}}\ .
\end{equation}
This procedure directly yields the Borel representation of the perturbative sum in the large--$\beta_0$ limit in the form:
\begin{equation}
  \left.R(Q^2)\right\vert_{\rm large \,\,\beta_0}\,
  =\,\frac{1}{\beta_0}\,\int\limits_0^{\infty}du \,
  T(u) \left(Q^2/\Lambda^2\right)^{-u} B(u)
  \ .
  \label{R_borel_}
\end{equation}
\item{}
{\bf The dispersive method:}
The dispersive representation of the dressed gluon~\cite{Ball:1995ni,DMW,Beneke:1994qe,Gardi:1999dq} takes the form
\begin{equation}
  \label{eq:simple-dispersive}
  \frac{\alpha_s^V(-k^2)}{\pi} \,=\,
  \frac{\alpha_s(\mu^2)}{\pi} \,\frac{1}{1+\Pi(k^2)} =\, \frac{1}{\beta_0}\left\{-\int\limits_0^\infty
\frac{\rho_V(m^2)\,dm^2}{m^2-k^2}
  \,-\,\frac{1}{1+k^2/\Lambda_V^2}\right\}
  \ ,
\end{equation}
where $\rho_V$ ($\rho_V<0$) is the spectral density function, defined by
the discontinuity of the coupling on the timelike
axis,
\begin{equation}
\label{eq:discontinuity}
\rho_V(m^2)\equiv  \frac{\beta_0}{\pi}\,{\rm
    Im}\left\{\alpha_s^V(-m^2-i0)/\pi\right\} =\frac{1}{\pi}
  \frac{\beta_0\alpha_s(\mu^2)}{\pi}\,\frac{ \text{Im}
    \left\{\Pi(m^2)\right\}}{\vert
    1+\Pi(m^2)\vert^2} \ .
\end{equation}
Taking the time--like discontinuity of the one--loop running coupling (\ref{one_loop})
one obtains:
\begin{equation}
\label{eq:discontinuity_1loop}
\left.\rho_V(m^2)\right\vert_{\rm one-loop}=\frac{-1}{\ln^2(m^2/\Lambda_V^2)+\pi^2}\,.
\end{equation}
The dispersive integral (the first term in the curly brackets in (\ref{eq:simple-dispersive})) then yields\footnote{The superscript APT on (\ref{APT}) stands for `Analytic Perturbation Theory'~\cite{Shirkov:1997wi}; it distinguishes this object from the ordinary one loop coupling. }
\begin{equation}
\label{APT}
  \frac{\alpha_s^{\APT}(-k^2)}{\pi} =\frac{1}{\beta_0}\int\limits_0^\infty \frac{dm^2}{m^2-k^2}\,\,
  \frac{1}{\ln^2(m^2/\Lambda_V^2)+\pi^2}\,
  =\frac{1}{\beta_0}\,\left\{\frac{1}{\ln\left(
-{k^2}/{\Lambda_V^2}\right)}+\frac{1}{1+k^2/\Lambda_V^2}\right\}
\end{equation}
that differs from the original one--loop coupling (\ref{one_loop}) by pure power terms that eliminate the Landau singularity. The second term in the curly brackets in (\ref{eq:simple-dispersive}) cancels this additional term, and thus restores the Landau pole, making (\ref{eq:simple-dispersive}) consistent with (\ref{one_loop}). Note that we shall be using (\ref{eq:simple-dispersive}) rather than (\ref{APT}): we will not assume anything about the way analytic properties of physical quantities are eventually restored in (non-perturbative) QCD.

All--order resummation can be achieved using (\ref{eq:simple-dispersive}) by performing the
momentum integration with a massive gluon propagator:
\begin{equation}
\label{eq:massive_gluon_replacement}
 \frac{1}{-k^2-i0} \,\to\,
\frac{1}{m^2-k^2-i0}.
\end{equation}
This results in the dispersive representation of the perturbative sum in the large--$\beta_0$ limit:
\begin{align}
  \label{RQ}
  \begin{split}
  \left.R(Q^2)\right\vert_{\rm large \,\,\beta_0}\!
  =  \,& \frac{1}{\beta_0}\Bigg\{\!\int_0^{\infty}\frac{dm^2}{m^2} \rho_V(m^2) \left[ {\cal
      F} (m^2/Q^2)-{\cal F} (0)\right]
  +\left[ {\cal
      F} (-\Lambda_V^2/Q^2)-{\cal F} (0)\right]
      \Bigg\}\\
    =
  \,& \frac{1}{\beta_0} \left\{\int_0^{\infty}\frac{dm^2}{m^2}
  a_V^{\Mink}(m^2) \dot{\cal F} (m^2/Q^2)
  +\left[ {\cal
      F} (-\Lambda_V^2/Q^2)-{\cal F} (0)\right]
   \right\}\,  ,
  \end{split}
\end{align}
where in the second line we applied integration--by--parts using (\ref{eq:int_discontinuity})
and
\[
\dot{\cal F} (m^2/Q^2)\equiv -m^2\frac{d}{dm^2} {\cal F} (m^2/Q^2).
\]
\end{itemize}

\section{Taking the large--$N$ limits of finite--$N$ characteristic functions\label{sec:taking_large_N}}

In Sec.~\ref{sec:kernel} we identified the Sudakov limit of the characteristic functions in momentum space.
A similar identification can be done in moment space.
Let us now show that the Sudakov characteristic function in Eq.~(\ref{G_r+v}) can be recovered as the appropriate $N\to\infty$ limit of the moment--space characteristic function of the corresponding physical evolution kernel.
We will also show that the formal `jet' and `soft' exponents (\ref{limit_Sud_exp_J}) and (\ref{limit_Sud_exp_S}), respectively, naturally emerge as limits of these kernels. Again we illustrate these statements in the examples of deep inelastic structure functions and the Drell--Yan cross section.

\vspace*{10pt}
\noindent
\underline{Deep inelastic structure functions}\\
The finite--$N$  moment--space characteristic function\footnote{In Ref.~\cite{DMW} ${\cal G}_{\DIS (F_2)}\left(\epsilon,N\right)$ is denoted by ${\cal F}_N\left(\epsilon\right)$, whereas
${\cal G}_{\DIS (F_2)}^{(v)}\left(\epsilon\right)$ and ${\cal G}^{(v)}_{\DY}\left(\epsilon\right)$ are denoted by ${\cal V}_s\left(\epsilon\right)$ and ${\cal V}_t\left(\epsilon\right)$, respectively. Note also that our normalization of the characteristic functions is half the one in this reference.}
\begin{equation}
\label{finite_N_char}
{\cal G}_{\DIS (F_2)}\left(\epsilon,N\right)=\int_0^1dx\,x^{N-1}{\cal F}_{\DIS (F_2)}(\epsilon,x)
\end{equation}
appears in the dispersive representation (valid in the large--$\beta_0$ limit) of the moment--space evolution kernel $\widetilde{K}_{\DIS (F_2)}(N,Q^2)$ (Eq.~(\ref{DIS_kernel_def})):
\begin{align}
  \label{finite-r-DIS-N}
  \left.\widetilde{K}_{\DIS (F_2)}(N,Q^2)\right\vert_{\rm large \,\,\beta_0}\,
  = &\,\frac{C_F}{\beta_0}\,\int_0^{\infty}\frac{d\mu^2}{\mu^2} \rho_V(\mu^2)  \left(\dot{\cal G}_{\DIS (F_2)}(\mu^2/Q^2,N)-\dot{\cal G}_{\DIS (F_2)}(0,N)\right)\,.
\end{align}
This is the moment-space version of Eq.~(\ref{finite-r-DIS}).
Let us now split ${\cal G}_{\DIS (F_2)}\left(\epsilon,N\right)$ into its real and virtual components:
\begin{equation}
\label{finite_N_char_r_v}
{\cal G}_{\DIS (F_2)}\left(\epsilon,N\right)={\cal G}_{\DIS (F_2)}^{(r)}\left(\epsilon,N\right)
+{\cal G}_{\DIS (F_2)}^{(v)}\left(\epsilon\right)\,.
\end{equation}
Considering first the real contribution, one finds \cite{Friot:2007fd}, using the explicit expression in \cite{DMW}
for characteristic function:
\begin{align}
\begin{split}
\label{cal-G-N-lim}
\lim_{\begin{array}{l}
N\rightarrow \infty \hspace*{-35pt}\\
N\epsilon \,\,\text{fixed}\hspace*{-80pt}
\end{array}
}
{\cal G}_{\DIS (F_2)}^{(r)}\left(\epsilon,N\right)&=
\int_0^{\infty}\,\frac{dr}{r}\,{\rm e}^{-Nr}\,
{\cal F}_{\cal J}\left(\epsilon/r\right)\equiv
{\cal G}_{\cal J}\left(N\epsilon\right)
\,,
\end{split}
\end{align}
which is the analogue of the momentum space relation, Eq.~(\ref{DIS_jet}).
Taking one derivative, we deduce:
\begin{align}
\begin{split}
\label{dot-cal-G-N-lim}
\lim_{\begin{array}{l}
N\rightarrow \infty \hspace*{-35pt}\\
N\epsilon \,\,\text{fixed}\hspace*{-80pt}
\end{array}
}
\dot{{\cal G}}_{\DIS (F_2)}^{(r)}\left(\epsilon,N\right)&=
\int_0^{\infty}\,\frac{dr}{r}\,{\rm e}^{-Nr}\,
\dot{{\cal F}}_{\cal J}\left(\epsilon/r\right)=
\dot{{\cal G}}_{\cal J}\left(N\epsilon\right)
\\&=
\left(1+\frac{N\epsilon}{2}-\frac{(N\epsilon)^2}{4}\right)\,{\rm Ei}(1,N\epsilon)
+\left(\frac{N\epsilon}{4}-\frac{3}{4}\right)\,{\rm e}^{-N\epsilon}
\,.
\end{split}
\end{align}
Let us turn now to the virtual corrections.
Considering the same limit the virtual pieces ${\cal G}^{(v)}\left(\epsilon\right)$ as well as its derivative,
$\dot{{\cal G}}^{(v)}\left(\epsilon\right)$ diverge. Indeed for $\epsilon\rightarrow0$ one gets,
using the explicit expressions in \cite{DMW}:
\begin{align}
\label{V_s}
\begin{split}
{\cal G}^{(v)}_{\DIS (F_2)}\left(\epsilon\right)
&\simeq -\frac{1}{2}\ln^2\left(\epsilon\right)-\frac{3}{2}\ln\left(\epsilon\right)-\frac{\pi^2}{3}-\frac{7}{4}
\end{split}
\end{align}
and thus
\begin{align}
\label{dot-cal-G-v-lim}
\lim_{\begin{array}{l}
N\rightarrow \infty \hspace*{-35pt}\\
N\epsilon \,\,\text{fixed}\hspace*{-80pt}
\end{array}
}
\dot{{\cal G}}_{\DIS (F_2)}^{(v)}\left(\epsilon\right)\simeq
-\ln\left(N\right)+\ln\left(N\epsilon\right)+\frac{3}{2}\,+\,{\cal O}(1/N)
\,.
\end{align}
Next, consider the combination $ \dot{\cal G}_{\DIS (F_2)}(\epsilon,N)-\dot{\cal G}_{\DIS (F_2)}(0,N)$ occurring in the dispersive representation Eq.~(\ref{finite-r-DIS-N}) and decompose it as
\begin{equation}
\label{split-G}
\dot{\cal G}_{\DIS (F_2)}(\epsilon,N)-\dot{\cal G}_{\DIS (F_2)}(0,N)=\dot{{\cal G}}_{\DIS (F_2)}^{(r)}\left(\epsilon,N\right)+\left(\dot{{\cal G}}_{\DIS (F_2)}^{(v)}\left(\epsilon\right)-\dot{\cal G}_{\DIS (F_2)}(0,N)\right)\,.
\end{equation}
Using now Eq.~(4.43) in \cite{DMW}, one gets
\begin{equation}
\label{dot-G-0}
\lim_{N\rightarrow\infty}\dot{\cal G}_{\DIS (F_2)}(0,N)\simeq -\ln N+\frac{3}{4}-\gamma_E\,,
\end{equation}
and thus, using Eq.~(\ref{dot-cal-G-v-lim}):
\begin{align}
\label{dot-cal-G-v-G-N-lim}
\lim_{\begin{array}{l}
N\rightarrow \infty \hspace*{-35pt}\\
N\epsilon \,\,\text{fixed}\hspace*{-80pt}
\end{array}
}
\left(\dot{{\cal G}}_{\DIS (F_2)}^{(v)}\left(\epsilon\right)-\dot{\cal G}_{\DIS (F_2)}(0,N)\right)=\ln\left(N\epsilon\right)+\frac{3}{4}+\gamma_E
\,.
\end{align}
Using Eq.~(\ref{dot-cal-G-N-lim}) we deduce:
\begin{align}
\begin{split}
\label{lim-finite-N-char}
\lim_{\begin{array}{l}
N\rightarrow \infty \hspace*{-35pt}\\
N\epsilon \,\,\text{fixed}\hspace*{-80pt}
\end{array}
}
\left(\dot{\cal G}_{\DIS (F_2)}(\epsilon,N)-\dot{\cal G}_{\DIS (F_2)}(0,N)\right)&=\dot{{\cal G}}_{\cal J}\left(N\epsilon\right)+\ln\left(N\epsilon\right)+\frac{3}{4}+\gamma_E\\&\hspace*{-120pt}
=
\left(1+\frac{N\epsilon}{2}-\frac{(N\epsilon)^2}{4}\right)\,{\rm Ei}(1,N\epsilon)
+\left(\frac{N\epsilon}{4}-\frac{3}{4}\right)\,{\rm e}^{-N\epsilon}
+\ln\left(N\epsilon\right)+\frac{3}{4}+\gamma_E
\,,
\end{split}
\end{align}
and, comparing with Eq.~(\ref{G_r+v}), we find:
\begin{align}
\label{lim-finite-N-char-2}
\lim_{\begin{array}{l}
N\rightarrow \infty \hspace*{-35pt}\\
N\epsilon \,\,\text{fixed}\hspace*{-80pt}
\end{array}
}
\left(\dot{\cal G}_{\DIS (F_2)}(\epsilon,N)-\dot{\cal G}_{\DIS (F_2)}(0,N)\right)=\Delta\dot{\cal G}_{\cal J}^{(r+v)}\left(N\epsilon\right)\,.
\end{align}
Thus, as announced, the `jet' Sudakov  characteristic function, $\Delta\dot{\cal G}_{\cal J}^{(r+v)}\left(N\epsilon\right)$, is identified as the large--$N$ limit with fixed $N \epsilon$ of the full finite--$N$ characteristic function of $F_2$.
Note also that, taking the derivative of Eq.~(\ref{lim-finite-N-char}), one gets \cite{Grunberg:2006gd}
\begin{align}
\label{lim-finite-N-dd-char}
\lim_{\begin{array}{l}
N\rightarrow \infty \hspace*{-35pt}\\
N\epsilon \,\,\text{fixed}\hspace*{-80pt}
\end{array}
}
\ddot{\cal G}_{\DIS (F_2)}(\epsilon,N)=\ddot{{\cal G}}_{\cal J}\left(N\epsilon\right)-1
\,,
\end{align}
where $-1$ is the virtual contribution. Finally, taking the limit $N\rightarrow\infty$ inside the integrand of Eq.~(\ref{finite-r-DIS-N}) (which yields an ultraviolet divergent integral), and comparing with Eq.~(\ref{limit_Sud_exp_JS_disp}), one obtains \cite{Friot:2007fd}:
\begin{align}
\label{lim-finite-N-EJ}
\lim_{\begin{array}{l}
N\rightarrow \infty \hspace*{-35pt}\\
Q^2/N \,\,\text{fixed}\hspace*{-80pt}
\end{array}
}
\left.\widetilde{K}_{\DIS (F_2)}(N,Q^2)\right\vert_{\rm large \,\,\beta_0}=\left.E_{\cal J}(Q^2/N)\right\vert_{\rm large \,\,\beta_0}
\,.
\end{align}

\vspace*{10pt}
\noindent
\underline{Drell--Yan}\\
In a similar way the finite--$N$ characteristic function in the Drell--Yan case
\begin{equation}
\label{finite_N_char_DY}
{\cal G}_{\DY}\left(\epsilon,N\right)=\int_0^1dx\,x^{N-1}{\cal F}_{\DY}(\epsilon,\tau)
\end{equation}
appears in the moment--space equivalent of (\ref{finite-r-DY}):
\begin{align}
  \label{finite-N-DY}
  \left.\widetilde{K}_{\DY}(\tau,Q^2)\right
  \vert_{\rm large \,\,\beta_0}\,
  =
  &\,\frac{C_F}{\beta_0}\,\int_0^{\infty}\frac{d\mu^2}{\mu^2}
  \rho_V(\mu^2)  \left[\left(\dot{\cal G}_{\DY}(\mu^2/Q^2,N)-\dot{\cal G}_{\DY}(0,N)\right)\right]\,.
\end{align}
Splitting the characteristic function into the real and virtual contributions we have:
\begin{equation}
\label{finite_N_char_r_v_DY}
{\cal G}_{\DY}\left(\epsilon,N\right)={\cal G}_{\DY}^{(r)}\left(\epsilon,N\right)
+{\cal G}_{\DY}^{(v)}\left(\epsilon\right)\,.
\end{equation}
For the real emission part, the large--$N$ limit with fixed $N^2\epsilon$ exists, and yields:
\begin{align}
\label{cal-F-DY-N-lim}
\lim_{\begin{array}{l}
N\rightarrow \infty \hspace*{-35pt}\\
N^2\epsilon \,\,\text{fixed}\hspace*{-80pt}
\end{array}
}
{\cal G}_{\DY}^{(r)}\left(\epsilon,N\right)=
2\int_0^{\infty}\,\frac{dr}{r}\,{\rm e}^{-Nr}\,
{\cal F}_{{\cal S}_{\DY}}\left(\epsilon/r^2\right)\equiv
2\,{\cal G}_{{\cal S}_{\DY}}\left(N^2\epsilon\right)
\,,
\end{align}
which is the analogue of the momentum space relation, Eq.~(\ref{Fr-x-scaling-DY}).
Taking a derivative we get:
\begin{align}
\label{dot-cal-F-DY-N-lim}
\lim_{\begin{array}{l}
N\rightarrow \infty \hspace*{-35pt}\\
N^2\epsilon \,\,\text{fixed}\hspace*{-80pt}
\end{array}
}
\dot{{\cal G}}_{\DY}^{(r)}\left(\epsilon,N\right)=
2\int_0^{\infty}\,\frac{dr}{r}\,{\rm e}^{-Nr}\,
\dot{{\cal F}}_{{\cal S}_{\DY}}\left(\epsilon/r^2\right)=
2\,\dot{{\cal G}}_{{\cal S}_{\DY}}\left(N^2\epsilon\right)=2K_0\left(2N\sqrt{\epsilon}\right)\,
\,.
\end{align}
The virtual contribution to the characteristic function ${\cal G}_{\DY}^{(v)}\left(\epsilon\right)$
as well as its derivative, $\dot{{\cal G}}^{(v)}\left(\epsilon\right)$ diverge in this limit.
The small--$\epsilon$ expansion is~\cite{DMW}:
\begin{align}
\label{V_t}
\begin{split}
{\cal G}^{(v)}_{\DY}\left(\epsilon\right)
&\simeq -\frac{1}{2}\ln^2\left(\epsilon\right)-\frac{3}{2}\ln\left(\epsilon\right)+\frac{\pi^2}{6}-\frac{7}{4}\,,
\end{split}
\end{align}
so
\begin{align}
\label{dot-cal-G-v-DY-lim}
\lim_{\begin{array}{l}
N\rightarrow \infty \hspace*{-35pt}\\
N^2\epsilon \,\,\text{fixed}\hspace*{-80pt}
\end{array}
}
\dot{{\cal G}}_{\DY}^{(v)}\left(\epsilon\right)\simeq
-2\ln\left(N\right)+\ln\left(N^2\epsilon\right)+\frac{3}{2}\,+\,{\cal O}(1/N)
\,.
\end{align}
Decomposing the difference entering (\ref{finite-N-DY}) according to
\begin{equation}
\label{split-G-DY}
\dot{\cal G}_{\DY}(\epsilon,N)-\dot{\cal G}_{\DY}(0,N)=\dot{{\cal G}}_{\DY}^{(r)}\left(\epsilon,N\right)+\left(\dot{{\cal G}}_{\DY}^{(v)}\left(\epsilon\right)-\dot{\cal G}_{\DY}(0,N)\right)\,,
\end{equation}
and using the relation
\begin{equation}
\label{dot-G-G-DY-0}
\dot{\cal G}_{\DY}(0,N)=2\,\dot{\cal G}_{\DIS (F_2)}(0,N)\,,
\end{equation}
which follows from the results in Sec. 4.6 of~\cite{DMW}
(and reflects the fact that the Drell--Yan cross section in the DIS scheme is
an infrared and collinear safe quantity) one gets:
\begin{equation}
\label{dot-G-DY-0}
\lim_{N\rightarrow\infty}\dot{\cal G}_{\DY}(0,N)\simeq -2\ln N+\frac{3}{2}-2\gamma_E\,+{\cal O}(1/N)
.
\end{equation}
Using Eq.~(\ref{dot-cal-G-v-DY-lim}), one thus finds
\begin{align}
\label{dot-cal-G-v-G-N-lim-DY}
\lim_{\begin{array}{l}
N\rightarrow \infty \hspace*{-35pt}\\
N\epsilon \,\,\text{fixed}\hspace*{-80pt}
\end{array}
}
\left(\dot{{\cal G}}_{\DY}^{(v)}\left(\epsilon\right)-\dot{\cal G}_{\DY}(0,N)\right)=\ln\left(N^2\epsilon\right)+2\,\gamma_E
\,,
\end{align}
which implies
\begin{align}
\label{lim-finite-N-char-DY}
\begin{split}
\lim_{\begin{array}{l}
N\rightarrow \infty \hspace*{-35pt}\\
N^2\epsilon \,\,\text{fixed}\hspace*{-80pt}
\end{array}
}
\left(\dot{\cal G}_{\DY}(\epsilon,N)-\dot{\cal G}_{\DY}(0,N)\right)&=2\left(\dot{\cal G}_{\cal S_{\DY}}\left(N^2\epsilon\right)+\frac{1}{2}\ln\left(N^2\epsilon\right)+\gamma_E\right)
\\&
=2\bigg(K_0\left(2N\sqrt{\epsilon}\right)+\ln\left(N\sqrt{\epsilon}\right)+\gamma_E\bigg)
\,,
\end{split}
\end{align}
and, comparing with Eq.~(\ref{G_r+v}), we find
\begin{align}
\label{lim-finite-N-char-DY1}
\lim_{\begin{array}{l}
N\rightarrow \infty \hspace*{-35pt}\\
N^2\epsilon \,\,\text{fixed}\hspace*{-80pt}
\end{array}
}
\left(\dot{\cal G}_{\DY}(\epsilon,N)-\dot{\cal G}_{\DY}(0,N)\right)=2\Delta\dot{\cal G}_{\cal S_{\DY}}^{(r+v)}\left(N^2\epsilon\right)\,.
\end{align}
Thus, as announced the large--$N$ limit of the full finite--$N$ characteristic function with fixed $N\epsilon^2$ reproduces the Sudakov characteristic function $\Delta\dot{\cal G}_{\cal S_{\DY}}^{(r+v)}\left(N^2\epsilon\right)$.
Note that in accordance with the discussion following Eq.~(\ref{DY_kernel_Mink}) above, the explicit result in Eq.~(\ref{lim-finite-N-char-DY}) is identical to the function appearing in Eq.~(54) in \cite{Laenen:2000ij} at $b=0$.
Note also that upon taking one derivative\footnote{Note that $\ddot{\cal G}_{\cal S_{\DY}}\left(N^2\epsilon\right)$ does not satisfy the analogue of the Laplace representation Eq.~(\ref{G_def}), owing to the singular behavior of the momentum space characteristic function in the Drell--Yan case.} of Eq.~(\ref{lim-finite-N-char-DY}) one obtains~\cite{Grunberg:2006gd}
\begin{align}
\label{lim-finite-N-dchar-DY}
\lim_{\begin{array}{l}
N\rightarrow \infty \hspace*{-35pt}\\
N^2\epsilon \,\,\text{fixed}\hspace*{-80pt}
\end{array}
}
\ddot{\cal G}_{\DY}(\epsilon,N)=2\,\ddot{\cal G}_{\cal S_{\DY}}\left(N^2\epsilon\right)-1
\,.
\end{align}
where $-1$ is the virtual contribution. Finally, taking the large--$N$ limit inside the integral in Eq.~(\ref{finite-N-DY}) (which is yields an ultraviolet divergent integral) and comparing with
Eq.~(\ref{limit_Sud_exp_JS_disp}) above one finds~\cite{Friot:2007fd}:
\begin{align}
\label{lim-finite-N-ES}
\lim_{\begin{array}{l}
N\rightarrow \infty \hspace*{-35pt}\\
Q^2/N^2 \,\,\text{fixed}\hspace*{-80pt}
\end{array}
}
\left.\widetilde{K}_{\DY}(N,Q^2)\right\vert_{\rm large \,\,\beta_0}=\left.E_{\cal S_{\DY}}(Q^2/N^2)\right\vert_{\rm large \,\,\beta_0}
\,.
\end{align}

\section{Renormalons and Landau singularities: example\label{sec:Mink_integral}}

Let us consider for example the power terms distinguishing between
(\ref{Dispersive_Exponent_Laplace_NLL}) and (\ref{Exponent_Laplace_Borel2}) in the case of
inclusive B decays under the assumption that the effective charges are given by (\ref{aSJ_one_loop}), namely they do have Landau singularities. We obtain:
\begin{align}
\label{power_corrections}
\begin{split}
&\left.\ln \Big( \overline{\rm Sud}(m^2,N) \Big)\right\vert_{\rm Dispersive}-
\left.\ln \Big( \overline{\rm Sud}(m^2,N) \Big)\right\vert_{\rm Borel}
\\&
=\left.\frac{C_F}{\beta_0}\int_0^{\mu^2/m^2}\frac{d\epsilon}{\epsilon} \,\Big[
\left(\dot{\cal G}_{\cal J}\left(\epsilon N\right)-
\dot{\cal G}_{\cal J}\left(\epsilon\right)\right)
-
\left(\dot{\cal G}_{\cal S}\left(\epsilon N^2\right)-
\dot{\cal G}_{\cal S}\left(\epsilon\right)\right)\Big]\right\vert_{\mu^2=-\Lambda_{\GB}^2}
\\
&=
\frac{C_F}{2\beta_0}\left[\underbrace{\frac{1}{2_{\,_{\,}}} (N^2-1)}_{\rm soft}\underbrace{- N \ln(N)-(N-1) \left(\ln\left(\frac{\Lambda_{\GB}^2}{m^2}\right)+\gamma_E-5\right)}_{\rm jet}\right] \left(\frac{\Lambda_{\GB}^2}{m^2}\right)+\,{\cal O}\left(\left(\frac{\Lambda_{\GB}^2}{m^2}\right)^2\right)
\\&
+\frac{i\pi C_F}{2\beta_0}\,
\left\{\underbrace{
(N-1)  \left(\frac{\Lambda_{\GB}^2}{m^2}\right)^{1/2}\!\!}_{\rm soft}
\underbrace{-(N-1)  \left(\frac{\Lambda_{\GB}^2}{m^2}\right)}_{\rm jet}
\underbrace{-\frac{1}{18}(N^3-1)  \left(\frac{\Lambda_{\GB}^2}{m^2}\right)^{3/2}\!\!}_{\rm soft}
+\,{\cal O}\left(\left(\frac{\Lambda_{\GB}^2}{m^2}\right)^2\right)\right\},
\end{split}
\end{align}
where the first few terms in the expansion have been computed explicitly using (\ref{WJ_IR}) and
(\ref{WS_IR}). The first line summarizes the leading \emph{real} power term that contributes to the (unambiguous) difference between (\ref{Dispersive_Exponent_Laplace_NLL}) and the \emph{Principal Value} of
the Borel sum (\ref{Exponent_Laplace_Borel2}); these terms are not related with renormalons. The second line summarizes the \emph{imaginary} power terms which represent the \emph{ambiguity} of the Borel sum owing to infrared renormalons. They provide an indication of the potential size of genuine non-perturbative effects. Note that these terms originate in the \emph{non-analytic terms} in the small $\mu^2$ expansion of the characteristic function. Obviously, there is one--to--one correspondence between these non-analyticities and the Borel singularities in (\ref{Exponent_Laplace_Borel2}). The latter have been discussed in~\cite{Gardi:2004ia,Andersen:2005bj,Andersen:2005mj,Andersen:2006hr,Gardi:2006jc}.

It is straightforward to identify the origin of the different terms in (\ref{power_corrections}), as indicated below each term:
terms that scale at large $N$ as $N\Lambda/m$ are associated uniquely with the `soft' (quark distribution) function, while those scaling as $N\Lambda^2/m^2$ with the jet function. It is evident that both these classes of corrections contribute to both the real and the imaginary parts of (\ref{power_corrections}). In practice, power corrections on the soft scale are important while those on the jet--mass scale can usually be neglected. Let us therefore consider the former in some detail: \begin{itemize}
\item{} The leading power term, ${\cal O}(\Lambda N /m)$, represents the $u=1/2$ renormalon ambiguity.  This ambiguity has been shown~\cite{Gardi:2004ia} to cancel in the physical spectra with the ambiguity in defining the $b$ quark pole mass, or equivalently $\bar{\Lambda}=M_B-m_b$, where $M_B$ is the meson mass and $m_b$ is the b quark pole mass.\\
    In Refs.~\cite{Gardi:2004ia,Andersen:2005bj,Andersen:2005mj,Andersen:2006hr,Gardi:2006jc}
    the Principal Value Borel sum was used to define both the Sudakov factor and the pole mass, eliminating the ambiguity.
    In Appendix~\ref{sec:pole_mass} we provide a definition of the pole mass that suites the dispersive approach.
\item{} As a consequence of the Landau singularity in $a_{\cal S}^{\Eucl}(k^2)$ assumed above\footnote{This correction does not appear if $a_{\cal S}^{\Eucl}(k^2)$ has a causal analyticity structure,
    see \cite{Grunberg:1998ix} and the discussion in Sec.~\ref{sec:PC} above.},
 a term of ${\cal O}\left((\Lambda N/m)^2\right)$ appears as a real contribution to the difference between (\ref{Dispersive_Exponent_Laplace_NLL}) and the Principal Value of the Borel sum (\ref{Exponent_Laplace_Borel2}). This is a parametrically large correction that directly influences the width of the computed spectra. It should be explicitly subtracted out when using the dispersive technique: it cannot be absorbed into the definition of any non--perturbative parameter, as there is no corresponding renormalon ambiguity.
\item{} Finally, power terms of ${\cal O}\left((\Lambda N/m)^p\right)$ where $p$ is an integer $p\geq3$,  appear as renormalon ambiguities.
    Corresponding non-perturbative corrections are expected to appear as a consequence of the interaction between the $b$ quark and the light degrees of freedom in the meson~\cite{Gardi:2004ia}, distinguishing the quark distribution is the meson from that in an on-shell $b$ quark. These power corrections have been parametrized in \cite{Andersen:2006hr}. As far as these corrections are concerned the dispersive representation of the exponent (or its quark distribution function part) provides an alternative definition of the perturbative component, which is a priori as good as the definition based on the Principal Value Borel sum.
\end{itemize}
To conclude, we have seen that
\begin{itemize}
\item{}
Genuine non-perturbative effects are related to renormalon singularities. As usual, the same conclusions with regards to such corrections can be reached using the dispersive and the Borel formulations.
As far as the regularization of the renormalons is concerned the two definitions are equivalent in principal: they differ by power corrections of the same parametric form one needs to introduce in order to parametrize genuine non-perturbative effects. Obviously, the non-perturbative parameters need to be defined according to the regularization used.
\item{}
Quite independently of this, when the effective charges have Landau singularities the dispersive integral differs from the Borel sum by an additional set of computable power terms that are not related with renormalons; these include parametrically important contributions, ${\cal O}(\Lambda^2N^2/m^2)$.
\end{itemize}

\section{Definition of the pole mass in the dispersive approach\label{sec:pole_mass}}

In Refs.~\cite{Gardi:2004ia,Andersen:2005bj,Andersen:2005mj,Andersen:2006hr,Gardi:2006jc}
    the Principal Value Borel sum was used to define both the Sudakov factor and the pole mass, eliminating the ambiguity (see e.g. Eq. (4.4) in \cite{Andersen:2005bj}).    The use of the dispersion integral to define the Sudakov exponent requires a corresponding regularization of the pole mass (or $\bar{\Lambda}$). Importantly, the relevant mass, which we shall refer to as the ``dispersive pole mass'', $m_b^{\rm disp.}$, is uniquely fixed by the formalism, and similarly to the Principal Value pole mass, it can be accurately determined based on some short distance mass (e.g. the $\overline{\rm MS}$ mass). The relation between the ``dispersive pole mass'' and the Principal Value pole mass becomes transparent upon recalling the cancellation mechanism of the $u=1/2$ renormalon.
    For concreteness let us refer to the $\bar{B}\to X_s \gamma$ example.
    
    By combining the Laplace weight of the inverse--Mellin transform in (\ref{inv_Mellin}) with the Sudakov factor (\ref{Dispersive_Exponent_Laplace_NLL}):
     \begin{align}
    \label{u12_renormalon_cancellation_disp}
    \begin{split}
     \left.\left(\frac{2E_{\gamma}}{m_b}\right)^{N-1}
     \,{\rm Sud}(m^2,N)\,\,\right\vert_{u=1/2} &\simeq  \left(\frac{2E_{\gamma}}{M_B}\right)^{N-1}\\
     &\hspace*{-60pt}\times \exp\left\{-\frac{N\bar{\Lambda}_{\rm disp.}}{m_b}-\frac{C_F\pi}{4}\int_0^{\mu_I^2}\frac{d\mu^2}{\mu^2}
    \,a_{{\cal S}_{\QD}}^{\Mink}(\mu^2) \sqrt{\frac{N^2 \mu^2}{m_b^2}}\right\}\,,
    \end{split}
    \end{align}
  where have used the relevant (square--root) term in the expansion of the characteristic function $\dot{\cal G}_{{\cal S}_{\QD}}(N^2{\epsilon})$  in (\ref{WS_IR}) and omitted all other terms, which are irrelevant for the $u=1/2$ renormalon. We consider the integration up to an arbitrary scale $\mu_I$ that should be above $\Lambda$; the specific scale choice will not affect the final result.
  
  The equivalent integral in the scheme--invariant Borel formulation can be obtained by inserting (\ref{eq:int_discontinuity_SJ}) into (\ref{u12_renormalon_cancellation_disp}), which yields:
      \begin{align}
    \label{u12_renormalon_cancellation_Borel}
    \begin{split}
     \left.\left(\frac{2E_{\gamma}}{m_b}\right)^{N-1}
     \,{\rm Sud}(m^2,N)\,\,\right\vert_{u=1/2} &\simeq  \left(\frac{2E_{\gamma}}{M_B}\right)^{N-1}\\
     &\hspace*{-177pt}\times \exp\left\{-\frac{N\bar{\Lambda}_{\PV}}{m_b}-\frac{\mu_I}{m_b}\frac{N\,C_F\pi}{2}\,\,{\rm PV}\int_0^{\infty}\frac{du}{1-2u} \left({\frac{\Lambda^2}{\mu_I^2}}\right)^u \,T(u)
     \,\frac{\sin \pi u}{\pi u}\, B\Big[a_{{\cal S}_{\QD}}^{\Eucl}\Big](u)
     \right\}\,,
    \end{split}
    \end{align}
 where we have chosen the Principal Value prescription for both $\bar{\Lambda}$ and the $u=1/2$ ambiguity in the quark distribution function.
 
The key point is that the object considered in (\ref{u12_renormalon_cancellation_disp}) and (\ref{u12_renormalon_cancellation_Borel}) is  unambiguous, so the relation between the two
    definitions of the pole mass can simply be read off comparing the two equations:
 \begin{align}
    \label{u12_disp_Borel}
    \begin{split}
    \bar{\Lambda}_{\rm disp.}-{\Lambda}_{\PV}&=m_b^{\rm PV}-m_b^{\rm disp.}\,=\,
    -m_b \frac{C_F\pi}{4}\int_0^{\mu_I^2}\frac{d\mu^2}{\mu^2}
    \,a_{{\cal S}_{\QD}}^{\Mink}(\mu^2) \sqrt{\frac{\mu^2}{m_b^2}}
    \\&+\mu_I \frac{C_F\pi}{2}\,\,{\rm PV}\int_0^{\infty}\frac{du}{1-2u} \left({\frac{\Lambda^2}{\mu_I^2}}\right)^u \,T(u)
     \,\frac{\sin \pi u}{\pi u}\, B\left[a_{{\cal S}_{\QD}}^{\Eucl}\right](u)\,.
    \end{split}
    \end{align}
    Moreover, the r.h.s. of  Eq.~(\ref{u12_disp_Borel}) does {\em not} depend on the cutoff $\mu_I$ (see Eq.~(B.19) in \cite{Grunberg:1998ix} and Sec.~3.2 in \cite{Gardi:1999dq}). We can thus set $\mu_I=\Lambda$ to get:
\begin{align}
    \label{u12_disp_Borel1}
    \begin{split}
    \bar{\Lambda}_{\rm disp.}-{\Lambda}_{\PV}&=m_b^{\rm PV}-m_b^{\rm disp.}\,=\,
   \Lambda\,\,\times\,\,\Bigg\{ - \frac{C_F\pi}{4}\int_0^{\Lambda^2}\frac{d\mu^2}{\mu^2}
    \,a_{{\cal S}_{\QD}}^{\Mink}(\mu^2) \sqrt{\frac{\mu^2}{\Lambda^2}}
    \\&
    + \frac{C_F\pi}{2}\,\,{\rm PV}\int_0^{\infty}\frac{du}{1-2u}  \,T(u)
     \,\frac{\sin \pi u}{\pi u}\, B\left[a_{{\cal S}_{\QD}}^{\Eucl}\right](u)\Bigg\}\,,
    \end{split}
    \end{align}
exhibiting the fact that this difference is a number of order $\Lambda$ which is independent of any hard scale such as $\mu_I$ or $m_b$.
Note that these equations are valid even if the Euclidean coupling has Landau singularities (such as the one loop coupling). The Minkowskian coupling is still infrared finite in this case, and the convergence of the Borel integral is ensured by the oscillations of the  factor ${\sin (\pi u)}/{\pi u}$.

It is also possible to give a \emph{Euclidean} version of the previous definition, \emph{iff} one assumes the Euclidean Sudakov effective charge is causal and infrared finite. It is straightforward to show that the analogues of Eqs. ~(\ref{u12_renormalon_cancellation_disp}), (\ref{u12_disp_Borel}) and (\ref{u12_disp_Borel1}) are:
  \begin{align}
    \label{u12_renormalon_cancellation_disp_Eucl}
    \begin{split}
     \left.\left(\frac{2E_{\gamma}}{m_b}\right)^{N-1}
     \,{\rm Sud}(m^2,N)\,\,\right\vert_{u=1/2,\Eucl} &\simeq  \left(\frac{2E_{\gamma}}{M_B}\right)^{N-1}\\
     &\hspace*{-60pt}\times \exp\left\{-\frac{N\bar{\Lambda}_{\rm disp.}}{m_b}-\frac{C_F}{4}\,\int_0^{\mu_I^2}\frac{dk^2}{k^2}
    \,a_{{\cal S}_{\QD}}^{\Eucl}(k^2) \sqrt{\frac{N^2 k^2}{m_b^2}}\right\}\,,
    \end{split}
    \end{align}
   \begin{align}
    \label{u12_disp_Borel_Eucl}
    \begin{split}
    \bar{\Lambda}_{\rm disp.}-{\Lambda}_{\PV}&=m_b^{\rm PV}-m_b^{\rm disp.}\,=\,
    -m_b\, \frac{C_F}{4}\,\int_0^{\mu_I^2}\frac{dk^2}{k^2}
    \,a_{{\cal S}_{\QD}}^{\Eucl}(k^2) \sqrt{\frac{k^2}{m_b^2}}
    \\&+\mu_I\, \frac{C_F}{2}\,{\rm PV}\int_0^{\infty}\frac{du}{1-2u} \left({\frac{\Lambda^2}{\mu_I^2}}\right)^u \,T(u)
     \, B\left[a_{{\cal S}_{\QD}}^{\Eucl}\right](u)\,,
    \end{split}
    \end{align}
  and,  setting  $\mu_I=\Lambda$, since  the r.h.s. of
  Eq.~(\ref{u12_disp_Borel_Eucl}) does {\em not} depend on the cutoff $\mu_I$ in the case of a causal coupling (see Eq.~(3.19) in \cite{Grunberg:1998ix}):
      \begin{align}
    \label{u12_disp_Borel_Eucl1}
    \begin{split}
    \bar{\Lambda}_{\rm disp.}-{\Lambda}_{\PV}&=m_b^{\rm PV}-m_b^{\rm disp.}\,=\,
    \Lambda\,\times \,\Bigg\{ -\frac{C_F}{4}\,\,\int_0^{\Lambda^2}\frac{dk^2}{k^2}
    \,a_{{\cal S}_{\QD}}^{\Eucl}(k^2) \sqrt{\frac{k^2}{\Lambda^2}}
    \\&+\, \frac{C_F}{2}\,\,\,{\rm PV}\int_0^{\infty}\frac{du}{1-2u}  \,T(u)
     \, B\left[a_{{\cal S}_{\QD}}^{\Eucl}\right](u)\Bigg\}\,.
    \end{split}
    \end{align}
Note also the convergence of the Borel integral has to be insured by oscillations in $B\left[a_{{\cal S}_{\QD}}^{\Eucl}\right](u)$ if $a_{{\cal S}_{\QD}}^{\Eucl}(k^2)$ has an infrared fixed point (similarly to the Minkowskian coupling above).
Finally, we note that, although Eqs.~(\ref{u12_renormalon_cancellation_disp}) and (\ref{u12_renormalon_cancellation_disp_Eucl}) define \emph{different} $u=1/2$ pieces of the Sudakov exponent, the corresponding $m_b^{\rm disp.}$ masses are the {\em same}: this result follows immediately comparing Eq.~(3.19) and (B.19) in Ref.~\cite{Grunberg:1998ix}.

\end{document}